\documentclass[floatfix,aps,prc,superscriptaddress,noeprint,preprint]{revtex4-1}
\usepackage{amsfonts}                            
\usepackage{amssymb}                             
\usepackage{amsmath}  
\usepackage{mathtools}                           
\usepackage{bm}
\usepackage{graphicx}                            
\usepackage{color}
\usepackage{latexsym}                            
\usepackage{dcolumn}                             
\usepackage[colorlinks=true,linkcolor=blue]{hyperref}
\usepackage{footnote}
\usepackage{slashed}
\usepackage{xfrac}
\usepackage{cleveref}
\crefname{figure}{Fig.}{Figs.}
\crefname{equation}{Eq.}{Eqs.}
\usepackage{natbib}
\usepackage{mathrsfs}
\usepackage{booktabs}
\usepackage{siunitx}

\newcommand\be{\begin{equation}}
\newcommand\ee{\end{equation}}
\newcommand\bea{\begin{eqnarray}}
\newcommand\eea{\end{eqnarray}}
\newcommand\bal{\begin{aligned}}
\newcommand\eal{\end{aligned}}
\newcommand\bes{\begin{subequations}}
\newcommand\ees{\end{subequations}}
\renewcommand{\Re}{\operatorname{Re}}
\renewcommand{\Im}{\operatorname{Im}}

\begin{document}
\title{Two-photon exchange from intermediate state resonances \\
in elastic electron-proton scattering}

\author{Jaseer Ahmed}
\thanks{Present address: Department of Physics, Shahjalal University of Science
and Technology, Sylhet-3114, Bangladesh.}
\affiliation{\mbox{Department of Physics and Astronomy,
  University of Manitoba}, Winnipeg, Manitoba, Canada R3T 2N2}
\author{P.~G.~Blunden}
\affiliation{\mbox{Department of Physics and Astronomy,
	University of Manitoba}, Winnipeg, Manitoba, Canada R3T 2N2}
\author{W.~Melnitchouk}
\affiliation{Jefferson Lab, Newport News, Virginia 23606, USA}

\begin{abstract} We use a recently developed dispersive approach to compute the
two-photon exchange (TPE) correction to elastic electron-proton scattering,
including contributions from hadronic $J^P=1/2^\pm$ and $3/2^\pm$ resonant
intermediate states below~1.8~GeV. For the transition amplitudes from the proton
ground state to the resonant excited states we employ new exclusive meson
electroproduction data from CLAS at $Q^2 \lesssim 5$~GeV$^2$, and we explore the
effects of both fixed and dynamic widths for the resonances. Among the resonant
states, the $N(1520)~\!3/2^-$ becomes dominant for $Q^2 \gtrsim 2$~GeV$^2$, with
a sign opposite to the comparably sized $\Delta(1232)~\!3/2^+$ contribution,
leading to an overall increase in the size of the TPE correction to the cross
section relative to the nucleon only contribution at higher $Q^2$ values. The
results are in good overall agreement with recent $e^+ p$ to $e^- p$ cross
section ratio and polarization transfer measurements, and provide compelling
evidence for a resolution of the electric to magnetic form factor ratio
discrepancy.
\end{abstract}

\date{\today}
\preprint{JLAB-THY-20-3212}
\maketitle

\section{Introduction}
\label{sec:intro}

Elastic electron--nucleon scattering has been one of the most indispensable
tools to probe the internal structure of nucleons through the determination of
their electromagnetic form factors. For many decades the proton's electric
($G_E(Q^2)$) and magnetic ($G_M(Q^2)$) elastic form factors have been measured
in unpolarized scattering experiments using the Rosenbluth
longitudinal-transverse (LT) separation technique~\cite{walker1994,
Andivahis1994, qattan2005}. These experiments found that the ratio $\mu_p\,
G_E/G_M$, where $\mu_p$ is the proton's magnetic moment, are consistent with 1
over a large range of the four-momentum transfer squared, $Q^2$, up to
8.83~GeV$^2$. More recently, measurements of the electric to magnetic form
factor ratio with significantly reduced uncertainties were performed at
Jefferson Lab using the polarization transfer (PT) technique~\cite{jones2000g,
gayou2002,punjabi2005, puckett2010, puckett2012}. These experiments found a
linear fall-off of the ratio $\mu_p\, G_E/G_M$ from 1 with increasing $Q^2$ in
the range up to 8.5~GeV$^2$.

Analysis of the LT separation electron scattering data has traditionally been
performed within the one-photon exchange (OPE) approximation.  The electric to
magnetic form factor ratio discrepancy motivated studies of hadron
structure-dependent two-photon exchange (TPE) radiative corrections, and it was
generally believed that the problem would be resolved with the inclusion of
these effects~\cite{blunden2003, guichon2003}. Subsequent years have seen a
growing sophistication in the theoretical efforts that have been made to better
understand the TPE phenomena using various approaches. These have included using
hadronic models to compute the real part of the TPE amplitude through loop
integrals with (on-shell) transition form factors~\cite{blunden2005,
kondratyuk2005, kondratyuk2007, zhou2015}, dispersive
approaches~\cite{borisyuk2008, borisyuk2012, borisyuk2015,
tomalak2015subtracted, blunden2017}, use of generalized parton distributions to
model the high-energy behavior of the intermediate state hadrons at the quark
level~\cite{chen2004, afanasev2005}, and QCD factorization
approaches~\cite{borisyuk2009,kivel2009}.

The use of hadronic degrees of freedom can be considered as a reasonable
approximation for low to moderate values of $Q^2 \lesssim 5$~GeV$^2$, where
hadrons are expected to retain their identity. However, for excited intermediate
states of higher spin, such as the $\Delta$ isobar, in the forward angle
limit~\cite{blunden2017} the direct loop-integral approach gives rise to
unphysical divergences in the TPE amplitude at forward angles. This problem of
unphysical behavior can be resolved using the dispersive method described in
Refs.~\cite{gorchtein2007, borisyuk2008, borisyuk2014, borisyuk2015,
tomalak2015subtracted, tomalak2017, blunden2017}, where the on-shell form
factors are used explicitly to calculate the imaginary part of the TPE amplitude
from unitarity, with the real part then obtained from a dispersion integral.

In this work we follow the dispersive approach for resonant intermediate states
developed in Ref.~\cite{blunden2017}. Unlike previous calculations which made
use of the narrow resonance approximation, here we allow a Breit-Wigner shape
with a nonzero width for each individual resonance, with either a fixed width or
a dynamical width that depends on the final state hadron mass. Furthermore, in
addition to the $\Delta(1232)~\!3/2^+$ resonance, we also compute the TPE
contribution from all the established $J^P=1/2^\pm$ and $3/2^\pm$ states below
1.8~GeV, including the
    $N(1440)~\!1/2^+$ Roper resonance,
    $N(1520)~\!3/2^-$,
    $N(1535)~\!1/2^-$, 
    $\Delta(1620)~\!1/2^-$, 
    $N(1650)~\!1/2^-$, 
    $\Delta(1700)~\!3/2^-$, 
    $N(1710)~\!1/2^+$,
and 
    $N(1720)~\!3/2^+$
resonances. With the exception of the $\Delta(1232)~\!3/2^+$, for which we use
the fit by Aznauryan and Burkert~\cite{blunden2017,aznauryan2012}, for the
resonance electrocouplings at the hadronic vertices we use the most recent
helicity amplitudes extracted from the analysis of CLAS meson electroproduction
data~\cite{HillerBlin:2019hhz, mokeev2012, mokeev2009}.

We begin in Sec.~\ref{sec.formalism} by describing the kinematics of the elastic
$e^- p$ scattering process, both in the one- and two-photon exchange
approximations. Details of the TPE calculations for the resonance states are
presented in Sec.~\ref{sec.GammaH}, where we summarize the formal relations for
the resonance transition current operators in terms of the form factors $G_i$
(Sec.~\ref{ssec.GammaG}), and relate the form factors to the helicity amplitudes
(Sec.~\ref{ssec.GammaAh}). The dispersive method and its practical
implementation are discussed in Sec.~\ref{ssec.dispersive}, where we express the
TPE amplitudes and cross sections in terms of the generalized TPE form factors.
Numerical results for the TPE corrections are presented in
Sec.~\ref{sec.results}, where the effects of finite widths
(Sec.~\ref{ssec.width}) and the role of spin, isospin and parity of the
intermediate states (Sec.~\ref{ssec.spin}) are discussed. Comparisons with
experimental observables sensitive to TPE contributions are made in
Sec.~\ref{sec.observables}. Finally, conclusions and future outlook of our work
are presented~in~Sec.~\ref{sec.conclusion}.

\section{Electron-Proton Elastic Scattering}
\label{sec.formalism}

\begin{figure}[t]
\graphicspath{{Images/}}
\includegraphics[width=5.5cm]{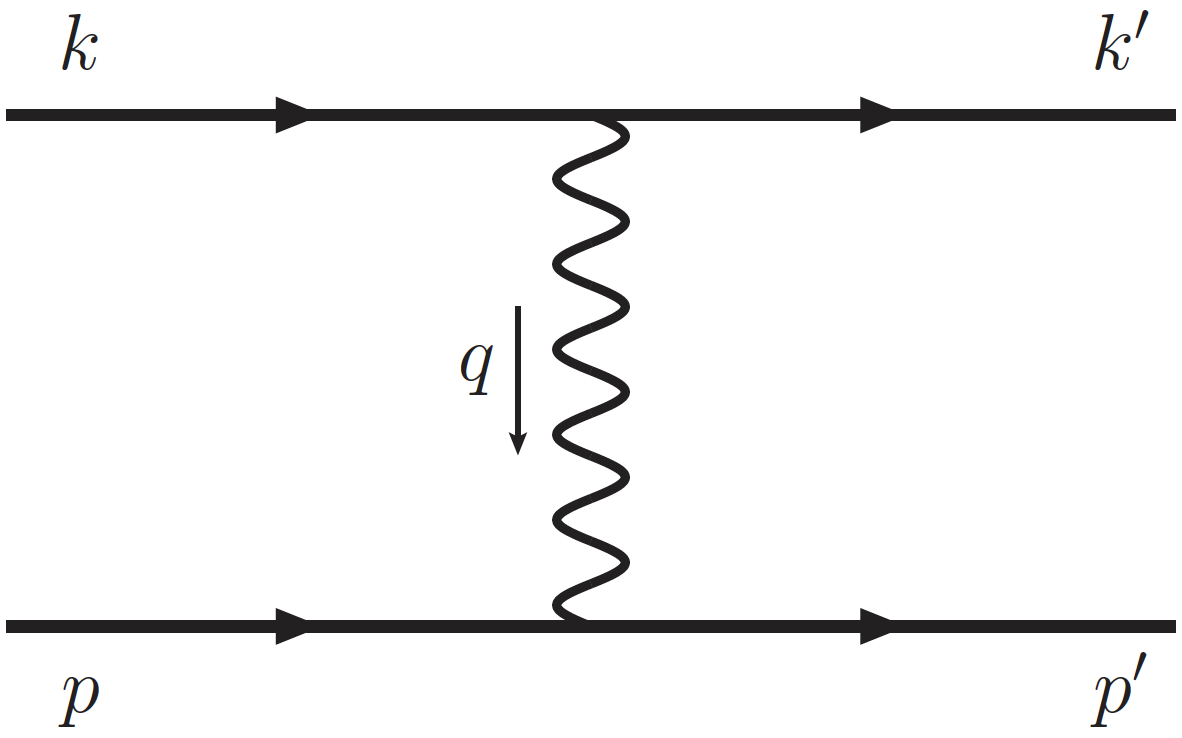} 
\hspace*{2cm}
\includegraphics[width=5.5cm]{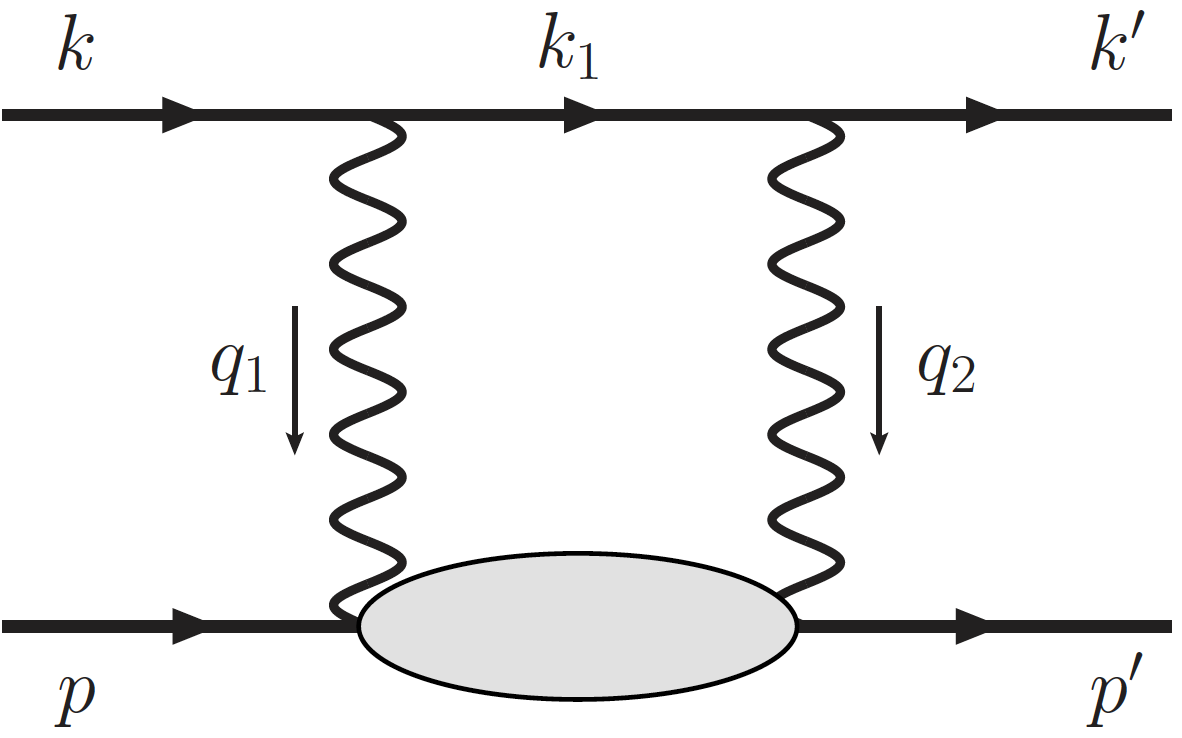}
\caption{Elastic scattering of an electron (four-momentum $k$) from a proton
($p$) to a final state electron ($k'$) and recoil proton ($p'$), with
$q=k-k'=p'-p$ the four-momentum transfer, in the Born approximation (left panel)
and for the two-photon exchange process (right panel). Only the $s$-channel box
diagram is shown for the TPE correction, in which the two photons carry momenta
$q_1$ and $q_2$, and the crossed box contribution can be obtained using the $s
\to u$ crossing symmetry.}     
\label{fig.tpe}
\end{figure}

The kinematics of the elastic electron--proton scattering process are shown in
Fig.~\ref{fig.tpe} for the one-photon exchange (or Born) approximation and for
the TPE contribution. Here an electron with four-momentum $k=(E,\bm{k})$ is
scattered from a proton initially at rest, $p=(M,0)$ in the target rest frame,
to an electron in the final state with four-momentum $k'=(E',\bm{k'})$. The
transferred four-momentum from the electron to the proton is $q=k-k'$, and the
proton recoils with four-momentum $p'=p+q$. For the TPE diagram, two virtual
photons of four-momenta $q_1$ and $q_2$ are exchanged, with $q=q_1+q_2$. 

\subsection{One-photon exchange}
\label{ssec.ope}

For the one-photon exchange approximation, the amplitude ${\cal M}_{\gamma}$ for
scattering an electron from a proton can be written as~\cite{arrington2011}
\be
{\cal M}_\gamma = e^2\, j_\mu\, \dfrac{1}{Q^2}\, J^{\mu},
\label{Mgamma}
\ee
where $e$ is the charge of the proton, and the total four-momentum transfer
squared $Q^2$ can be written in terms of the electron energy $E$ and scattering
angle $\theta$ as $Q^2 \equiv -q^2 = 4 E E' \sin^2(\theta/2)$. The electron
transition current is given by $j_\mu = \bar{u}_e(k')\, \gamma_\mu\, u_e(k)$,
while the proton transition current is $J^\mu = \bar{u}_N(p')\, \Gamma^\mu(q)\,
u_N(p)$, where the hadronic current operator $\Gamma^\mu$ is parameterized
through form factors that take into account the proton's internal structure. For
on-shell particles, current conservation at the hadron vertex allows for two
independent Lorentz vectors, so the hadronic current operator is typically
parameterized in terms of the Dirac $F_1$ and Pauli $F_2$ form factors,
\be
\Gamma^{\mu}(q) = F_1(Q^2)\gamma^{\mu} + F_2(Q^2)\dfrac{\textit{i}
\sigma^{\mu\nu} q_{\nu}}{2M},
\label{Mgamma2}
\ee 
where $M$ is the proton mass. It is often convenient to use the Sachs electric
and magnetic form factors $G_E$ and $G_M$, which are defined as linear
combinations of the Dirac and Pauli form factors,
\be
G_E(Q^2) = F_1(Q^2) - \tau F_2(Q^2),\ \ \ \ \ \ \
G_M(Q^2) = F_1(Q^2) + F_2(Q^2),
\ee
where $\tau = Q^2/4M^2$. The differential cross section for single photon
exchange is proportional to the square of the scattering amplitude ${\cal
M}_\gamma$, and can be expressed in terms of the electric and magnetic form
factors as
\bea
\bigg(\dfrac{d\sigma}{d\Omega}\bigg)_0
&=& \left(\dfrac{\alpha E'}{4M Q^2 E}\right)^2 
    \left|{\cal M}_{\gamma}\right|^2\,
 =\, \dfrac{\sigma_{\rm Mott}}{\varepsilon (1+\tau)}\,
    \sigma_R^{\rm{Born}},
\label{eq.ope}
\eea
where $\alpha = e^2/4\pi$ is the electromagnetic fine structure constant, and 
\be
\varepsilon
= \left[1+2\left(1+\tau\right)\tan^2\left(\theta/2 \right)\right]^{-1}
\ee
is the virtual photon polarization. In Eq.~(\ref{eq.ope}) the reduced cross
section $\sigma_R^{\rm{Born}}$ is given by
\be
\sigma_R^{\rm{Born}}
= \varepsilon G_E^2(Q^2) + \tau G_M^2(Q^2),
\label{eq.sigred}
\ee
and the Mott cross section,
\be
\sigma_{\rm Mott}
= \dfrac{4\alpha^2 E'^3\cos^2(\theta/2)}{E Q^4},
\ee
gives the cross section for scattering an electron from a point target.

\subsection{Two-photon exchange}
\label{ssec.tpe}

The two-photon exchange amplitude, ${\cal M}_{\gamma \gamma}$, is the sum of
contributions from the box diagram of Fig.~\ref{fig.tpe} and the corresponding
crossed-box diagram (not shown),
\be
{\cal M}_{\gamma \gamma}
= {\cal M}_{\gamma \gamma}^{\rm{box}}
+ {\cal M}_{\gamma \gamma}^{\rm{xbox}}.
\ee
In general, the box diagram amplitude ${\cal M}_{\gamma \gamma}^{\rm{box}}$ can
be written as an integral over loop momenta $q_1$ or $q_2$ of the exchanged
photons~\cite{arrington2011},   
\be
{\cal M}_{\gamma \gamma}^{\rm{box}}
= -i e^4 \int\!\dfrac{d^4 q_1}{(2\pi)^4}
    \dfrac{L_{\mu\nu} H^{\mu\nu}}{(q_{1}^2-\lambda^2)(q_2^2-\lambda^2)},
\label{eq.M}
\ee
where an infinitesimal photon mass $\lambda$ is introduced to regulate infrared
divergences. The leptonic tensor $L_{\mu\nu}$ in Eq.~(\ref{eq.M}) is given by
\be
L_{\mu\nu} = \bar{u}_e(k') \gamma_{\mu} S_F(k_1,m_e) \gamma_{\nu} u_e(k),
\ee
where $k_1=k-q_{1}$ is the intermediate lepton four-momentum, $m_e$ is the
electron mass (which can in practice be taken to zero at the kinematics
considered here), and $S_F$ is the electron propagator defined by
\be
S_F(k_1,m_e) = \dfrac{(\slashed{k}_1 + m_e)}{k_1^2 - m_e^2 + i 0^+}.
\ee
The hadronic tensor $H^{\mu\nu}$ can be expressed as
\be
H^{\mu\nu}
= \bar{u}_N(p')\,
  \Gamma_{R\to \gamma N}^{\mu\alpha}(p_R,-q_2)\,
  S_{\alpha\beta}(p_R,W)\,
  \Gamma_{\gamma N\to R}^{\beta\nu}(p_R,q_1)\,
  u_N(p),
\ee
in terms of the transition operators $\Gamma^{\beta\nu}_{\gamma N\to R}$ and
$\Gamma^{\mu\alpha}_{R\to \gamma N}$ between the initial nucleon and
intermediate resonance $R$ states, where $p_R = p + q_1 = p' - q_2$ is the
four-momentum of the resonance and $W$ its (in principle running) mass.

For spin $1/2$ baryon intermediate states the propagator
$S_{\alpha\beta}(p_R,W)$ reduces to the usual spin-$1/2$ propagator,
\be
S_{\alpha \beta}(p_{R},W)
= \delta_{\alpha\beta}
  \dfrac{(\slashed{p}_{R}+W)}{p_{R}^2-W^2+i 0^+}
= \delta_{\alpha\beta}\, S_F(p_{R},W),
\ee
for a particle with mass $W$. 
The hadronic tensor for spin-1/2 baryons can then be written
\be
H^{\mu \nu}
= \bar{u}_N(p')\,
  \Gamma_{R \to \gamma N}^\mu(p_R, -q_2)\,
  S_F(p_R,W)\,
  \Gamma_{\gamma N \to R}^\nu(p_R, q_1)\,
  u_N(p),
\ee
where the operator $\Gamma_{\gamma N \to R}$ describes the transition to a
baryon resonance with spin 1/2.

For the hadronic propagator of spin-3/2 states we use the form
\be
S_{\alpha\beta}(p_R, W)
= -{\cal P}_{\alpha\beta}^{3/2}(p_R)\,
  \dfrac{(\slashed{p}_R + W)}{p_R^2 - W^2 + i 0^+},
\ee
where the spin-3/2 projection operator ${\cal P}_{\alpha\beta}^{3/2}$ is defined by
\be
{\cal P}_{\alpha\beta}^{3/2}(p_R)
= g_{\alpha\beta}
- \dfrac{1}{3} \gamma_\alpha \gamma_\beta 
- \dfrac{1}{3 p_R^2}
  \left( \slashed{p}_R \gamma_\alpha (p_R)_\beta
        + (p_R)_\alpha \gamma_\beta \slashed{p}_R
  \right),
\ee
The resonance transition currents $\Gamma_{\gamma N \to R}^{\beta\nu}(p_R, q_1)$
and $\Gamma_{R \to \gamma N}^{\mu\alpha}(p_R,-q_2)$ at the two hadron vertices
can be parameterized using the form factors $G_1, G_2$ and $G_3$. Details of the
transition current are discussed in Sec.~\ref{sec.GammaH}.

The TPE crossed-box amplitude ${\cal M}_{\gamma \gamma}^{\rm{xbox}}$ can be
calculated by replacing the lepton tensor $L_{\mu \nu}$ in \cref{eq.M} by the
tensor
\be
L_{\mu \nu}^{\rm{xbox}}
= \bar{u}_e(k')\, \gamma_\nu\, S_F(k_2,m_e)\, \gamma_\mu u_e(k),
\ee
where the intermediate lepton momentum is $k_2=k-q_2$. The crossed-box amplitude
can also be obtained from the crossing symmetry relation~\cite{blunden2017}
\be
{\cal M}_{\gamma\gamma}^{\rm{xbox}}(u,t)
= - M_{\gamma\gamma}^{\rm{box}}(s,t)\Big\vert_{s\to u},
\ee
where the Mandelstam variables $s$, $t$, and $u$ are defined by
\bea
\bal
s &= (k+p)^2 =(k'+p')^2,\\
t &=(k-k')^2 =q^2,\\
u &=(p-k')^2 =(p'-k)^2.
\eal
\eea 
Note, however, that, unlike the box amplitude, which is complex, the crossed-box
amplitude is purely real. In the dispersive approach it is therefore not
necessary to consider the crossed-box term explicitly.
Including the one- and two-photon exchange contributions, the total squared
amplitude can be written
\bea
\left\vert {\cal M}_\gamma + {\cal M}_{\gamma\gamma} \right\vert^2
&\approx& \left\vert {\cal M}_\gamma \right\vert^2
                +2 \Re({\cal M}_\gamma^\dagger {\cal M}_{\gamma\gamma})  \nonumber\\
&\equiv& \left\vert {\cal M}_\gamma \right\vert^2 (1 + \delta_{\gamma\gamma}),
\eea
where terms of order $\alpha^4$ have been neglected, and we have defined the
relative two-photon exchange correction to the cross section as
\be
\delta_{\gamma\gamma}
= \dfrac{2 \Re({\cal M}_\gamma^\dagger {\cal M}_{\gamma \gamma})}
        {\vert {\cal M}_\gamma \vert^2}.
\ee

For the nucleon intermediate state the TPE cross section correction is infrared
(IR) divergent in the soft photon limit, but this divergence is exactly
cancelled by a corresponding divergence in the real photon emission from the
electron and proton \cite{maximon2000, arrington2011}. It is useful, however, to
define a finite TPE correction which has the IR divergent contribution
subtracted. This correction will not be unique, as it depends on the
prescription used for the regularization~\cite{mo1969, maximon2000,
arrington2011, blunden2017}. For most of the theoretical results presented in
this analysis we use the prescription of Maximon and Tjon~\cite{maximon2000},
\be
\delta = \delta_{\gamma\gamma}(\rm{unsubtracted}) - \delta_{\rm{IR}}(\rm{MTj}).
\label{eq.delta_finite}
\ee
However, most experimental analyses use the prescription of Mo and
Tsai~\cite{mo1969}, and for comparison with experimental data we incorporate the
additional correction $\delta_{\rm{IR}}(\rm{MTj})-\delta_{\rm{IR}}(\rm{MTs})$. A
discussion of the differences between $\delta_{\rm{IR}}(\rm{MTj})$ of
Maximon-Tjon~\cite{maximon2000} and $\delta_{\rm{IR}}(\rm{MTs})$ of
Mo-Tsai~\cite{mo1969} can be found in Ref.~\cite{arrington2011}.

\section{Two-photon exchange from resonant intermediate states}
\label{sec.GammaH}

In this work we present the general decomposition of the hadronic transition
current operators and parameterize these in terms of the transition form factors
$G_i$ ($i=1,2,3$) as defined in
Ref.~\cite{jones1973,devenish1976,aznauryan2012}. For our numerical calculation
of the imaginary part of the TPE amplitude we use input on the electromagnetic
helicity amplitudes $A_{1/2}$, $A_{3/2}$ and $S_{1/2}$ from the analysis of
exclusive meson electroproduction data from CLAS at Jefferson
Lab~\cite{HillerBlin:2019hhz}. We provide the explicit relations between the
form factors and the helicity amplitudes for the proton to resonance transitions
for the spin-parity $1/2^\pm$ and $3/2^\pm$ excited states in
Sec.~\ref{ssec.GammaAh}. Following this, in Sec.~\ref{ssec.dispersive} we
describe the details of the dispersive method utilized in our study, including
the effects of finite resonance widths.

\subsection{Resonance transition current operators}
\label{ssec.GammaG}

We begin this section by parameterizing the transition current operator
$\Gamma_{\gamma N \to R}$ describing the absorption of a virtual photon with
momentum $q_1$ on a nucleon $N$ with momentum $p$, producing a resonant state
$R$ with momentum $p_R=p+q_1$. Following Refs.~\cite{devenish1976,
aznauryan2012}, we decompose $\Gamma_{\gamma N \to R}$ into several terms with
coefficients defining the form factors $G_1$, $G_2$ and $G_3$. Specifically, for
spin-$3/2$ resonant states the current operator has the decomposition
\be
\Gamma_{\gamma N\to R}^{\beta\nu}(p_R,q_1)
 = G_1(Q_1^2)\, \Theta_1^{\beta\nu}(p_R,q_1)
 + G_2(Q_1^2)\, \Theta_2^{\beta\nu}(p_R,q_1)
 + G_3(Q_1^2)\, \Theta_3^{\beta\nu}(p_R,q_1),
\label{eq:Gam32def}
\ee
where the $\Theta_i^{\beta\nu}$ operators are defined as
\begin{subequations}
\bea
\Theta_1^{\beta\nu}
&=& \left(\begin{array}{cc}\gamma_5\\{\rm I}\end{array}\right)
    (\slashed{q}_1 g^{\beta\nu} - q_1^\beta \gamma^\nu),       \\
\Theta_2^{\beta\nu}
&=& \left(\begin{array}{cc}\gamma_5\\{\rm I}\end{array}\right)
    (q_1^{\beta} p_R^\nu - q_1\cdot p_R\, g^{\beta\nu}),   \\
\Theta_3^{\beta\nu}
&=& \left(\begin{array}{cc}\gamma_5\\{\rm I}\end{array}\right)
    (q_1^\beta q_1^\nu - q_1^2 g^{\beta\nu}),
\eea
\end{subequations}
and the upper and lower rows refer to positive and negative parity states,
respectively. For spin-$1/2$ resonances, we define the current operator
$\Gamma_{\gamma N \to R}^\nu$ as
\be
\Gamma_{\gamma N \to R}^\nu(p_R,q_1)
= G_1(Q_1^2)
  \left(\begin{array}{cc}\rm I \\ {\gamma_5}\end{array}\right)
  \left( \slashed{q}_1 q_1^\nu - q_1^2 \gamma^\nu \right)
+  G_2(Q_1^2)
  \left(\begin{array}{cc}\rm I \\ {\gamma_5}\end{array}\right)
  \left( \slashed{q}_1 P^\nu - P\cdot q_1\, \gamma^\nu \right),
\ee
where $P = (p + p_R)/2 = p_R - q_1/2$, and again the upper and lower rows refer
to positive and negative parity states, respectively. For the inverse transition
$R \to \gamma N$ in \cref{fig.tpe}, the current operator $\Gamma_{R \to \gamma
N}^{\mu\alpha }(p_R,q_1)$ for the spin-3/2 states can be obtained using the
Hermitian property of the transition matrix element,
\be
\Gamma_{R \to \gamma N}^{\mu\alpha}(p_R,q_1)
= \gamma_0\,
  \big[ \Gamma_{\gamma N\to R}^{\alpha\mu}(p_R,q_1) \big]^\dagger\, \gamma_0,
\label{eq:conjugation3/2}
\ee
where $q_1$ is now the momentum of the outgoing photon. A similar relation also
holding for the spin-1/2 operator $\Gamma_{R \to \gamma N}^\nu(p_R,q_1)$.

\subsection{Form factors from electrocouplings}
\label{ssec.GammaAh}

Since the electroproduction of resonance states $R$ is often parameterized in
terms of resonance electrocouplings $A_h$ \cite{HillerBlin:2019hhz}, we define
here the transition form factors $G_i$ in terms of the amplitudes for specific
helicity configurations.

\subsubsection{Resonance electrocouplings} 
\label{ssec.Ah}

The resonance electrocouplings at the hadronic vertices are defined in terms of
the matrix elements of the hadron electromagnetic current operator
as~\cite{aznauryan2012}
\begin{subequations}
\bea
A_{1/2}
&=& \sqrt{\dfrac{2\pi \alpha}{K}}\dfrac{1}{e}\,
    \Big\langle
        R, S_z^R=\tfrac12\,
        \Big\vert\, \epsilon_\mu^+ J^{\mu}_{\rm{em}}\, \Big\vert\,
        N, S_z=-\tfrac12
    \Big\rangle,           \\
A_{3/2}
&=& \sqrt{\dfrac{2\pi \alpha}{K}}\dfrac{1}{e}\,
    \Big\langle
        R, S_z^R=\tfrac32\,
        \Big\vert\, \epsilon_\mu^+ J^{\mu}_{\rm{em}}\, \Big\vert\,
        N, S_z=\tfrac12
    \Big\rangle,            \\
S_{1/2}
&=& \sqrt{\dfrac{2\pi \alpha}{K}}\dfrac{1}{e}\,
    \Big\langle
        R, S_z^R=\tfrac12\,
        \Big\vert\, \dfrac{|\bm{q}_1|}{Q_1} \epsilon_\mu^0 J^{\mu}_{\rm{em}} \Big\vert\,
        N, S_z=\tfrac12
    \Big\rangle,
\eea
\end{subequations}
where $\alpha = e^2/4\pi$ is the fine structure constant, and $K$ is the
equivalent photon energy at the real photon point, $K = (W^2-M^2)/2W$. The spin
projections of the nucleon and resonances $R$ on the $z$-axis are labeled by
$S_z$ and $S_z^R$, and $\epsilon_\mu^{+,0}$ is the photon polarization vector
for transversely or longitudinally polarized photons,
\begin{subequations}
\bea
\epsilon_\mu^+ &=& (0; -\bm{\epsilon}^+),\ \ \ \ \
    \bm{\epsilon}^+ = -\dfrac{1}{\sqrt2}(1,i,0),    \\
\epsilon_\mu^0 &=& \dfrac{1}{Q_1}(|\bm{q}_1|; 0,0,-q_1^0).
\eea
\end{subequations}
The virtual photon three-momentum $\bm{q}_1$ is taken to be along the $z$-axis
in the rest frame of the resonance $R$, and its magnitude is given in terms of
the final state hadron mass $W$ and the photon virtuality
$Q_1^2=|\bm{q}_1|^2-(q_1^0)^2$,
\bea
|\bm{q}_1| &=& \sqrt{Q_1^2+ \left( \dfrac{W^2-M^2-Q_1^2}{2 W}\right)^2}.
\eea

The parameterizations of the resonance electrocouplings obtained from the
analysis of CLAS meson electroproduction data at Jefferson
Lab~\cite{HillerBlin:2019hhz} are at the sharp resonance point. To generate the
electrocouplings as a function of the running invariant mass $W$ of the
intermediate state, we use the $W$-dependent electrocoupling $A_h(W,Q_1^2)$
defined as
\be
A_h(W,Q_1^2)
= \dfrac{W}{W_R} \dfrac{|\bm{q}_{1,R}|}{|\bm{q}_1|} A_h^R(Q_1^2),
\label{eq.AhW}
\ee
which is consistent with the prescription in the JM model of
Ref.~\cite{Ripani:2000va, HillerBlin:2019hhz}. In Eq.~(\ref{eq.AhW}), $A_h^R$
represents the electrocouplings $A_{1/2}$, $A_{3/2}$ or $S_{1/2}$ at the
resonance point, $W_R$ is the invariant mass $W$ at the resonance point, and
$\bm{q}_{1,R}$ is defined as
\bea
|\bm{q}_{1,R}|
&=& \sqrt{Q_1^2 + \left(\dfrac{W_R^2-M^2-Q_1^2}{2 W_R}\right)^2}.
\eea

\subsubsection{Relations between form factors and electrocouplings} 

Following Devenish {\it et al.}~\cite{devenish1976}, the hadronic transition
current operator $\Gamma_{\gamma N\to R}^{\beta\nu}$ for spin-3/2 resonances can
also be parameterized in terms of helicity form factors $h_1$, $h_2$ and $h_3$,
which are given in terms of the helicity amplitudes $A_{1/2}$, $A_{3/2}$ and
$S_{1/2}$ by
\be
h_1 = \dfrac{\sqrt{3}\, W}{b\, |\bm{q}_1|}\, S_{1/2},\ \ \ \
h_2 = \pm \dfrac{1}{\sqrt{2}\, b}\, A_{3/2},\ \ \ \
h_3 = \dfrac{\sqrt{3}}{\sqrt{2}\, b}\, A_{1/2},
\label{eq.hA}
\ee
where
\be
b \equiv \sqrt{\pi \alpha \dfrac{ \left(W \mp M \right)^2+Q_1^2 }{24 M W K}}
\label{eq:b}
\ee
and the upper (lower) sign corresponds to even (odd) parity states. (Note that
the expressions for the helicity form factors $h_i$ in terms of the
electrocouplings $A_h$ of Ref.~\cite{aznauryan2012} are off by a factor of
$\sqrt{2/3}$, which has been corrected in Eq.~(\ref{eq.hA}).) For spin-parity
$3/2^+$ excitations, the current operator then can be written as
\bea
\Gamma_{\gamma N\to R}^{\beta\nu}(p,q_1)
&=& \frac{h_1}{C}\, q_1^\beta \left[p \cdot q_1\, q_1^\nu - q_1^2\, p^\nu \right] \gamma_5\,
 +\, \frac{h_2}{C}\, \left[ 2\epsilon^{\beta\sigma}(q_1 p)\, \epsilon^{\nu\sigma}(q_1 p)
 \gamma_5 + i\, W q_1^\beta \epsilon^\nu (q_1 p\gamma) \right]   \nonumber\\
&+& i\, \frac{h_3}{C}\, W q_1^\beta \epsilon^\nu(q_1 p\gamma),
\label{eq:Gam32hi}
\eea
while for spin-parity $3/2^-$ states it is given by
\bea
\Gamma_{\gamma N\to R}^{\beta\nu}(p,q_1)
&=& \frac{h_1}{C}\, \gamma_5 q_1^\beta \left[ p \cdot q_1\, q_1^\nu - q_1^2\, p^\nu \right] \gamma_5\,
 +\, \frac{h_2}{C}\, \gamma_5 \left[ 2\epsilon^{\beta\sigma}(q_1 p)\,
 \epsilon^{\nu\sigma}(q_1p) \gamma_5 - i\, W q_1^\beta \epsilon^\nu(q_1 p \gamma) \right] \nonumber\\
&-& i\, \frac{h_3}{C}\, W \gamma_5\, q_1^\beta \epsilon^\nu(q_1 p\gamma),
\label{eq:Gam12hi}
\eea
where 
\be
C = \left[ (W+M)^2+Q_1^2\right] \left[ (W-M)^2+Q_1^2\right].
\ee
Note that in Eqs.~(\ref{eq:Gam32hi}) and (\ref{eq:Gam12hi})  we use the shorthand notation
  $\epsilon^{\beta\sigma}(q_1 p) \equiv \epsilon^{\beta\sigma\rho\lambda} (q_1)_\rho p_\lambda$
and
  $\epsilon^\nu(q_1 p \gamma) \equiv \epsilon^{\nu\rho\lambda\alpha} (q_1)_\rho p_\lambda \gamma_\alpha$,
where ``$\gamma$'' in the Levi-Civita tensor denotes the Dirac $\gamma$-matrix. 
Equating the expressions in Eqs.~(\ref{eq:Gam32def}) and (\ref{eq:Gam32hi}), the
form factors $G_i$ can be expressed in terms of the helicity form factors, and
hence in terms of the electrocouplings $A_h$, as
\begin{subequations}
\bea
G_1(W,Q_1^2)
&=& \mp\dfrac{W (h_2 + h_3)}{2 \big[ (W \pm M)^2+Q_1^2\big]},     \\
G_2(W,Q_1^2)
&=& \dfrac{Q_1^2 h_1 + ( M^2 \mp M W + Q_1^2 ) h_2 + W (W \mp M) h_3}
          {C},\\
G_3(W,Q_1^2)
&=& \dfrac{2 W^2 (h_2 - h_3) - (M^2-W^2+Q_1^2) h_1}
          {2\,C}.
\eea
\label{eq:GiAh32}
\end{subequations}

For spin-parity $1/2^\pm$ resonant intermediate states, the spin-1/2 transition
form factors $G_1$ and $G_2$ can be related to the electrocouplings $A_h$
according to~\cite{aznauryan2012}
\begin{subequations}
\bea
G_1(W,Q_1^2)
&=& \dfrac{|\bm{q}_1| A_{1/2} + \sqrt2 (M \pm W) S_{1/2}}
          {2\, b'\, |\bm{q}_1| \left[ (M \pm W)^2 + Q_1^2 \right]},  \\
G_2(W,Q_1^2)
&=& \dfrac{|\bm{q}_1| (M \pm W) A_{1/2} - \sqrt2\, Q_1^2 S_{1/2}}
          {b'\, |\bm{q}_1| (M \mp W) \left[ (M \pm W)^2 + Q_1^2 \right]},
\eea
\label{eq:GiAh12}
\end{subequations}
where
\be
b' \equiv \sqrt{\pi \alpha \dfrac{ (W \mp M)^2+Q_1^2 }{4 M W K}},
\ee
in analogy with Eq.~(\ref{eq:b}). Note that the relations between the form
factors and electrocouplings in Eqs.~(\ref{eq.hA}), (\ref{eq:GiAh32}) and
(\ref{eq:GiAh12}) also hold at the second vertex of the TPE box diagram in
Fig.~\ref{fig.tpe}, with the substitution $q_1 \to q_2$.

\subsection{Dispersive method}
\label{ssec.dispersive}

Before describing the details of the dispersive method for calculating the TPE
amplitude ${\cal M}_{\gamma\gamma}$, it will be convenient to define the
amplitude in terms of the generalized TPE form factors $F'_1$, $F'_2$ and
$G'_a$, in analogy to the single-photon exchange amplitude ${\cal M}_\gamma$ of
\cref{Mgamma}. After presenting the formal results for the generalized TPE form
factors in the narrow resonance approximation, we then consider the effects of
finite resonance widths.

\subsubsection{Generalized TPE form factors}
\label{sssec.genFF}

In the massless electron limit, the TPE amplitude can be represented in terms of
the generalized TPE form factors $F'_1$, $F'_2$ and $G'_a$
as~\cite{blunden2017,guichon2003} 
\bea
{\cal M}_{\gamma\gamma}
&=& -\dfrac{e^2}{q^2} \bar{u}_e(k') \gamma_\mu u_e(k)\,
    \bar{u}_N(p')
      \left[ F'_1(Q^2,\nu) \gamma^\mu + F'_2(Q^2,\nu) \dfrac{i\sigma^{\mu\nu} q_\nu}{2M} \right]
    u_N(p)        \nonumber\\
& & -\dfrac{e^2}{q^2} \bar{u}_e(k') \gamma_\mu \gamma_5 u_e(k)\,
    \bar{u}_N(p') G'_a(Q^2,\nu) \gamma^\mu \gamma_5 u_N(p),
\label{eq:Mprime}
\eea
where these are functions of $Q^2$ and the dimensionless variable
\bea \bal
\nu \equiv \dfrac{s-u}{4M^2} = \sqrt{\dfrac{\tau(1+\tau)(1+\varepsilon)}{1-\varepsilon}}.  
\eal
\eea
The TPE cross section can then be expressed in terms of the generalized TPE form
factors as~\cite{blunden2017}
\bea
\delta_{\gamma\gamma}
= 2 {\rm Re}\dfrac{\varepsilon G_E (F'_1-\tau F'_2) + \tau G_M (F'_1+F'_2)
                    + \nu(1-\varepsilon) G_M G'_a}
                  {\varepsilon G_E^2+\tau G_M^2}.
\label{eq.geTPE}
\eea
An alternative representation for the TPE cross section combines the $F'_1$,
$F'_2$ and $G'_a$ generalized TPE form factors into combinations that resemble
the electric and magnetic Sachs form factors at the Born level. Namely,
defining~\cite{borisyuk2008}
\bes
\label{eq.calG}
\bea
\mathcal{G}_E &\equiv& F'_1 - \tau F'_2,
\label{eq.calGE} \\
\mathcal{G}_M &\equiv& F'_1 + F'_2 + \frac{\nu}{\tau} (1-\varepsilon) G'_a,
\label{eq.calGM}
\eea
\ees
the TPE cross section can be written in a simplified form analogous to the
diagonal structure of the Born cross section of Eq.~(\ref{eq.sigred}),
\bea
\delta_{\gamma\gamma}
= 2 {\rm Re}\dfrac{\varepsilon G_E \mathcal{G}_E + \tau G_M \mathcal{G}_M}
                  {\varepsilon G_E^2 + \tau G_M^2}.
\label{eq.geTPE_delG}
\eea

The generalized TPE form factors $F'_1$, $F'_2$ and $G'_a$ can be expressed in
terms of the resonance transition form factors $G_1$, $G_2$ and $G_3$ by mapping
the TPE amplitude of Eq.~(\ref{eq.M}) onto the generalized TPE amplitude ${\cal
M}_{\gamma\gamma}$ in Eq.~(\ref{eq:Mprime}) \cite{blunden2017}. As input, we use
the CLAS parameterization~\cite{HillerBlin:2019hhz} of the electrocouplings
$A_h$ for all the resonance states, except the $\Delta(1232)~\!3/2^+$ resonance,
for which we instead use the parameterization from Refs.~\cite{blunden2017,
aznauryan2012} that is constrained by the well-established PDG value of the
magnetic transition form factor, $G_M(0) \approx 3.00$~\cite{pdg2018}.

While each of the individual generalized TPE form factors $F'_1$ and $F'_2$ is
formally infrared-divergent, we construct finite ratios $\delta$ by subtracting
from the total TPE amplitudes the infrared-divergent part, as discussed in
Eq.~(\ref{eq.delta_finite}) above. The axial form factor, $G'_a$, on the other
hand, has no infrared-divergent part.  

Note that the TPE amplitude corresponding to the box diagram of
Fig.~\ref{fig.tpe} has both real and imaginary parts, whereas the corresponding
crossed box part of amplitude is purely real. Using the Cutkosky cutting
rules~\cite{cutkosky1960}, one can put the intermediate lepton and hadron states
on-shell by substituting the propagator factors as
\begin{subequations}
\bea
\frac{1}{p_R^2-W^2+i0^+}\
&\to\ & -2\pi i\, \theta(p_R^0)\, \delta({p_R^2-W^2}),    \\
\frac{1}{k_1^2-m_e^2+i0^+}\
&\to\ & -2\pi i\, \theta(k_1^0)\, \delta({k_1^2-m_e^2}),
\eea
\end{subequations}
to obtain the imaginary part of the TPE amplitude 
    ${\cal M}_{\gamma\gamma}$,
and hence the imaginary part of the generalized TPE form factors $F'_1$, $F'_2$ and $G'_a$.
The advantage of this approach is that one can use the on-shell parameterization
of the hadronic transition current operator at the two hadronic vertices of
\cref{fig.tpe} without introducing any ambiguities about the off-shell behavior
of the amplitudes.

The use of the Cutkosky rules allows the integration for the imaginary part of
the generalized TPE form factors to be expressed in terms of an integration over
the solid angle $\Omega_{k_1}$ of the intermediate state lepton,
\be
I_\delta
= \dfrac{s-W^2}{4s} \int d\Omega_{k_1}
  \dfrac{G_i(Q_1^2)\, G_j(Q_2^2)\, f_{ij}(Q_1^2,Q_2^2)}{(Q_1^2 + \lambda^2)(Q_2^2 + \lambda^2)},
\label{eq:intI}
\ee
where $G_i(Q_1^2)$ and $G_j(Q_2^2)$ are the form factors at the two respective
$\gamma N R$ vertices ($i,j=1,2,3$), and the function $f_{ij}(Q_1^2,Q_2^2)$ is a
polynomial of combined degree 4 in $Q_{1,2}^2$. The imaginary part of the
generalized TPE form factors can be computed from \Cref{eq:intI} for each
resonance state at a specific value of $W$, such as at the peak of the
resonance, $W^2 = W_R^2$. The numerical evaluation of the integral $I_\delta$ in
\cref{eq:intI} at $W^2 = W_R^2$ gives the imaginary part of the generalized TPE
form factors, and hence the amplitude, as a function of electron energy $E$, at
fixed values of the four momentum transfer squared $Q^2$. 

The real parts of the TPE amplitudes can then be computed from the dispersion
relations~\cite{borisyuk2008, tomalak2015subtracted, blunden2017} according to,
\bes
\bea
\Re F_1'(Q^2,\nu)
&=& \frac2{\pi} {\cal P} \int_{\nu_{\rm min}}^\infty d\nu'\ 
    \frac{\nu}{\nu'^2-\nu^2}\, \Im F_1'(Q^2,\nu'),   \\
\Re F_2'(Q^2,\nu)
&=& \frac2{\pi} {\cal P} \int_{\nu_{\rm min}}^\infty d\nu'\ 
    \frac{\nu}{\nu'^2-\nu^2}\, \Im F_2'(Q^2,\nu'),   \\
\Re G_a'(Q^2,\nu)
&=& \frac2{\pi} {\cal P} \int_{\nu_{\rm min}}^\infty d\nu'\ 
    \frac{\nu'}{\nu'^2-\nu^2}\, \Im G_a'(Q^2,\nu'),
\eea
\label{eq:dispI}%
\ees
where $\cal{P}$ refers to the Cauchy principal value integral, with
    $\nu_{\rm min} = E_{\rm{min}}/M - \tau$
and
    $E_{\rm{min}} = (W^2-M^2)/2M$
is the minimum energy required to excite a state of invariant mass $W$.

For elastic nucleon intermediate states, the minimum energy is
    $E_{\rm{min}} = 0$,
so that one has
    $\nu_{\rm min} = -\tau$.
The physical threshold for electron scattering at $\varepsilon = 0$, or backward
angles, $\cos\theta = -1$, is $\nu_{\rm th} \equiv \sqrt{\tau(1+\tau)}$. In
other words, the threshold energy for physical scattering to take place is
    $E_{\rm{th}} = M (\tau + \nu_{\rm th})$.
At a certain limit of the values of $W$ and $Q^2$, the integrals in
Eqs.~(\ref{eq:dispI}) extend into the unphysical region; for example, for the
$\Delta(1232)~\!3/2^+$ resonance, at $Q^2=0.5$~GeV$^2$ and $W=1.232$~GeV the
physical threshold $\nu_{\rm th} \cong 0.4,$ whereas the integration runs from
$\nu_{\rm min} \cong 0.22$. The analytic continuation of the integral $I_\delta$
in \cref{eq:intI} into the unphysical region was discussed in detail in
Ref.~\cite{blunden2017}.

\subsubsection{Finite widths}
\label{sssec.finitewidth}

As the imaginary part of the TPE box diagram corresponds to real excitation,
there is a discontinuity in the imaginary part of the TPE amplitudes for
resonance intermediate states with zero width, at sharp $W = W_R$, such that
they vanish for $E < E_{\rm min}(W_R)$. When put into a dispersion integral,
this will translate into a cusp in the real part of the amplitude at the same
energy. If the threshold energy is above the minimum energy, $E_{\rm th} \ge
E_{\rm min}$, then this cusp is of no concern. However, if $ E_{\rm th}< E_{\rm
min}$, then there exists some physical energy $E$ for which one may have $E =
E_{\rm min}$. Equivalently, there is a cusp if the four-momentum transfer
squared goes below a threshold value, $Q^2 < Q_{\rm th}^2$, where
\bea
Q_{\rm th}^2 = \frac{(W^2-M^2)^2}{W^2}.
\eea
In terms of the photon polarization variable $\varepsilon$, the cusp will occur for
\bea
\varepsilon_{\rm cusp}(Q^2) = \frac{2 W^2 \left(Q_{\rm th}^2 - Q^2\right)}
{2 W^2 \left(Q_{\rm th}^2 - Q^2\right) + Q^2 \left(4 M^2 + Q^2\right)}.
\eea
In Table~\ref{tab.cusp} we show the values of $Q_{\rm th}^2$ and
$\varepsilon_{\rm cusp}(Q^2)$ for several physically relevant examples that
illustrate the effect, specifically, the $\Delta(1232)$, $N(1520)$ and $N(1720)$
states.

\begin{table}[t]
\caption{Kinematics at which threshold cusp effects appear for the
$\Delta(1232)~\!3/2^+$, $N(1520)~\!3/2^-$ and $N(1720)~\!3/2^+$ resonances, at
several typical $Q^2$ values relevant phenomenologically.\\}
\centering
\begin{tabular}{S[table-format=1.3] S[table-format=1.2] S[table-format=1.2]
S[table-format=1.2] S[table-format=1.2]} \hline
~{$W_R$~(GeV)}~~ & ~~{$Q_{\rm th}^2$~(GeV$^2$)}~~ & \multicolumn{3}{c}{$\varepsilon_{\rm cusp}(Q^2)$}\\
 \cmidrule{3-5}
 \cmidrule{3-5}
 & & ~~{$Q^2=0.2$~GeV$^2$}~~ & ~~{$Q^2=0.5$~GeV$^2$}~~ & ~~{$Q^2=1.0$~GeV$^2$}~~ \\ \hline
1.232 & 0.27 & 0.06 & {---} & {---}\\
1.520 & 0.87 & 0.81 & 0.46 & {---}\\
1.720 & 1.46 & 0.91 & 0.74 & 0.37\\[5pt]    \hline
\end{tabular}
\label{tab.cusp}
\end{table}

For the case of a resonance of finite width $\Gamma(W)$ that is centred at $W =
W_R$ and governed by a Breit-Wigner distribution,
\be 
f(W^2) = \frac{\cal N}{\pi} \dfrac{\Gamma(W) W_R}{(W^2-W_R^2)^2 + \Gamma^2(W) W_R^2},
\label{eq.BW}
\ee
the cusp behavior is smoothed out. Here ${\cal N}$ is a normalization constant,
defined so that
\be
\int_{(M+m_\pi)^2}^{W_{\rm max}^2} dW^2\, f(W^2) = 1.
\ee
In our numerical calculations, we take $W_{\rm max}=2$~GeV for all the resonance
states except the $\Delta(1232)~\!3/2^+$ and $N(1440)~\!1/2^+$, for which we
restrict the integration to $W_{\rm{max}}=1.7$~GeV.

To consider a finite width, we assume the continuum of the invariant mass
squared $W^2$ as an infinite set of Dirac $\delta$ functions,
$\delta(W^2-W_i^2)$, and evaluate the integral of \cref{eq:intI} at a set of
discrete values of $W$ ranging from $M+m_\pi$ to 2~GeV for each resonance
intermediate state. The corresponding real parts are calculated from
Eqs.~(\ref{eq:dispI}). The set of generated real parts of the generalized TPE
from factors are then interpolated using a spline fit to obtain a smooth
function $F(W^2)$ for the generalized TPE form factors at fixed values of $Q^2$
and electron energy $E$.

While the total decay widths $\Gamma(W)$ of the resonances are in general energy
dependent, for the default calculations in this work we restrict ourselves to
the case $\Gamma(W) = \Gamma(W_R) = \Gamma_R$, the constant total decay width.
The numerical values of $\Gamma_R$ and the Breit-Wigner resonance masses $W_R$
for each of the resonance states are taken from Ref.~\cite{HillerBlin:2019hhz}.
In Sec.~\ref{ssec.width} below we will discuss the effect of the nonzero width,
both constant and dynamic, on the total TPE cross section in more detail.

\section{Numerical TPE effects}
\label{sec.results}

In this section we present detailed numerical results for the TPE corrections to
the elastic scattering cross section from excited intermediate state resonances
(Sec.~\ref{ssec.tpesigma}). In particular, we study the effect of nonzero widths
for the resonances (Sec.~\ref{ssec.width}), and identify the dependence of the
TPE corrections on the spin, isospin and parity of the intermediate states
(Sec.~\ref{ssec.spin}). For completeness we also present (Sec.~\ref{ssec.GFF})
the results for the TPE contribution to the generalized electric and magnetic
TPE form factors defined in Sec.~\ref{sssec.genFF}. 

\subsection{TPE correction to the elastic cross section}
\label{ssec.tpesigma}

\begin{figure}[t]%
\graphicspath{{Images/}}
\hspace*{-8.4cm}\vspace*{-3.5cm}\includegraphics[width=8.4cm]{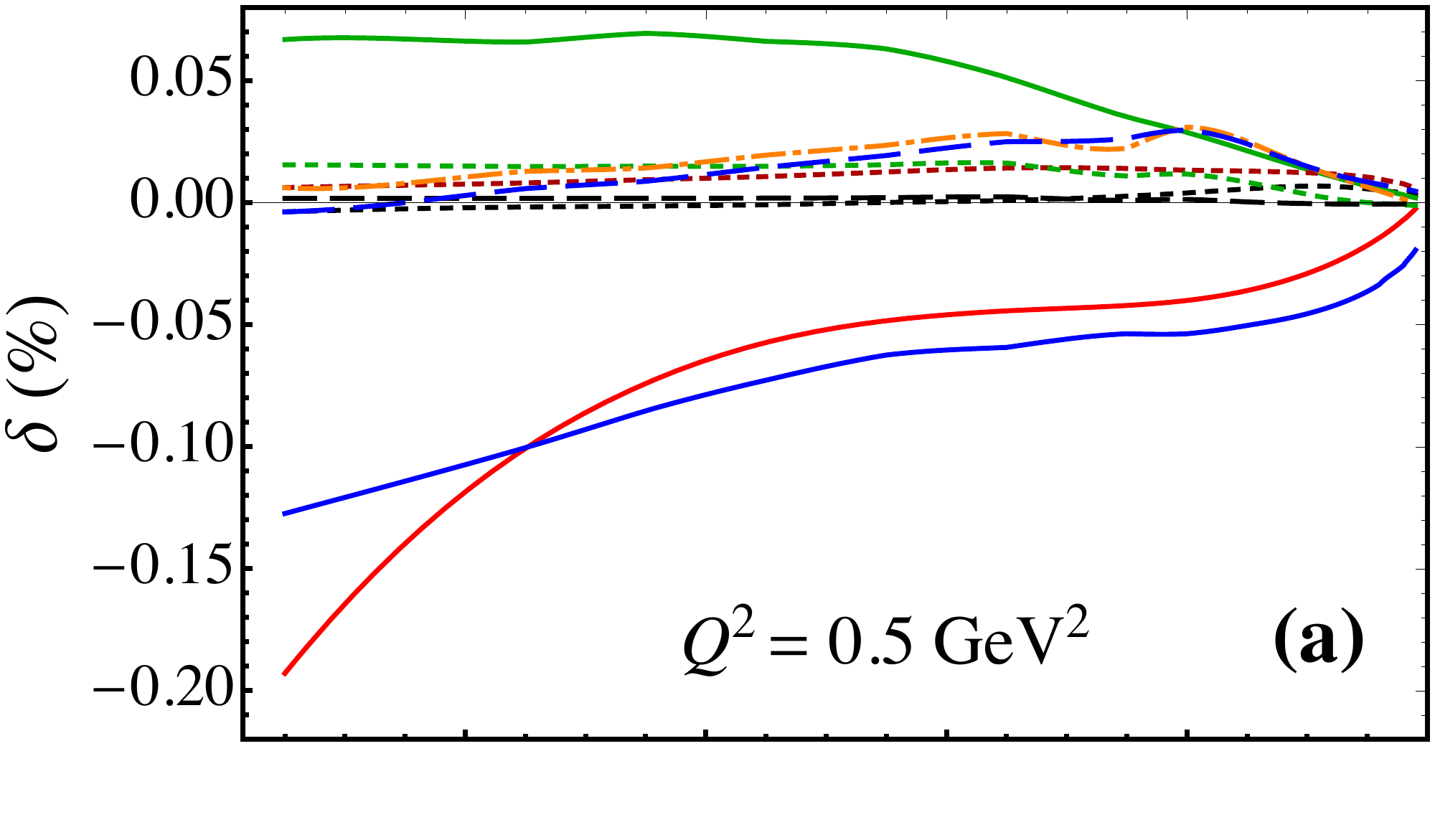} \\
\hspace*{-0.33cm}\includegraphics[width=8.2cm]{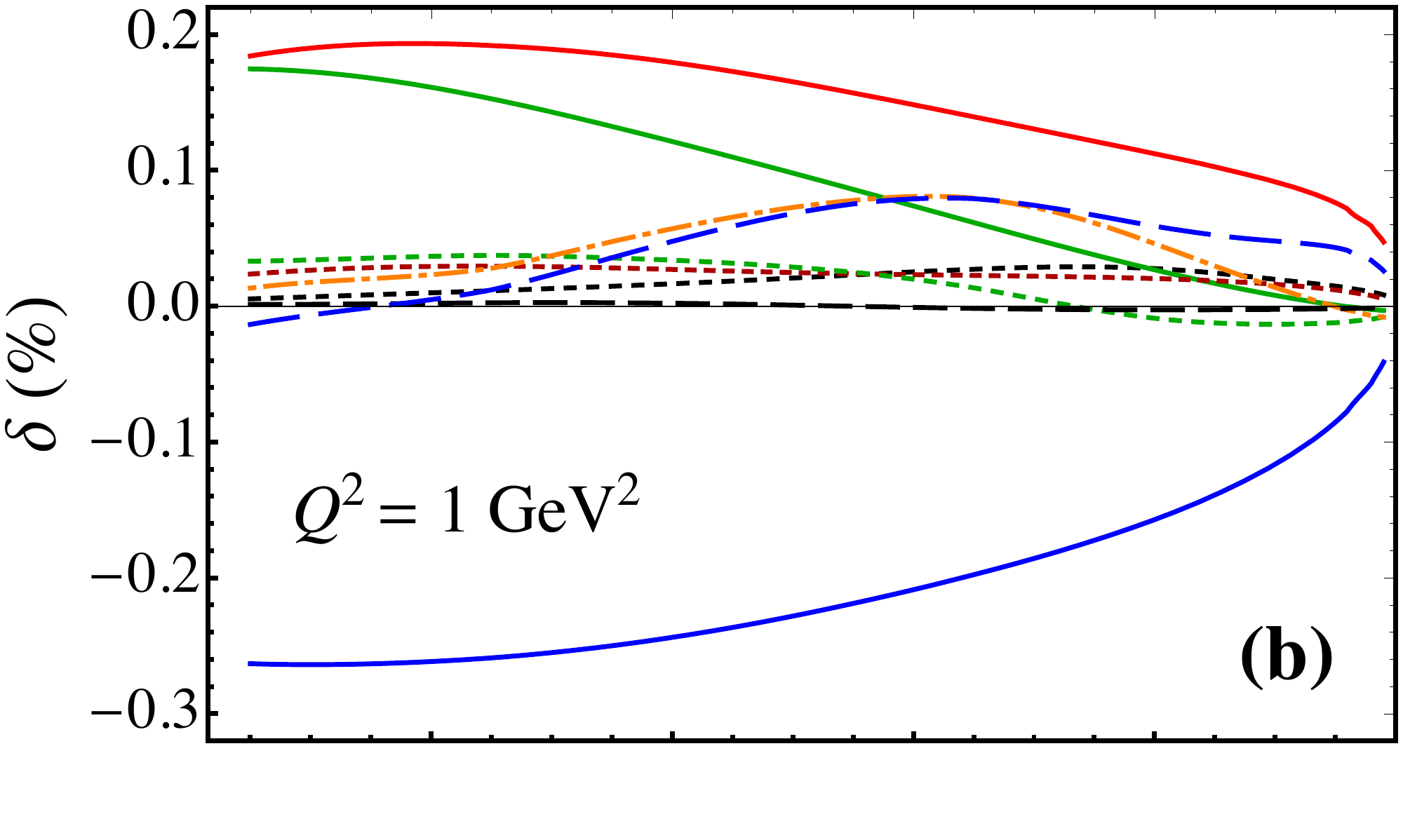}
\hspace*{0.0cm}\includegraphics[width=7.7cm]{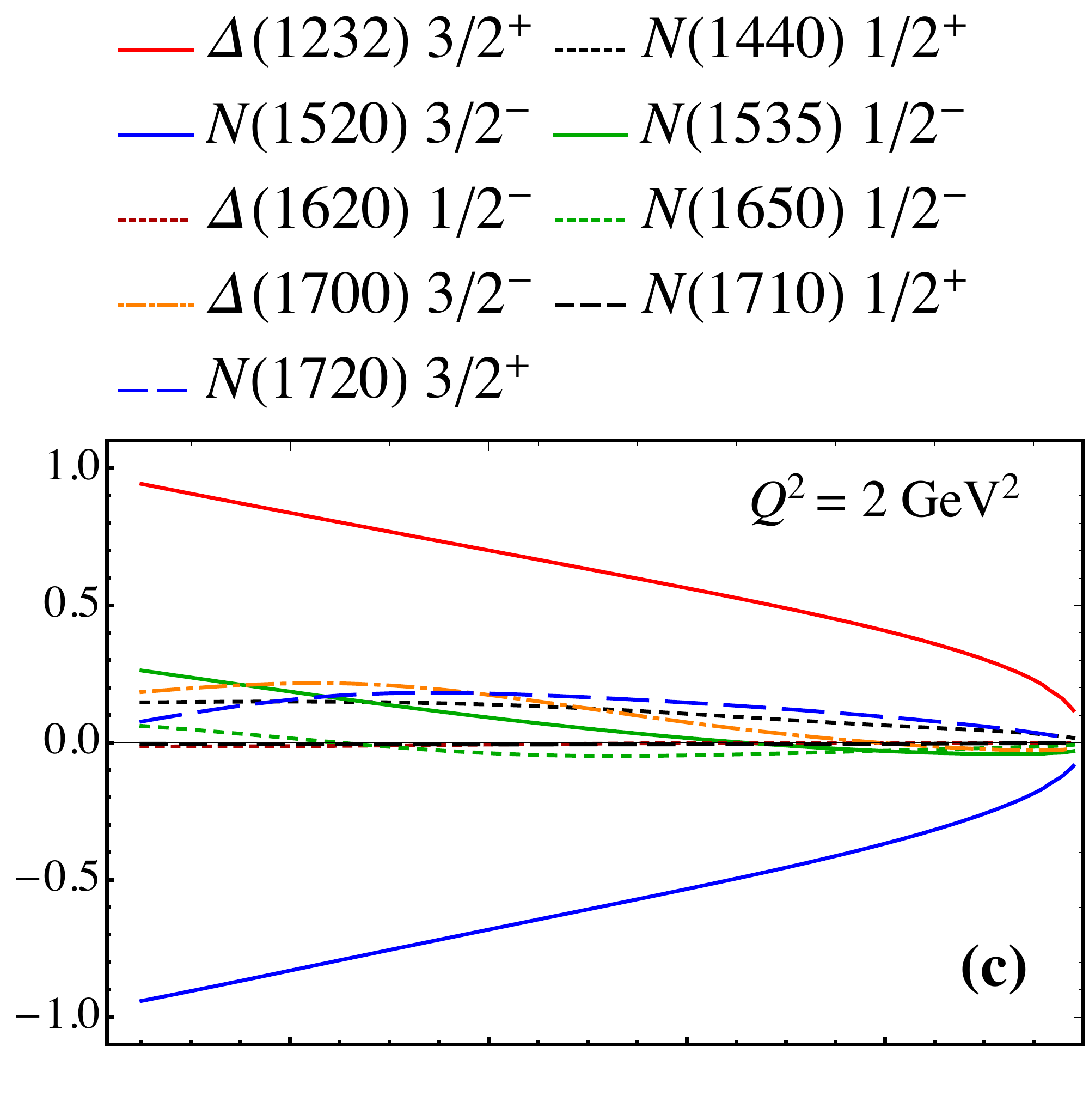}\vspace*{-0.5cm} \\
\hspace*{0.14cm}\includegraphics[width=8.11cm]{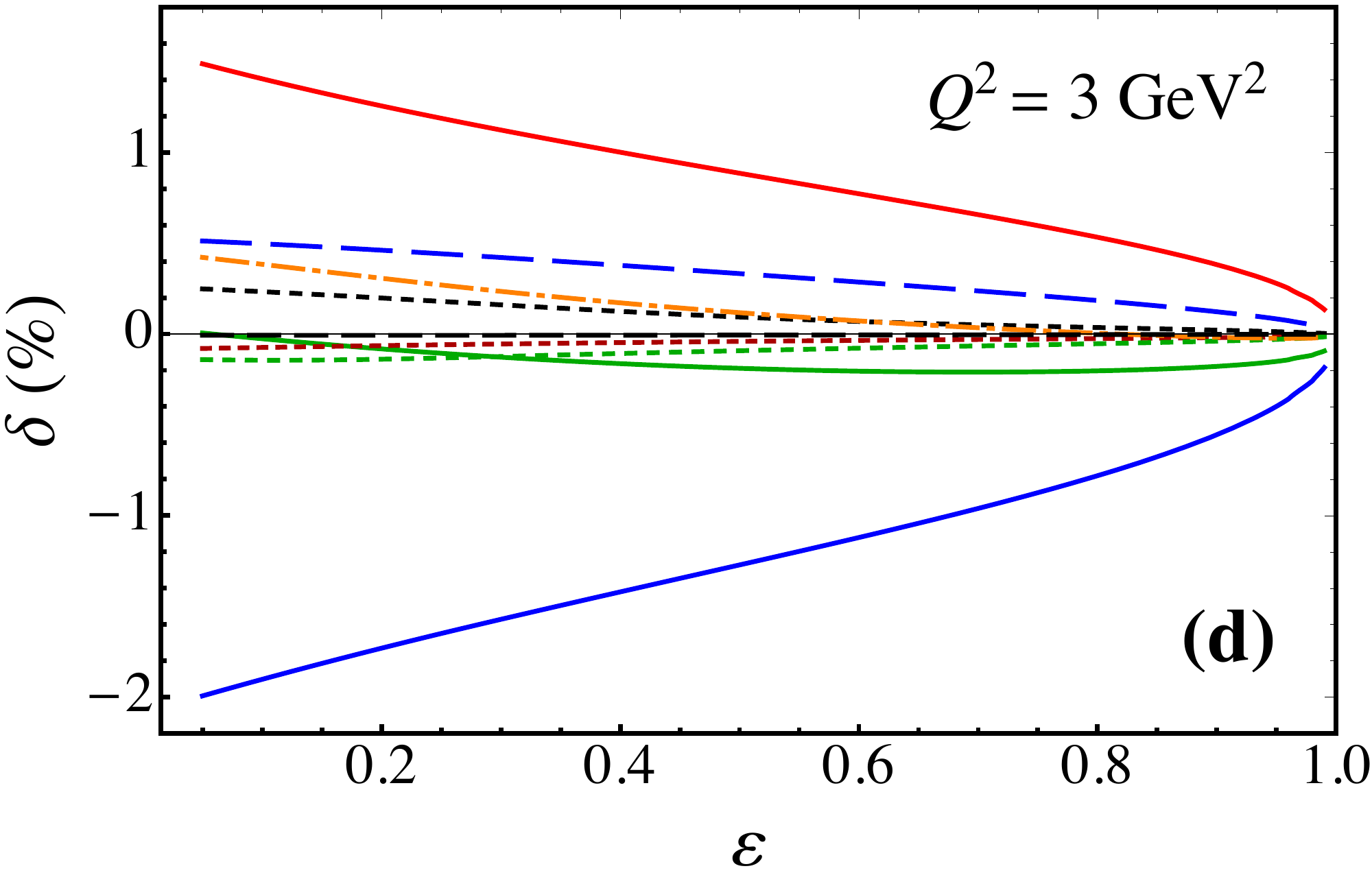}
\hspace*{-0.4cm}\includegraphics[width=8.1cm]{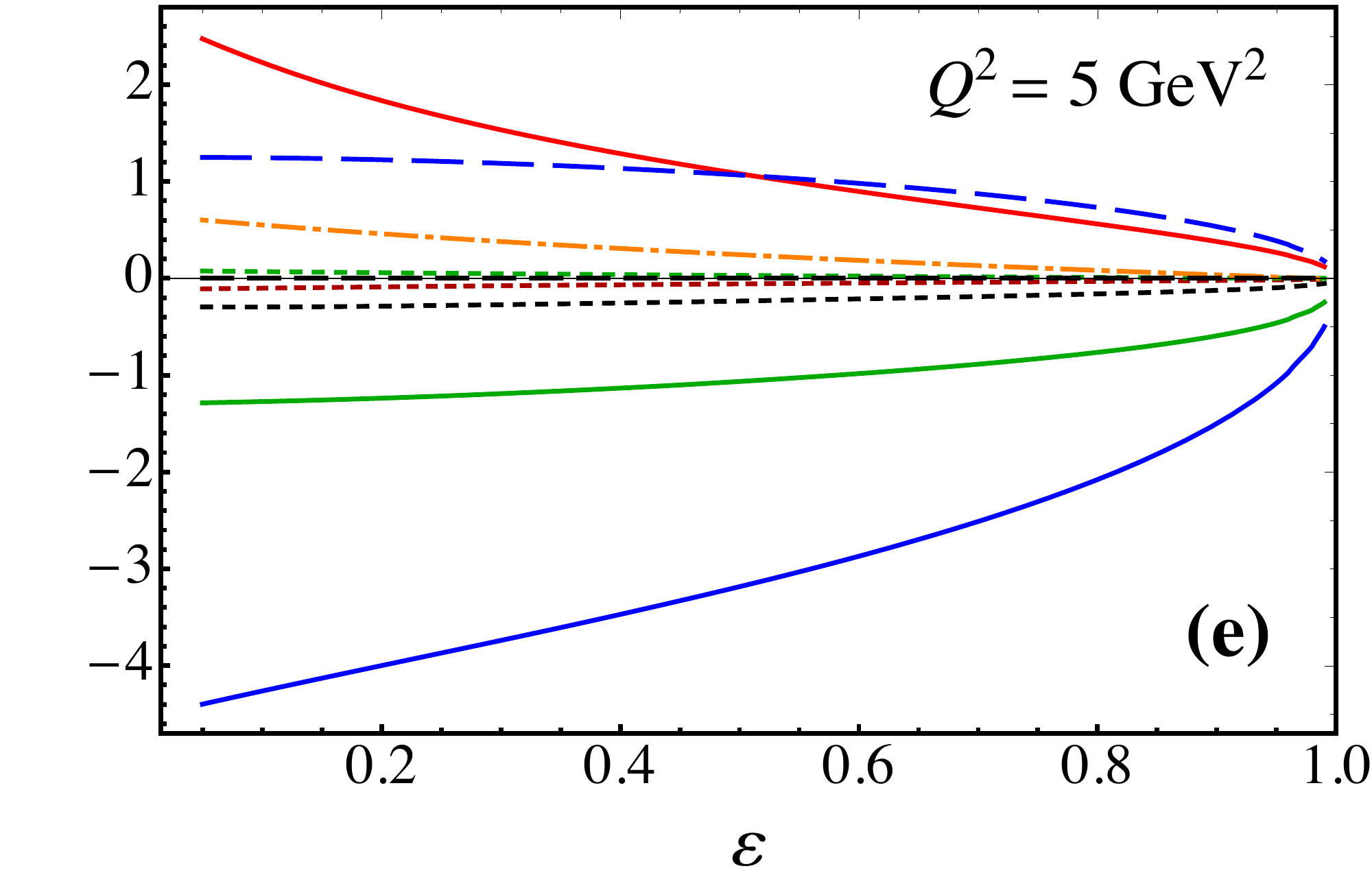}
\caption{Relative contributions $\delta$ (in percent) to the TPE cross section
for the nine spin-1/2 and spin-3/2 nucleon and $\Delta$ intermediate state
resonances, as indicated in the legend, versus the virtual photon polarization
$\varepsilon$ for fixed $Q^2$ values:
{\bf (a)} 0.5~GeV$^2$,
{\bf (b)} 1~GeV$^2$,
{\bf (c)} 2~GeV$^2$,
{\bf (d)} 3~GeV$^2$ and
{\bf (e)} 5~GeV$^2$.
Note the vertical scale is different in each panel.}
\label{fig.sig}
\end{figure}

The contributions to the TPE correction $\delta$ from the individual
intermediate state resonances are shown in Fig.~\ref{fig.sig} versus
$\varepsilon$, for fixed values of $Q^2 = 0.5$, 1, 2, 3 and 5~GeV$^2$. As
mentioned earlier, we account for all 4 and 3-star spin-1/2 and spin-3/2
resonances with mass below 1.8~GeV from the Particle Data Group~\cite{pdg2018},
which include the six isospin-1/2 states
    $N(1440)\, 1/2^+$,
    $N(1520)\, 3/2^-$,
    $N(1535)\, 1/2^-$, 
    $N(1650)\, 1/2^-$, 
    $N(1710)\, 1/2^+$ and
    $N(1720)\, 3/2^+$,
and the three isospin-3/2 states
    $\Delta(1232)\, 3/2^+$,
    $\Delta(1620)\, 1/2^-$ and
    $\Delta(1700)\, 3/2^-$.
In our numerical calculations, for the resonance electrocouplings at the
hadronic vertices we use the most recent helicity amplitudes extracted from the
analysis of CLAS electroproduction data~\cite{mokeev2012, HillerBlin:2019hhz},
except for the $\Delta(1232)\, 3/2^+$ resonance, for which we use the fit by
Aznauryan and Burkert~\cite{blunden2017,aznauryan2012}. For the elastic
intermediate state contribution, care must be taken to avoid poles in the
spacelike region of the proton electric and magnetic form factors $G_E(Q^2)$ and
$G_M(Q^2)$, which would be problematic in the dispersive framework. Some
commonly used parameterizations, such as from Venkat {\it et
al.}~\cite{venkat2011} or Arrington {\it et al.}~\cite{arrington2007}, have
poles for $Q^2 \gtrsim 4.5$~GeV$^2$. In order to compute the TPE corrections
that include contributions from intermediate states with larger $Q^2$, we use
the parameterization from Kelly~\cite{kelly2004}, which has poles only in the
timelike region.

In the low-$Q^2$ region, for $Q^2$ up to $\sim 1$~GeV$^2$, the $N(1520)~\!3/2^-$
and $N(1535)~\!1/2^-$ resonances give the most significant contributions, aside
from the $\Delta(1232)~\!3/2^+$ resonance, although the largest correction from
the $\Delta(1232)~\!3/2^+$ ranges within only $0.2\%$ of the Born level cross
section. We find an almost complete cancellation of the $N(1520)~\!3/2^-$ state
contribution by that from the sum of other higher-mass resonances, leaving a net
correction that is well approximated by that from the $\Delta(1232)~\!3/2^+$
alone. In this $Q^2$ range the $\Delta(1232)~\!3/2^+$ contribution flips in sign
and suppresses the elastic nucleon intermediate state correction. At higher
$Q^2$ values, $Q^2 \gtrsim 2$~GeV$^2$, the $N(1520)~\!3/2^-$ overtakes the
$\Delta(1232)~\!3/2^+$ contribution to $\delta$, but with opposite sign.
Moreover, in the high-$Q^2$ region the $N(1535)~\!1/2^-$ contribution flips sign
from positive to negative, however, this effect is somewhat negated by the
growth of the $N(1720)~\!3/2^+$ and $\Delta(1700)~\!3/2^-$ corrections. The
overall effect is that the suppression of the TPE cross section (relative to the
nucleon elastic contribution) by the $\Delta(1232)~\!3/2^+$ is largely nullified
by the $N(1520)~\!3/2^-$, leaving a small increase in the total TPE correction
over that from the nucleon intermediate state alone.

\begin{figure}[t]
\graphicspath{{Images/}}
\includegraphics[width=8cm]{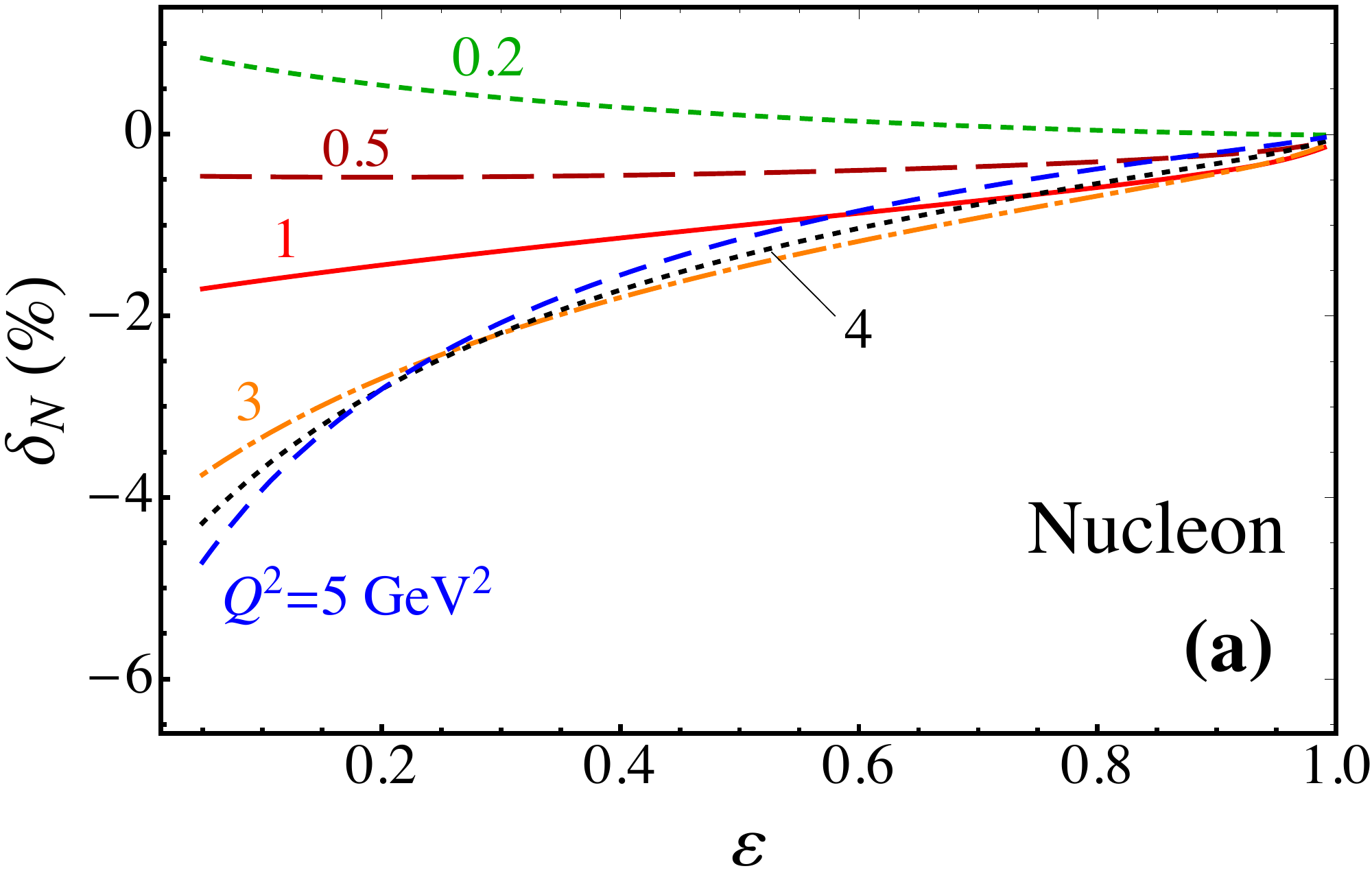}
\includegraphics[width=8cm]{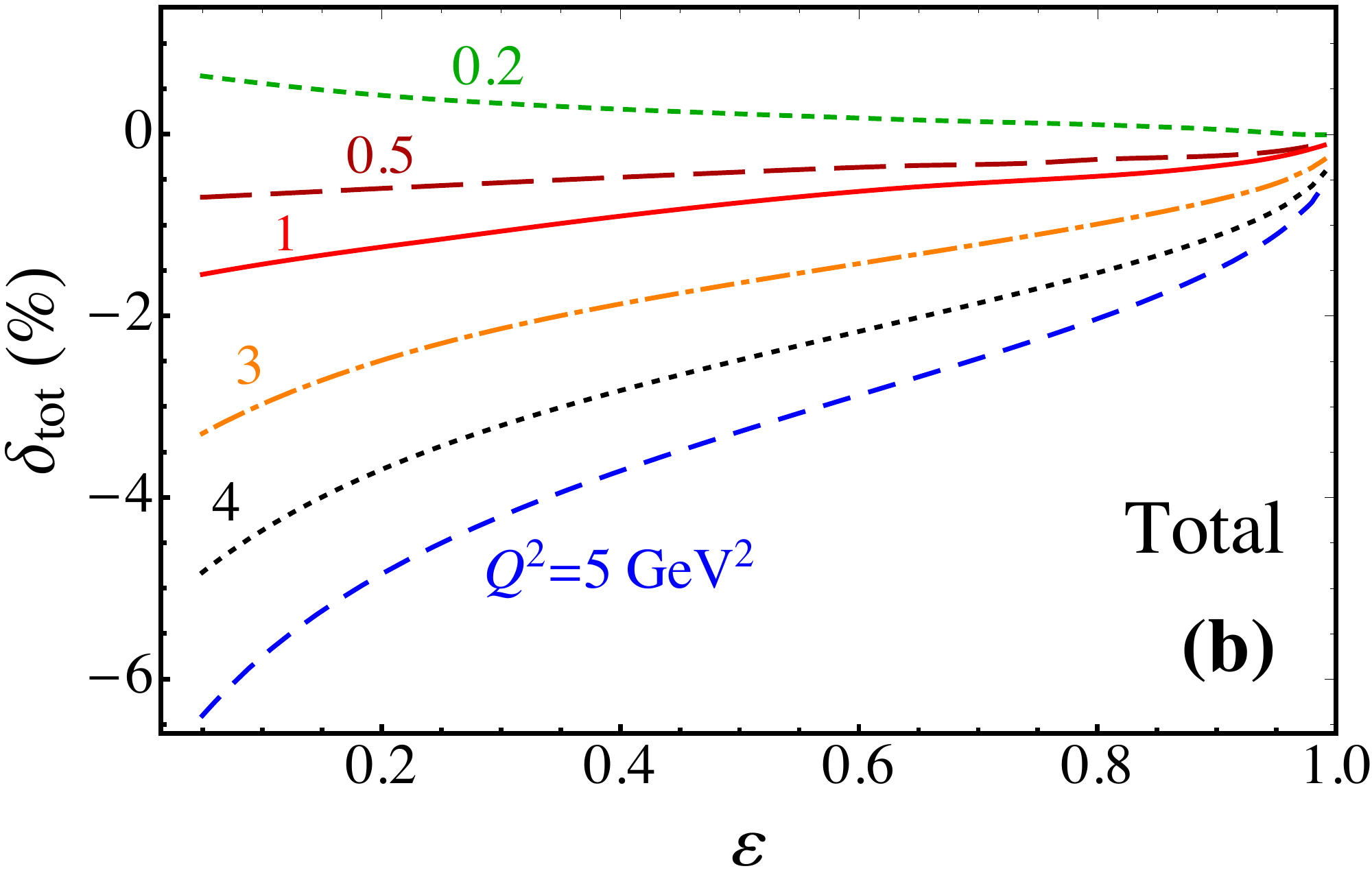}
\caption{Contributions to the TPE correction $\delta$ (in \%) versus the virtual
photon polarization $\varepsilon$ for {\bf (a)} nucleon only and {\bf (b)} all
spin-parity $1/2^\pm$ and $3/2^\pm$ states including the nucleon, at
    $Q^2 = 0.2$~GeV$^2$ (green dashed line),
    0.5~GeV$^2$ (dark red long-dashed),
    1~GeV$^2$ (red solid),
    3~GeV$^2$ (orange dot-dashed), and
    5~GeV$^2$ (blue dashed).}
\label{fig.Dsig}
\end{figure}

The combined effect on the TPE correction $\delta$ from all the spin-parity
$1/2^\pm$ and $3/2^\pm$ resonances is illustrated in Fig.~\ref{fig.Dsig} as a
function of virtual photon polarization, $\varepsilon$, for a range of fixed
$Q^2$ values between 0.2 and 5~GeV$^2$. For contrast, the contribution from the
nucleon elastic intermediate state alone is also shown at the same kinematics.
At low $Q^2$ the excited state resonance contributions are found to be
negligible, and the total correction is dominated by the nucleon elastic
intermediate state. Note that the elastic contribution is positive at the lowest
$Q^2$, $Q^2=0.2$~GeV$^2$, but rapidly changes sign and becomes increasingly more
negative at higher $Q^2$. At $Q^2=5$~GeV$^2$ the nucleon contribution becomes as
large as $4\%-5\%$ at low values of $\varepsilon \approx 0.1-0.2$. There is also
a trend toward increasing nonlinearity at higher $Q^2$ values, $Q^2 \gtrsim
3$~GeV$^2$, especially at low $\varepsilon$.

The net effect of the higher mass resonances is to increase the magnitude of the
TPE correction at $Q^2 \gtrsim 3$~GeV$^2$, due primarily to the growth of the
(negative) odd-parity $N(1520)~\!3/2^-$ and $N(1535)~\!1/2^-$ resonances which
overcompensates the (positive) contributions from the $\Delta(1232)~\!3/2^+$. At
the highest $Q^2=5$~GeV$^2$ value shown in Fig.~\ref{fig.Dsig}, the total TPE
correction $\delta_{\rm tot}$ reaches $\approx$~6-7\% at low $\varepsilon$.

\begin{figure}[t]
\graphicspath{{Images/}}
\includegraphics[width=8cm]{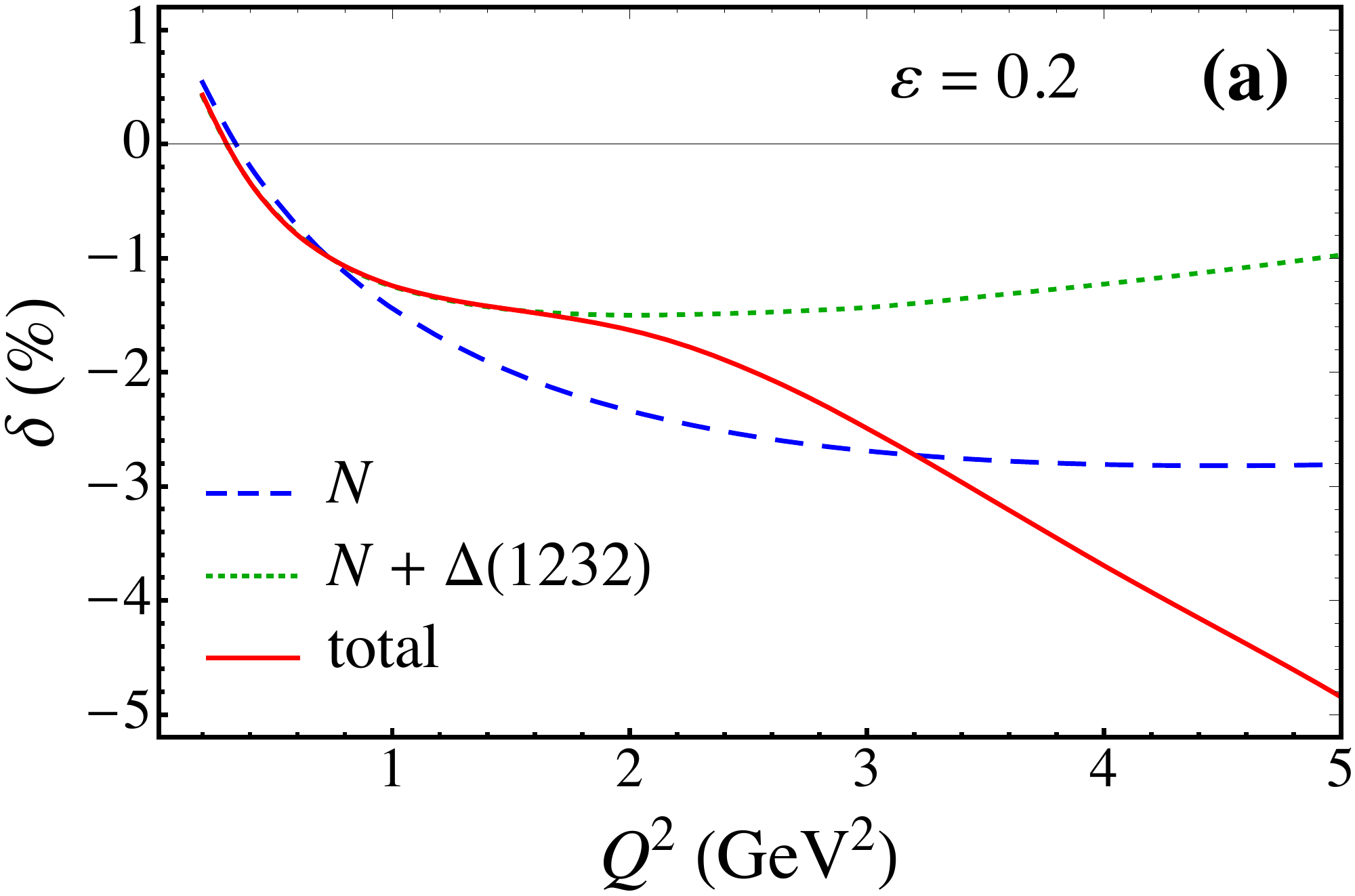}
\includegraphics[width=8cm]{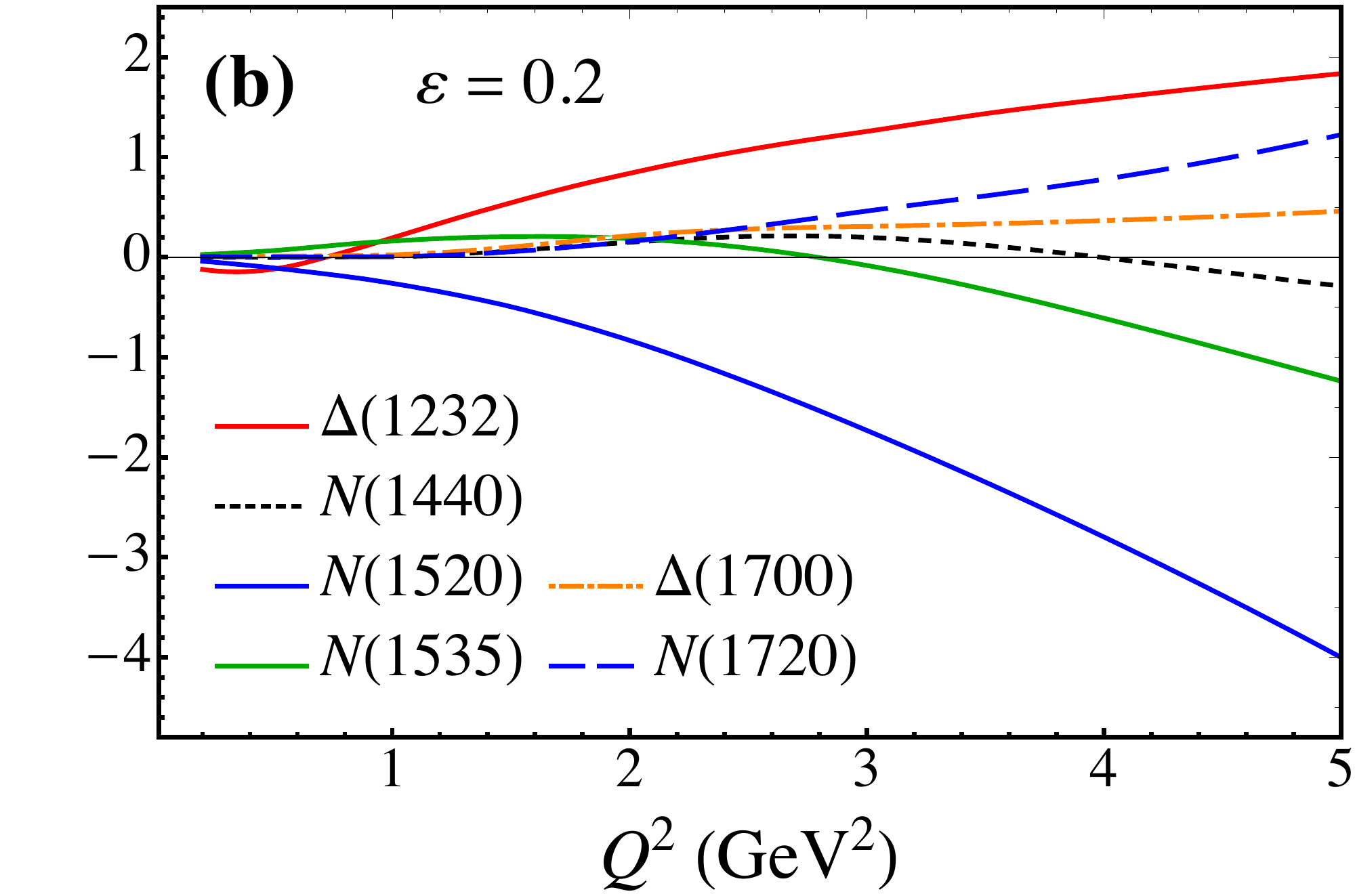}
\caption{Contributions to the TPE correction $\delta$ (in \%) versus $Q^2$ at
backward scattering angles, $\varepsilon=0$, for
    {\bf (a)} nucleon only (blue dashed), $N+\Delta(1232)$ (green dotted) and
    the sum of all resonances (red solid), and
    {\bf (b)} the major individual contributors at the same kinematics, including the
    $\Delta(1232)$ (red solid),
    $N(1440)$ (black dashed),
    $N(1520)$ (blue solid), 
    $N(1535)$ (green solid),     
    $\Delta(1700)$ (orange dot-dashed), and
    $N(1720)$ (blue dashed).}
\label{fig.DsigQ}
\end{figure} 

To provide a more graphic illustration of the $Q^2$ dependence of the
intermediate state resonance contributions to the cross section, we show in
Fig.~\ref{fig.DsigQ} the TPE corrections from the major individual contributors
for $Q^2$ up to 5~GeV$^2$. We choose a nominal value for the virtual photon
polarization of $\varepsilon=0.2$ in order to emphasize the largest effect on
$\delta$ at backward angles. One of the prominent effects is the cancellation of
part of the nucleon elastic contribution by the $\Delta(1232)~\!3/2^+$ resonance
across the entire $Q^2$ range. On the other hand, the sum of the higher-mass
resonances has a mixed impact on $\delta$. In the low-$Q^2$ region, $Q^2
\lesssim 1.8$~GeV$^2$, the higher resonance state corrections largely cancel,
leaving an approximately zero net contribution. As $Q^2$ increases, the role of
the $\Delta(1232)~\!3/2^+$ is partially nullified by contributions from the
higher mass resonances, and eventually is outweighed by the heavier states. An
overall increase in the total TPE cross section over that from the nucleon alone
is thus observed for $Q^2 \gtrsim 3$~GeV$^2$.

In the low-$Q^2$ range, the odd parity $N(1520)~\!3/2^-$ resonance state gives a
comparable cross section to that from the $\Delta(1232)~\!3/2^+$ state, but with
opposite sign. The TPE correction from the $N(1520)~\!3/2^-$ state keeps rising
with $Q^2$ and becomes the largest contributor at $Q^2 \gtrsim 4$~GeV$^2$,
outweighing even the elastic nucleon component. The other resonances largely
cancel each other, leaving behind a negligible net contribution.

As noted previously, for the default numerical calculations presented here the
resonance width has been taken to be the constant total decay width, $\Gamma_R$,
for each resonance $R$. To explore the sensitivity of the TPE corrections to the
assumptions about the width, in the next section we consider other cases,
including the zero-width approximation and an energy-dependant dynamical-width.

\subsection{Nonzero resonance widths}
\label{ssec.width}

As discussed in Sec.~\ref{ssec.dispersive} above, the discontinuity in the
imaginary part of the TPE amplitude for the case of zero-width resonances gives
rise to cusps in the real part of the amplitude from physical threshold effects
at specific kinematics. In this section we consider the threshold effect on the
TPE correction for the three representative resonance states
$\Delta(1232)~\!3/2^+$, $N(1520)~\!3/2^-$ and $N(1720)~\!3/2^+$ discussed in
Table~\ref{tab.cusp}.

\begin{figure}[t]
\graphicspath{{Images/}}
\includegraphics[width=5.6cm]{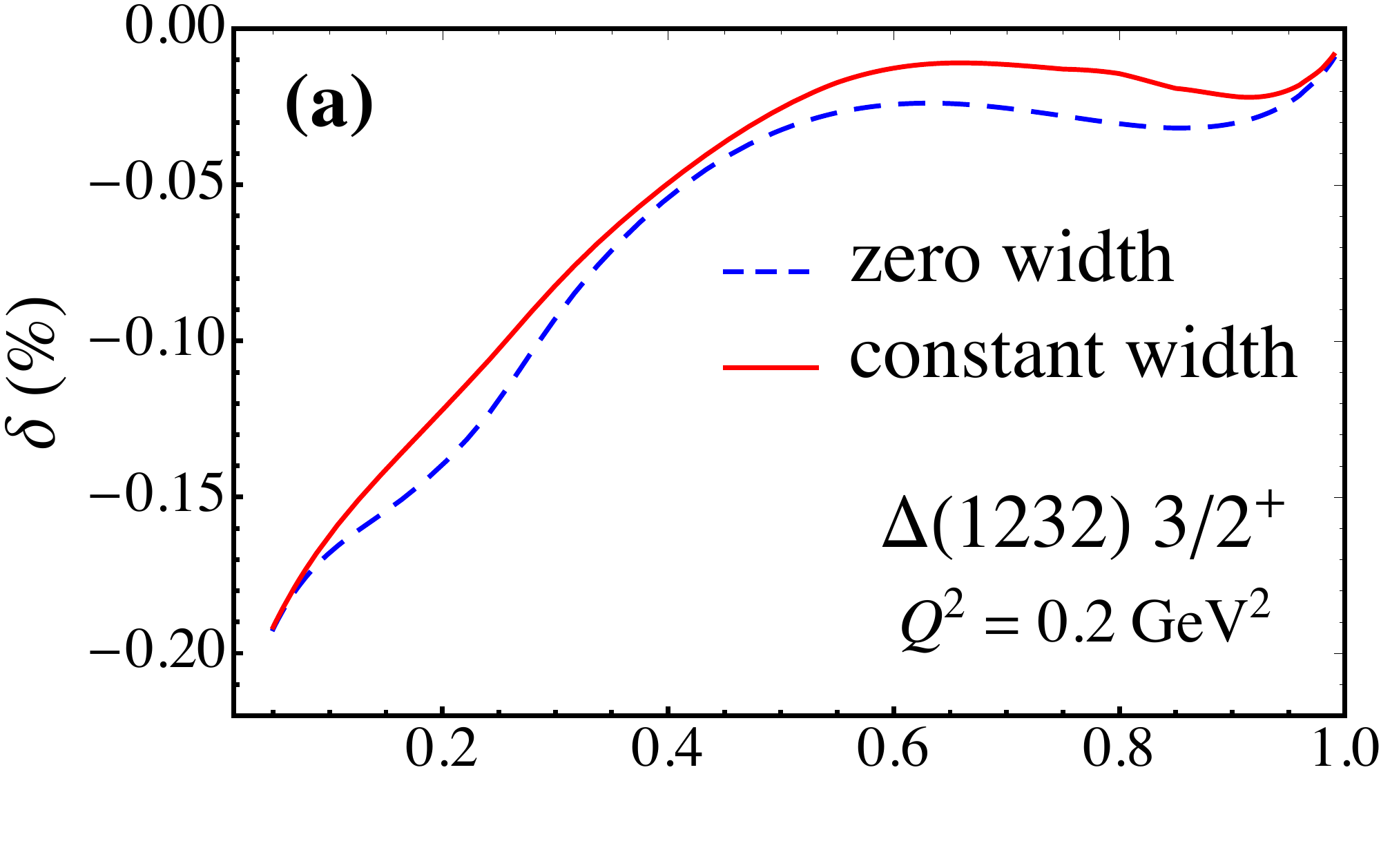} \hspace*{-0.5cm}
\includegraphics[width=5.6cm]{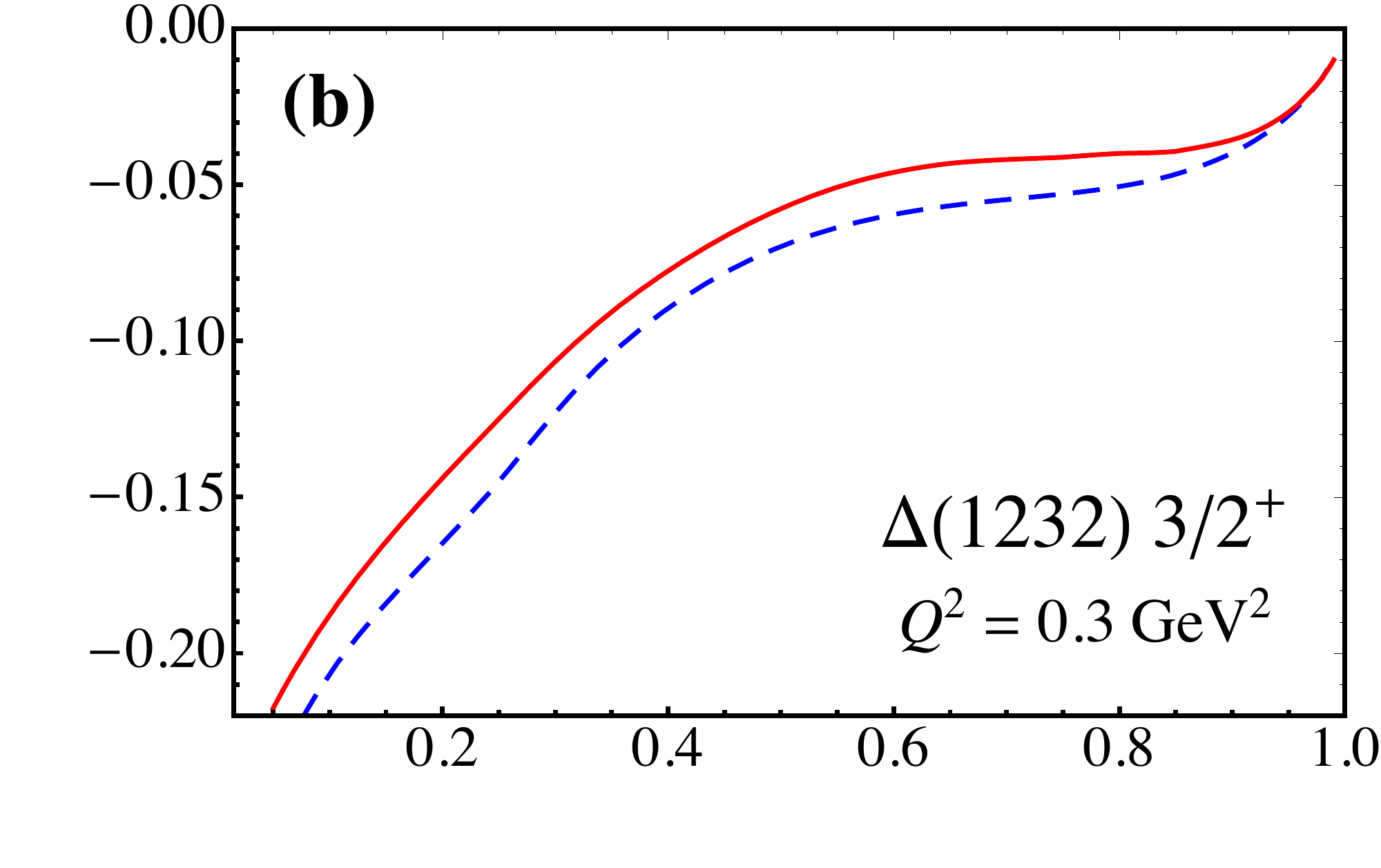} \hspace*{-0.5cm}
\includegraphics[width=5.6cm]{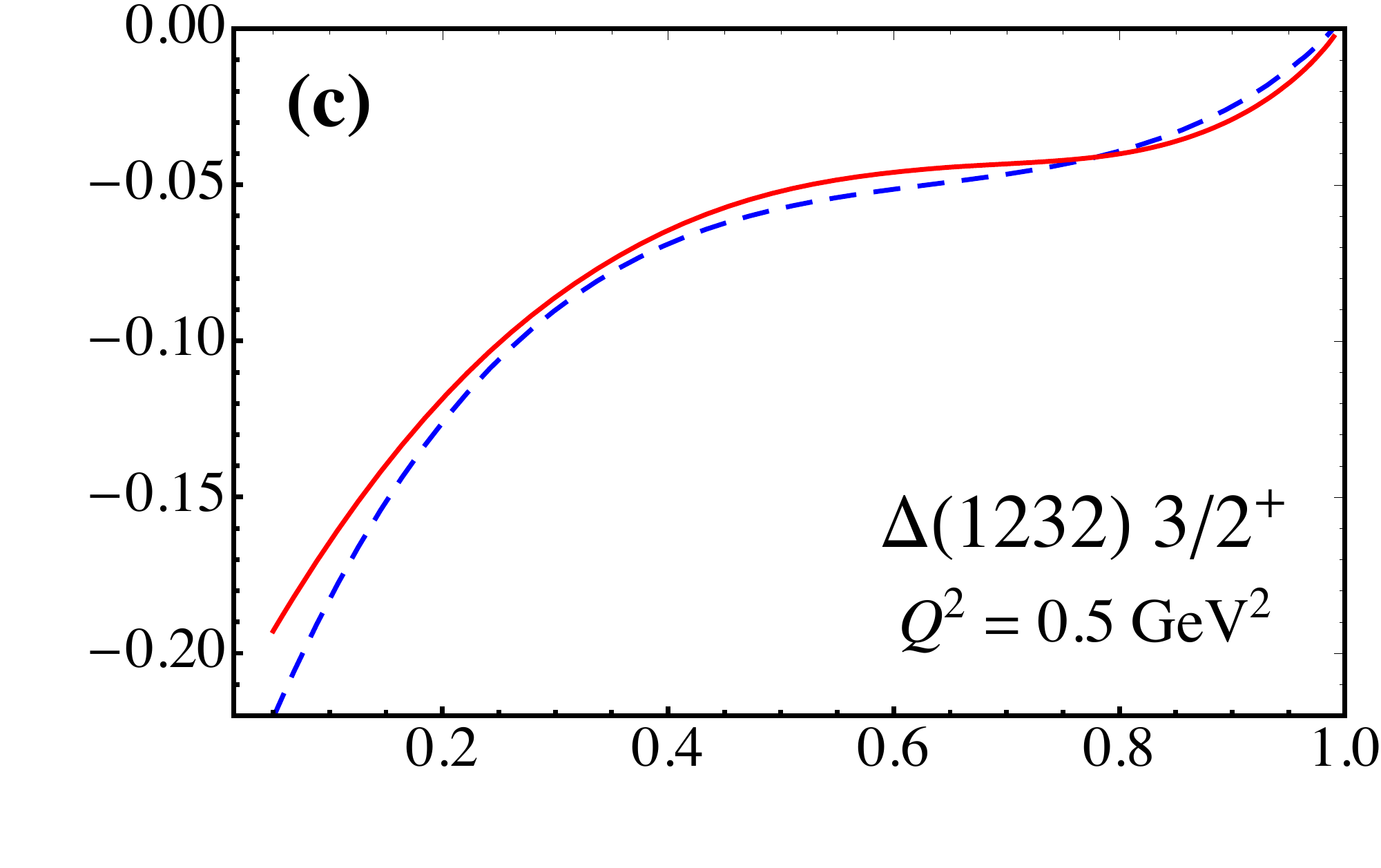}\\ [-0.5em]
\includegraphics[width=5.6cm]{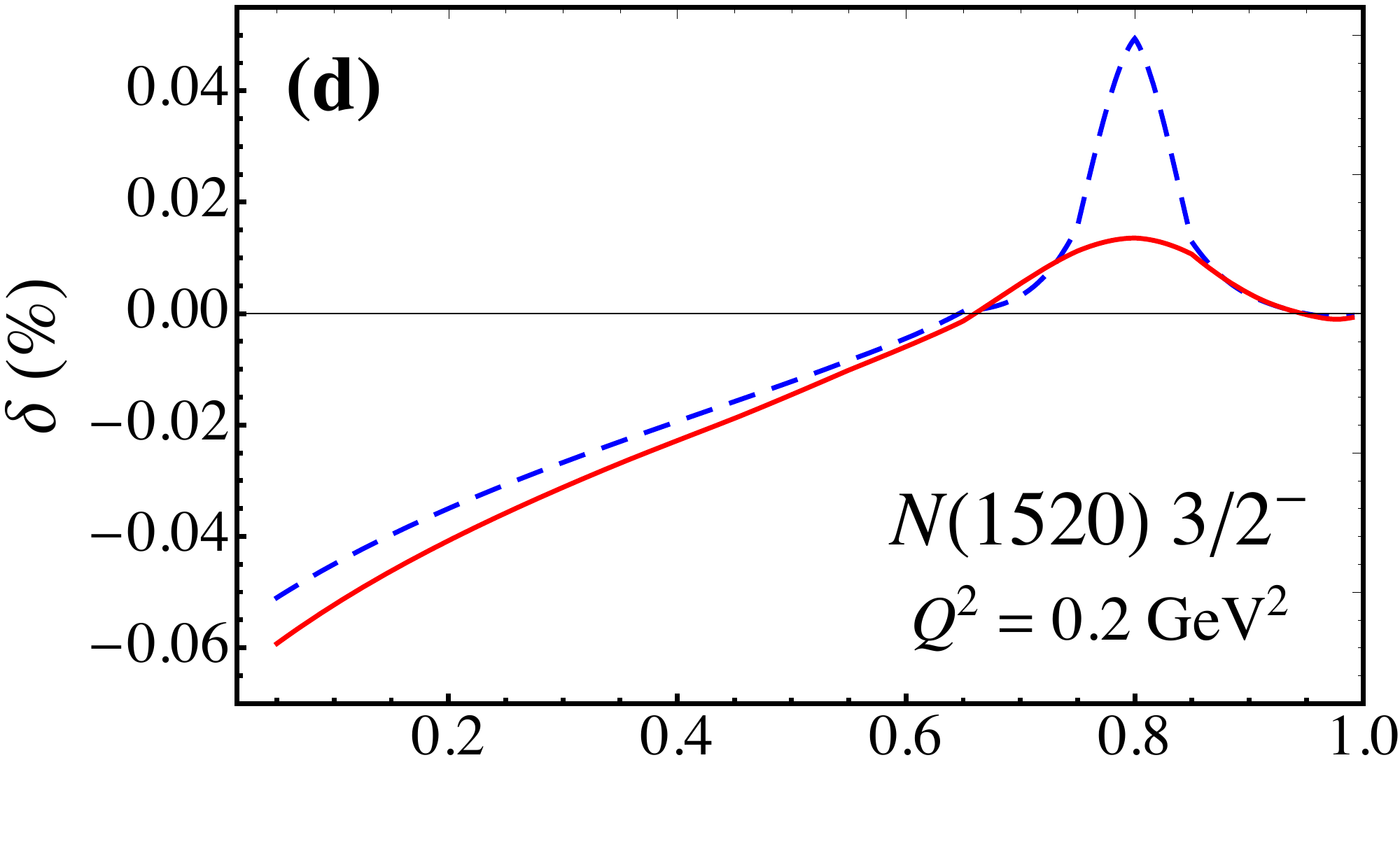} \hspace*{-0.5cm}
\includegraphics[width=5.6cm]{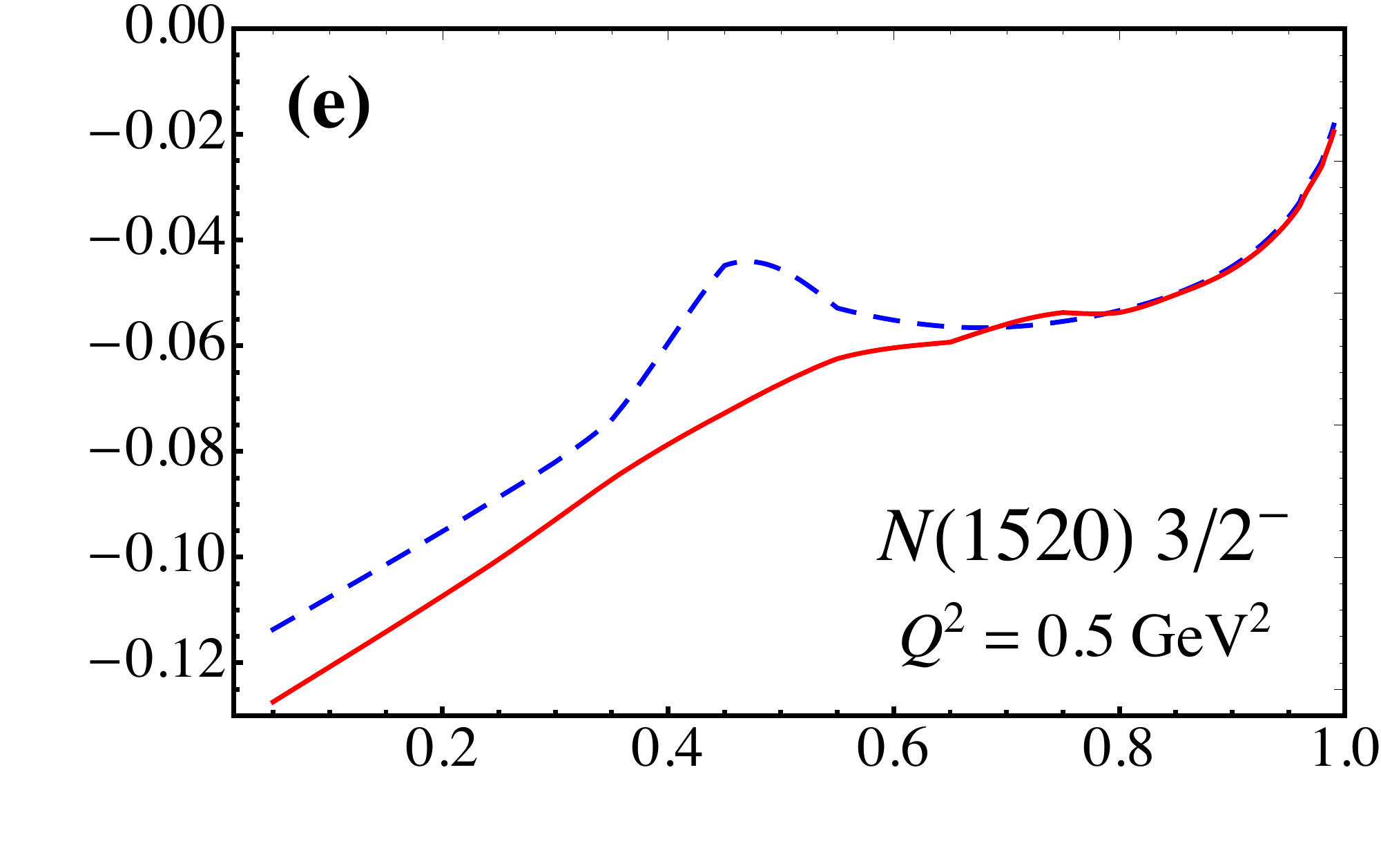} \hspace*{-0.5cm}
\includegraphics[width=5.6cm]{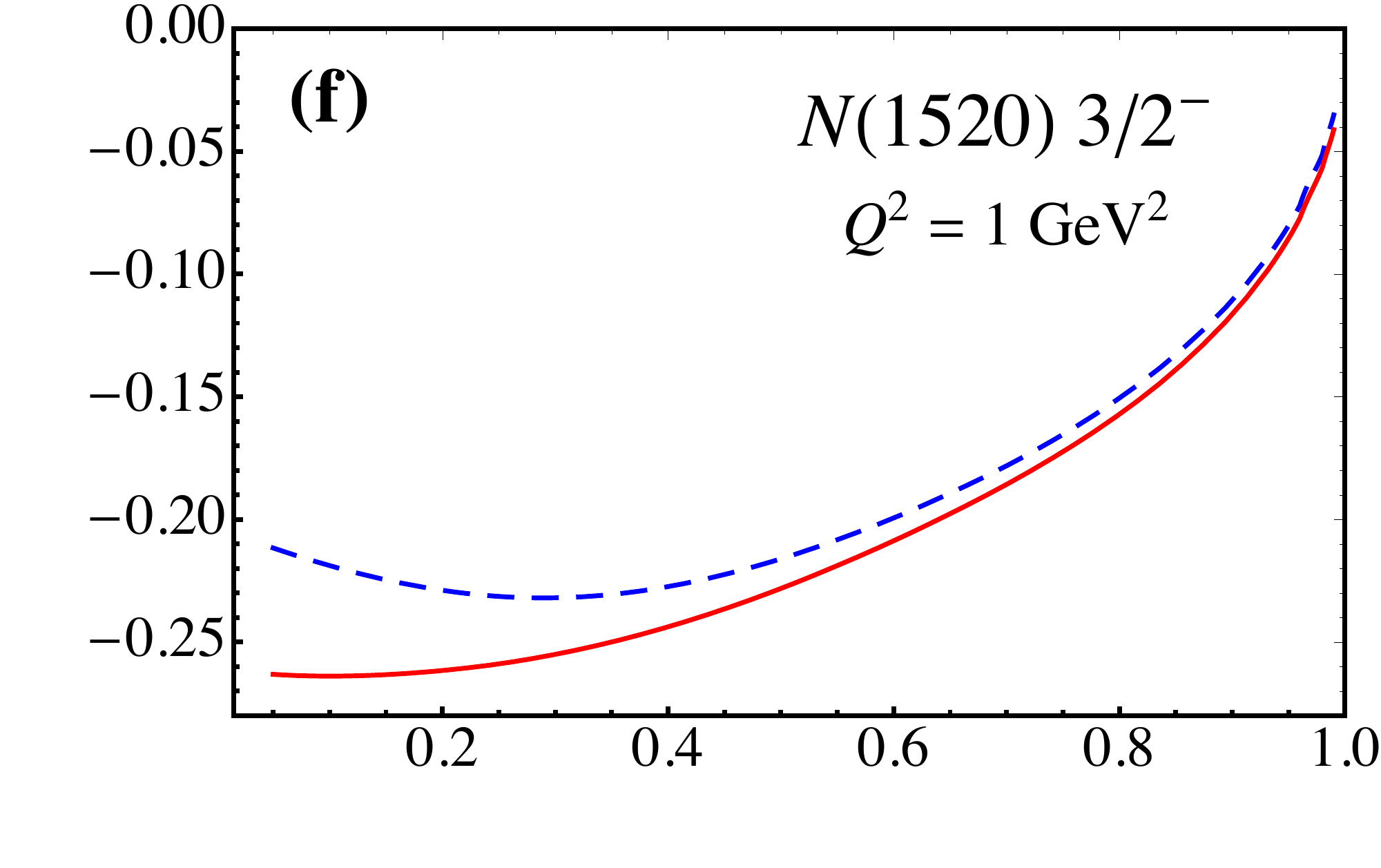}\\ [-0.5em]
\includegraphics[width=5.6cm]{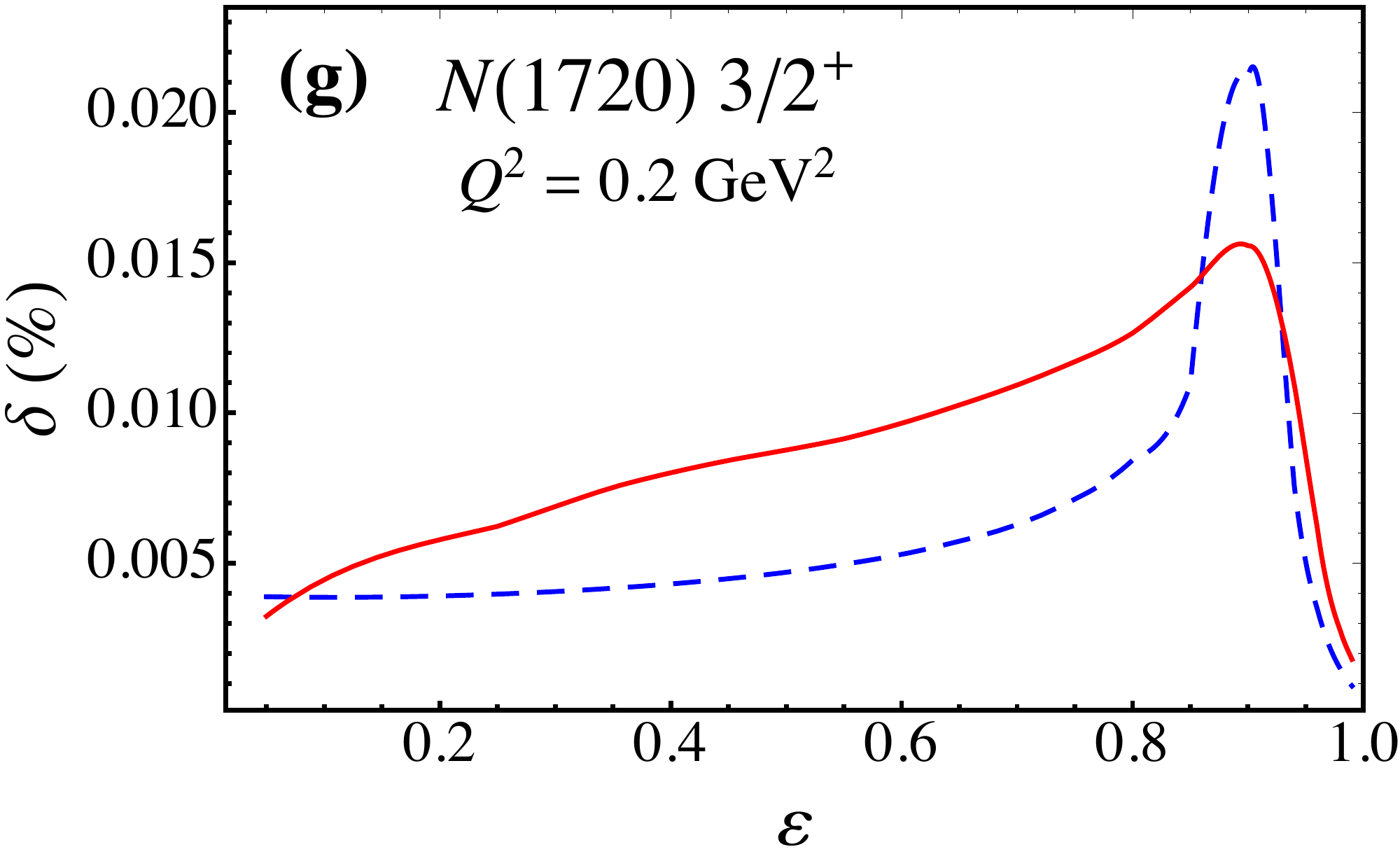} \hspace*{-0.5cm}
\includegraphics[width=5.6cm]{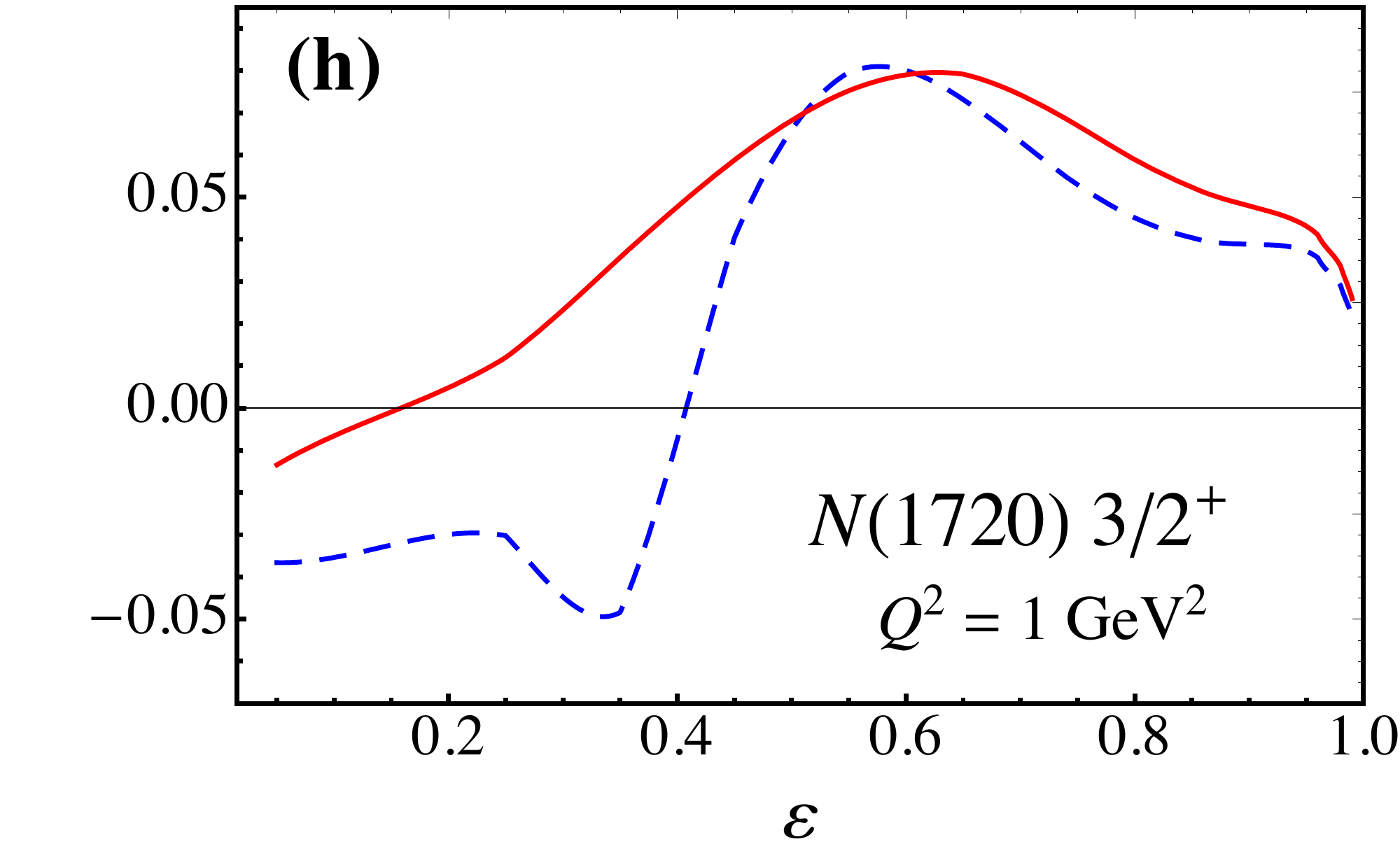} \hspace*{-0.5cm}
\includegraphics[width=5.6cm]{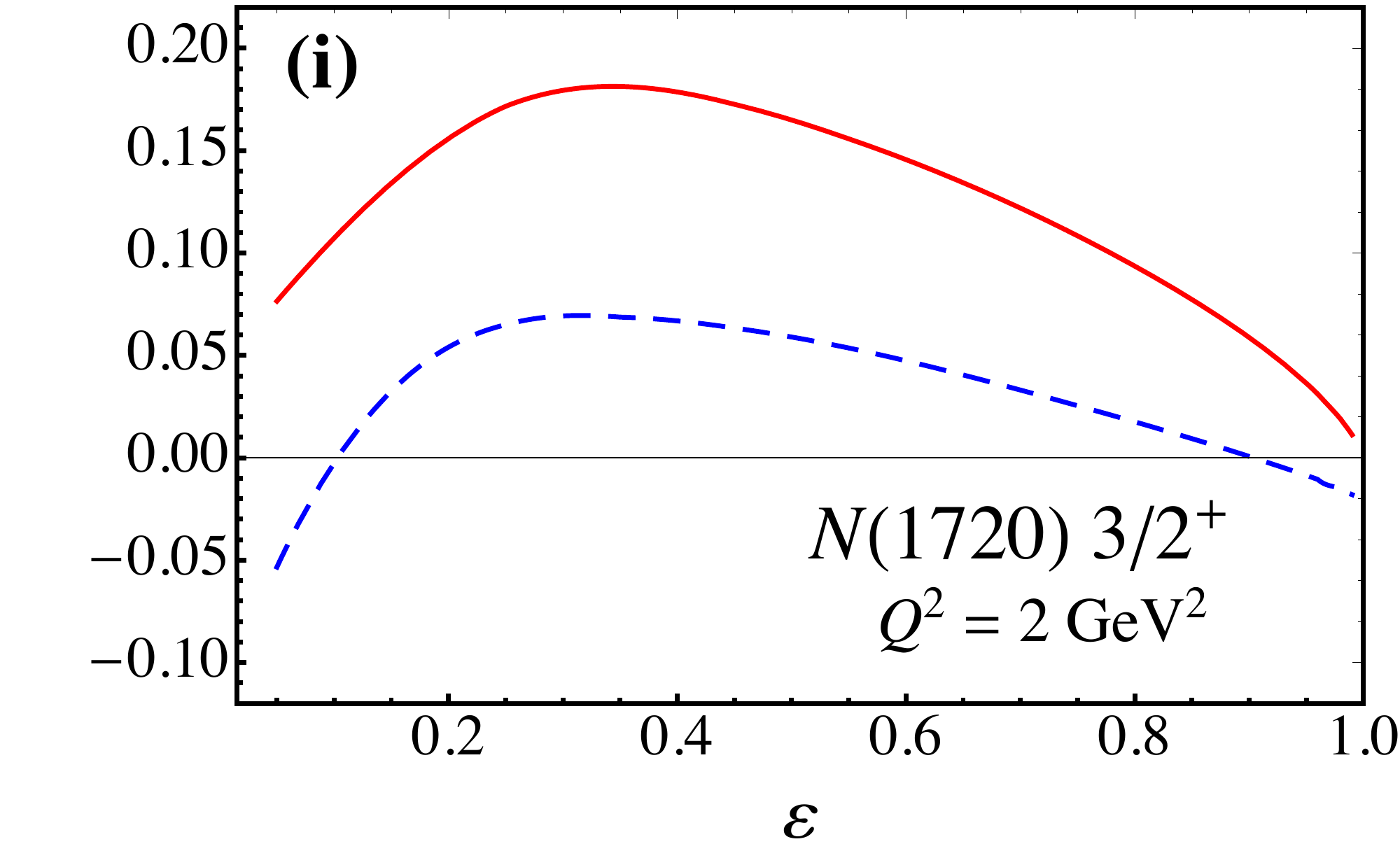}
\caption{Effect of a finite resonance width on the TPE correction $\delta$ (in
\%) from three significant resonance intermediate states,
    $\Delta(1232)~\!3/2^+$ [{\bf (a)}--{\bf (c)}], 
    $N(1520)~\!3/2^-$ [{\bf (d)}--{\bf (f)}], and
    $N(1720)~\!3/2^+$ [{\bf (g)}--{\bf (i)}],
as a function of $\varepsilon$ at fixed $Q^2$ values. The kinematical kinks in
the zero-width results (blue dashed lines) are smoothed out by the effect of the
nonzero, constant width (red solid lines).}
\label{fig.Width1}
\end{figure}

The interplay between the resonance mass and the $Q^2$ and $\varepsilon$ values
at which the threshold effect appears is illustrated in \Cref{fig.Width1}, where
the TPE correction $\delta$ is shown as a function of $\varepsilon$ at several
fixed values of $Q^2$. One observes that the higher the resonance mass, the
higher the $Q^2$ value at which the cusp comes in.
For the lowest-mass $\Delta(1232)$ excitation, the cusp at the lowest
$Q^2=0.2$~GeV$^2$ value occurs at $\varepsilon \approx 0.06$, as indicated by
the wiggle in \Cref{fig.Width1}(a). The effect of the constant, nonzero width,
with a Breit-Wigner distribution centered at the resonance mass, is to smooth
out the wiggles in the calculated $\delta$, although the effect overall is not
dramatic here. At higher $Q^2$, above the kinematic threshold, both curves are
smooth, and the finite width has little impact on the TPE correction
[\Cref{fig.Width1}(b) and (c)].

For the intermediate-mass $N(1520)~\!3/2^-$ resonance, the effect of the
kinematical threshold is more dramatic, with a prominent cusp visible for the
zero-width result at $\varepsilon \approx 0.8$ for $Q^2=0.2$~GeV$^2$
[\Cref{fig.Width1}(d)], and a smaller cusp at $\varepsilon \approx 0.5$ for
$Q^2=0.5$~GeV$^2$ [\Cref{fig.Width1}(e)]. In both cases the finite width of the
resonance washes out the cusps, leaving a smooth function across the threshold.
Above the threshold the contribution to $\delta$ is smooth
[\Cref{fig.Width1}(f)], and the finite width has little impact. The most
dramatic effect is seen for the heaviest $N(1720)~\!3/2^+$ resonance, where the
kinematic threshold produces strong cusps at $\varepsilon \approx 0.9$ for
$Q^2=0.2$~GeV$^2$ [\Cref{fig.Width1}(g)] and $\varepsilon \approx 0.4$ for
$Q^2=1$~GeV$^2$ [\Cref{fig.Width1}(h)]. Once again the finite, constant width
modulates the cusps and leads to considerably smoother results. At
$Q^2=2$~GeV$^2$, above the kinematic threshold for this state, both the
zero-width and finite-width results produce smooth curves, but the effect of the
latter is still numerically significant [\Cref{fig.Width1}(i)].

To test the model dependence of the TPE correction on the resonance width
prescription, we also consider the effect of including an energy-dependent
dynamic decay width, $\Gamma(W)$, of Eq.~(\ref{eq.BW}) for each resonant
intermediate state. We consider the energy-dependant $\Gamma(W)$ to have
contributions from three different decay channels for each resonances, namely,
$\pi N$, $\pi \pi N$ and $\eta N$,
\be
\bal
\Gamma(W) = \Gamma_{\pi N}(W)+ \Gamma_{\pi \pi N}(W)+ \Gamma_{\eta N}(W).
\eal
\label{eq.width1}
\ee
Following Ref.~\cite{HillerBlin:2019hhz}, the partial decay widths
$\Gamma_{\pi(\eta) N}(W)$ and $\Gamma_{\pi \pi N}(W)$ are parameterized as
\begin{subequations}
\label{eq.width23}
\bea
\Gamma_{\pi(\eta) N}(W)
&=& \Gamma_{R}\, \beta_{\pi(\eta)N}
    \left( \dfrac{p_{\pi(\eta)}(W)}{p_{\pi(\eta)}(W_R)} \right)^{2L_{R}+1}
    \left( \dfrac{X^2+p_{\pi(\eta)}^2(W_R)}{X^2+p_{\pi(\eta)}^2(W)} \right)^{L_R}, \\
\label{eq.width2}
\Gamma_{\pi\pi N}(W)
&=& \Gamma_{R}\, \beta_{\pi\pi N}
    \left( \dfrac{p_{\pi\pi}(W)}{p_{\pi\pi}(W_R)} \right)^{2L_R+4}
    \left( \dfrac{X^2+p_{\pi\pi}^2(W_R)}{X^2+p_{\pi\pi}^2(W)} \right)^{L_R+2},
\label{eq.width3}
\eea
\end{subequations}
where the constant total decay width $\Gamma_R$ of each resonance state is taken
from Ref.~\cite{HillerBlin:2019hhz}, and we have assumed the centrifugal barrier
penetration factors to be the major contributors to the off-shell behavior of
the resonances. Here the energy and momentum factors for the two-body channels
are given by
\begin{subequations}
\bea
p_{\pi(\eta)}(W)
&=& \sqrt{E_{\pi(\eta)}^2(W) - m_{\pi(\eta)}^2},    \\
E_{\pi(\eta)}(W)
&=& \dfrac{W^2 + m_{\pi(\eta)}^2 - M^2}{2W},
\eea
\end{subequations}
and for the three-body channel is given by
\begin{subequations}
\bea
p_{\pi\pi}(W)
&=& \sqrt{E_{\pi\pi}^2(W) - 4 m_\pi^2},             \\
E_{\pi\pi}(W)
&=& \dfrac{W^2 + 4m_{\pi}^2 - M^2}{2W},
\eea
\end{subequations}
where $m_{\pi(\eta)}$ is the mass of pion ($\eta$ meson).
The branching fractions for the resonance decays into the $\pi N$, $\pi \pi N$
and $\eta N$ channels are given by $\beta_{\pi N}$, $\beta_{\pi \pi N}$ and
$\beta_{\eta N}$, respectively, and satisfy the relation $\beta_{\pi N} +
\beta_{\pi \pi N} + \beta_{\eta N} = 1$. The values of the other parameters in
Eqs.~(\ref{eq.width23}) --- $X$, $L_R$, $\beta_{\pi N}$, $\beta_{\pi \pi N}$ and
$\beta_{\eta N}$ --- are taken from Ref.~\cite{HillerBlin:2019hhz}. 

\begin{figure}[h!]%
\graphicspath{{Images/}}
\includegraphics[width=8.0cm]{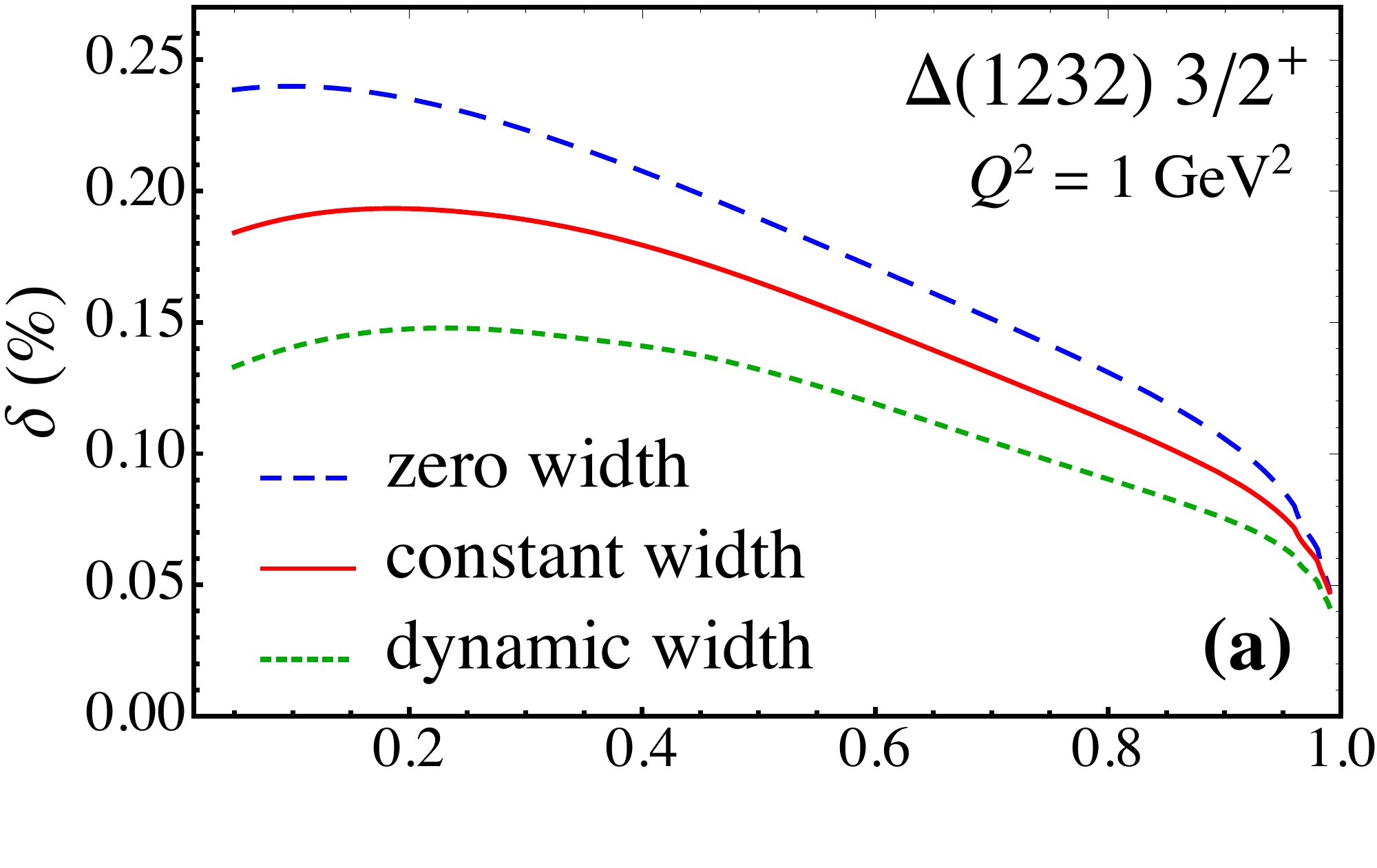} \hspace*{-0.3cm}
 \includegraphics[width=7.85cm]{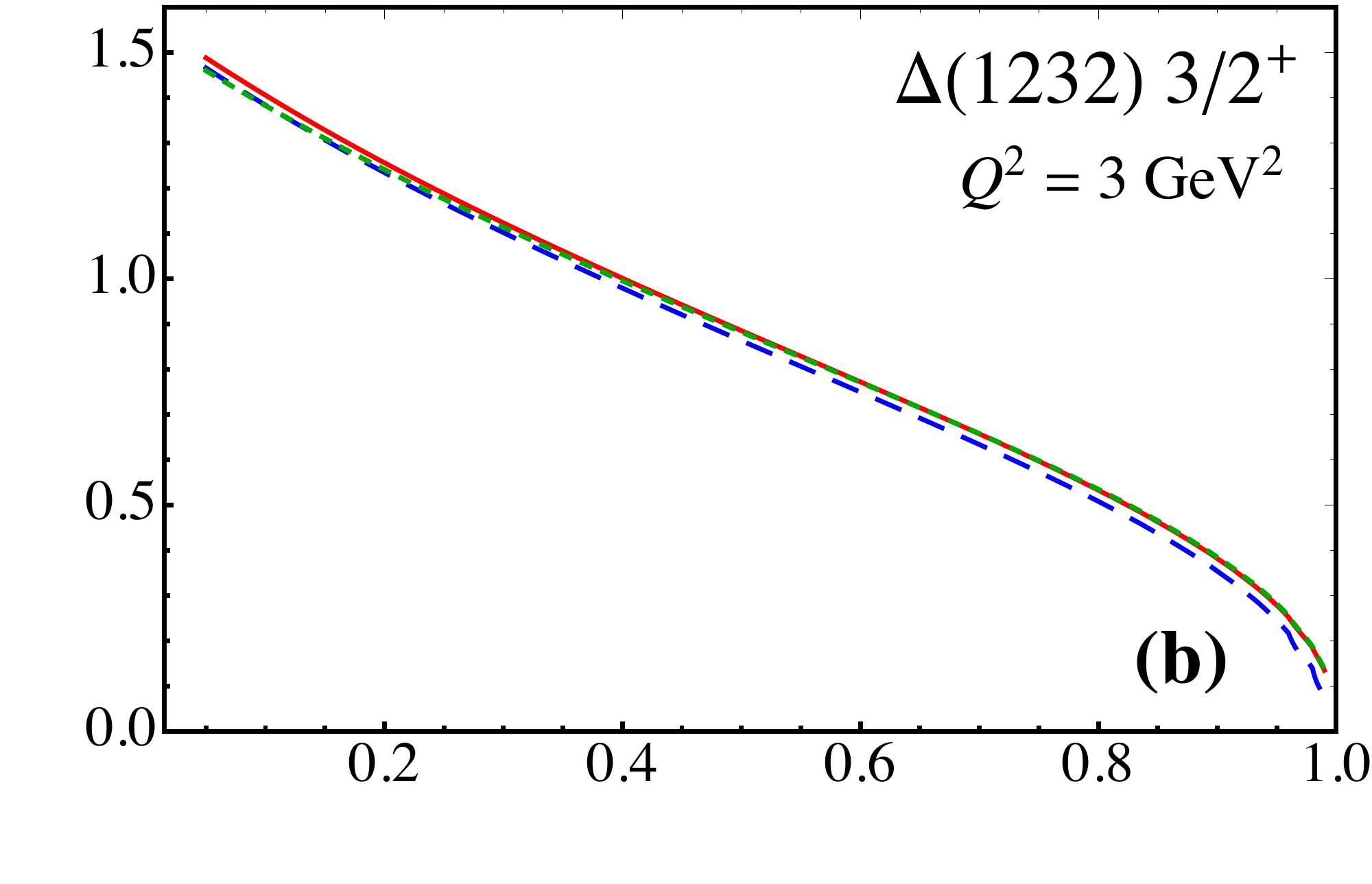} \\ [-0.7em] \hspace*{-0.3cm} 
\includegraphics[width=8.3cm]{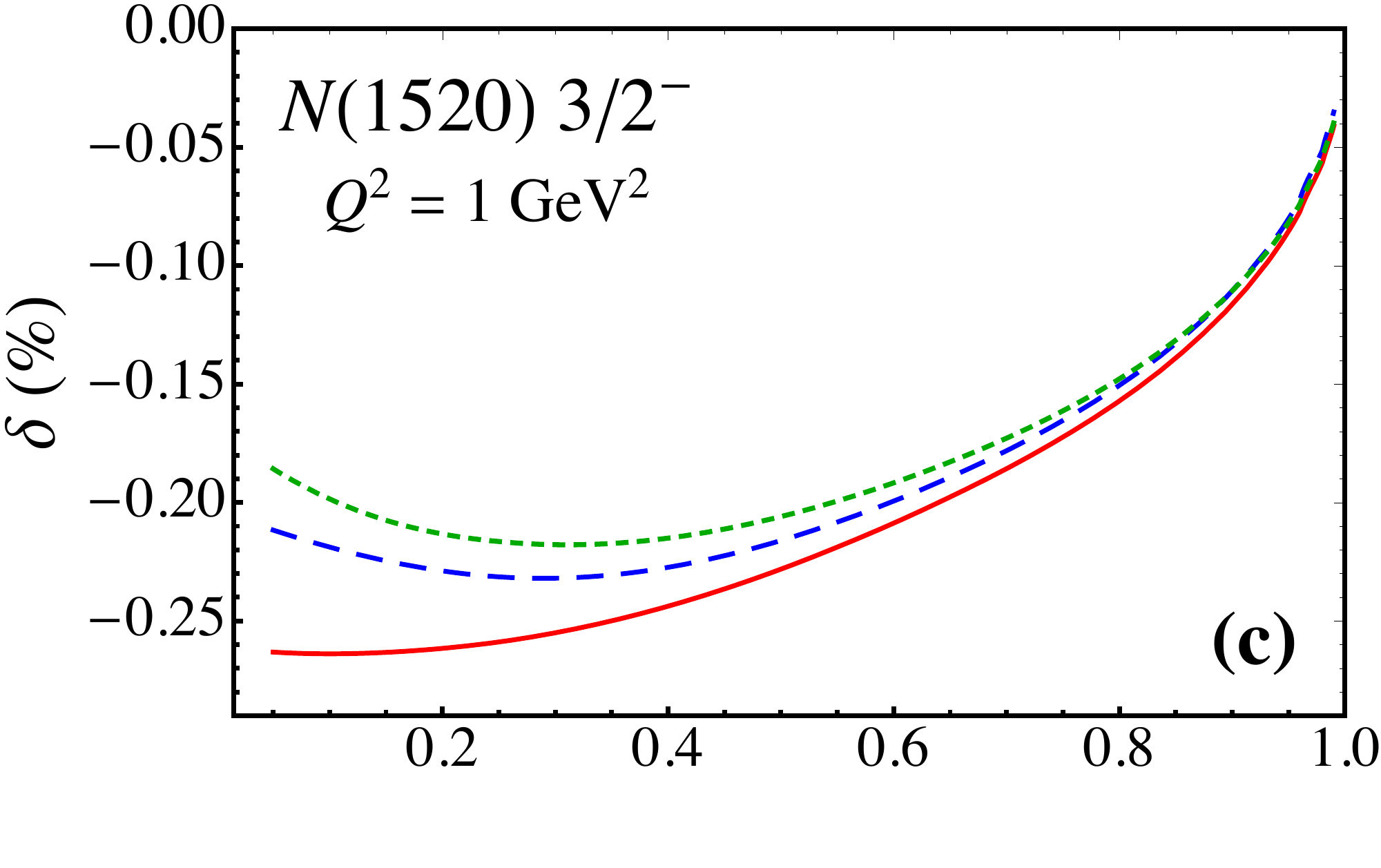} \hspace*{-0.5cm}
\includegraphics[width=8.15cm]{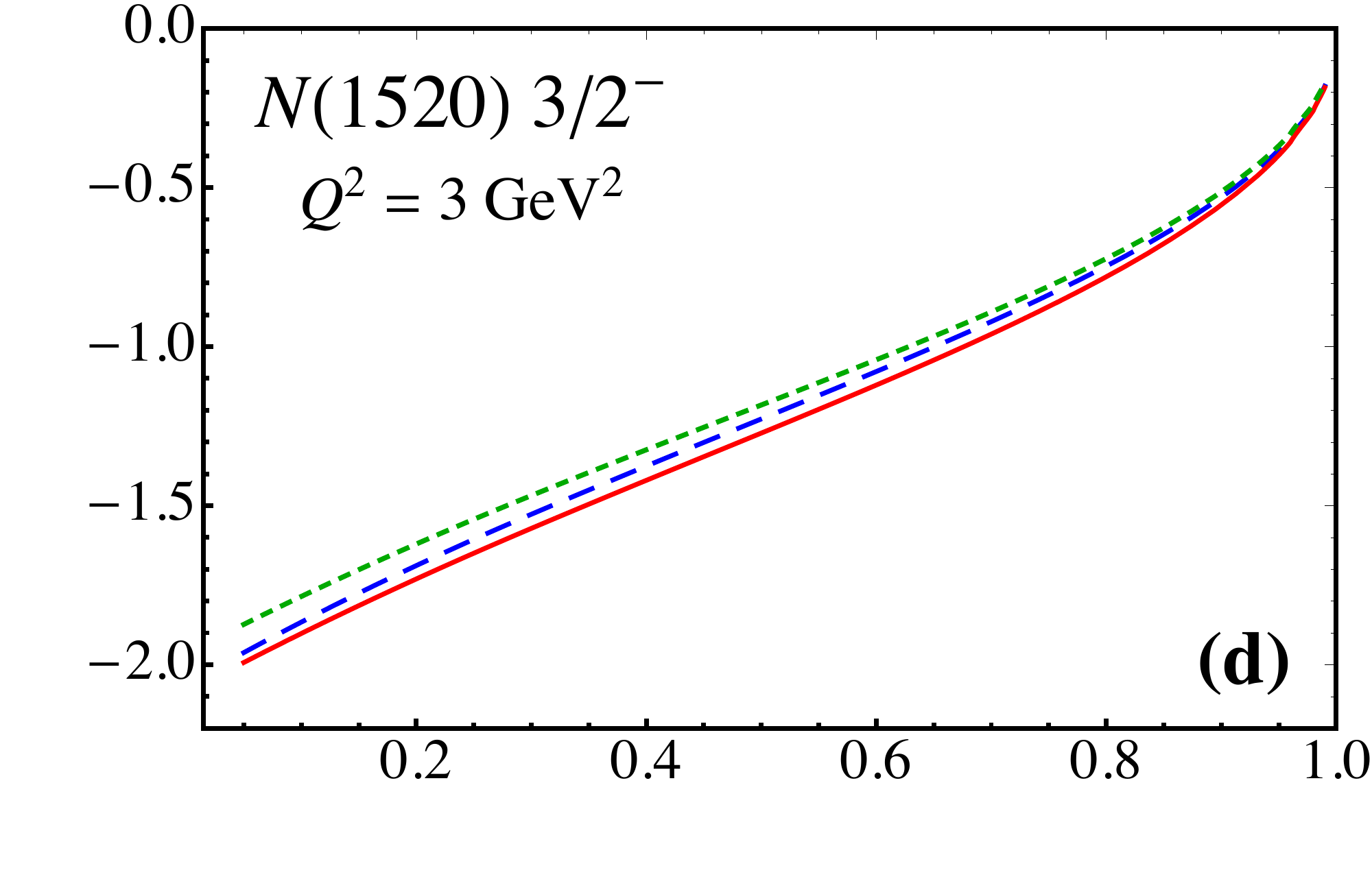} \\ [-0.7em] \hspace*{-0.1cm} 
\includegraphics[width=8.15cm]{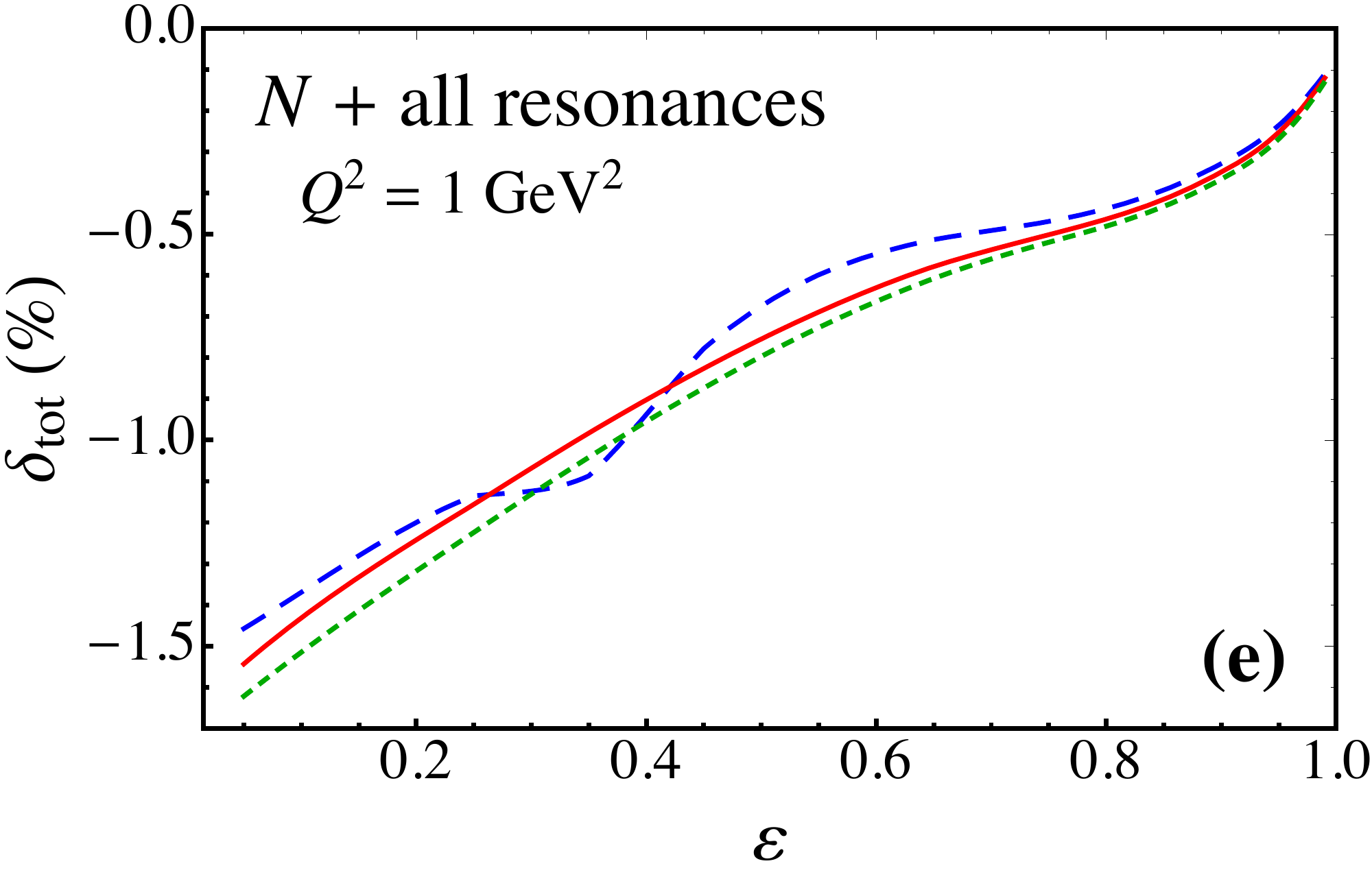} \hspace*{-0.5cm}
\includegraphics[width=8.15cm]{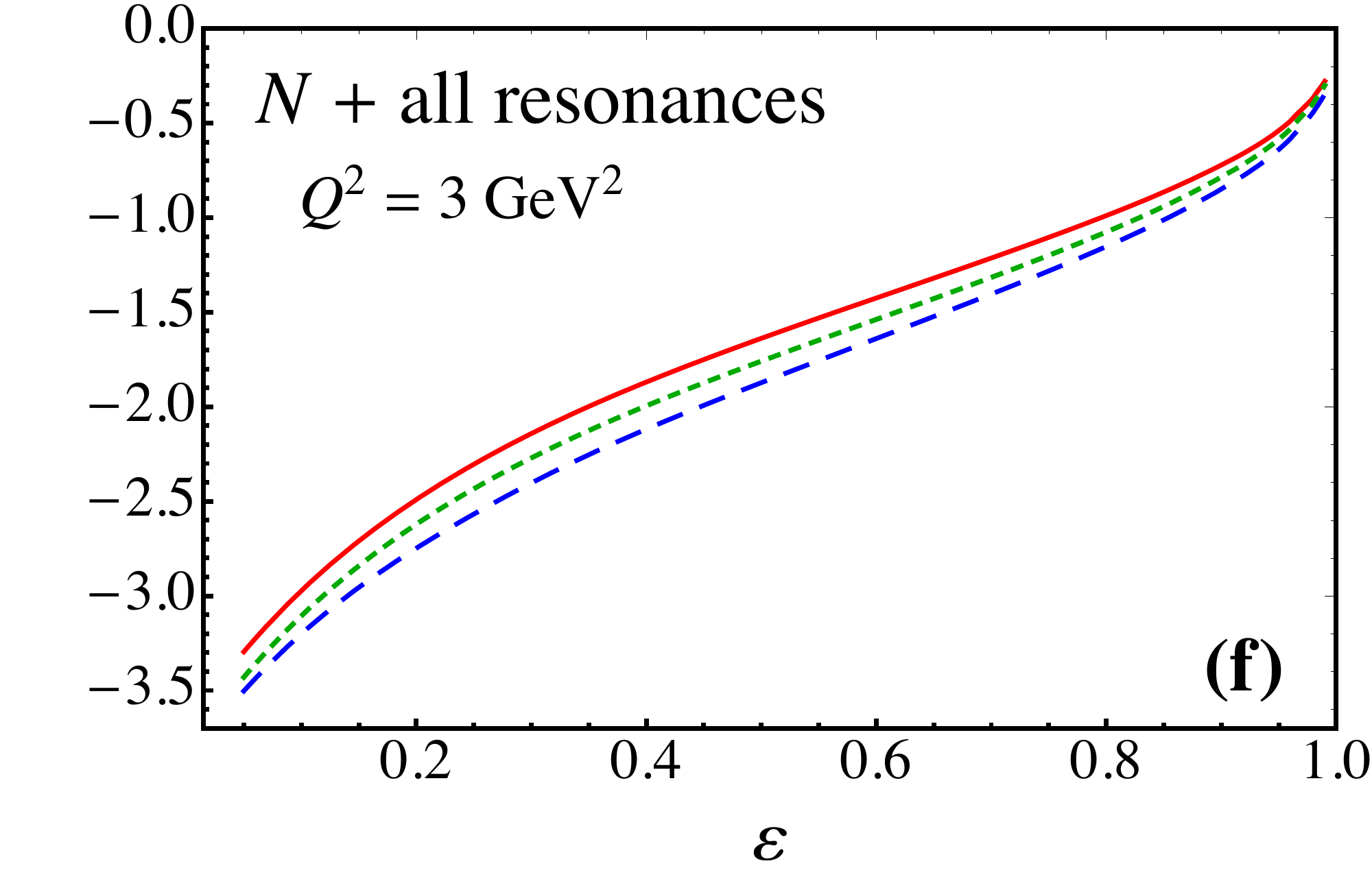}
\caption{Comparison of the TPE correction $\delta$ (in \%) computed for
resonances with zero width (blue dashed lines), constant width (red solid lines)
and a dynamical width (green dotted lines) for $Q^2=1$~GeV$^2$ (left panels) and
3~GeV$^2$ (right panels). Contributions from the $\Delta(1232)~\!3/2^+$ [{\bf
(a), (b)}] and $N(1520)~\!3/2^-$ [{\bf (c), (d)}] states are shown separately,
along with the sum of all resonance contributions [{\bf (e), (f)}].}
\label{fig.Width2}
\end{figure}

To illustrate the effect of the dynamical width, we select the two major
resonance contributors to the total cross section, namely, the
$\Delta(1232)~\!3/2^+$ and $N(1520)~\!3/2^-$ states. In \Cref{fig.Width2}(a)-(d)
we compare the TPE correction $\delta$ using the dynamic, energy-dependent width
with the results of the zero-width and constant-width calculations at fixed
$Q^2=1$ and 3~GeV$^2$. At the higher $Q^2=3$~GeV$^2$ value, well above the
kinematic thresholds, the dependence on the prescription for the width is
negligibly small, with the dynamic- and constant-width results very similar to
those for the zero-width case. On the other hand, at $Q^2=1$~GeV$^2$ the details
of the treatment of the widths are more important. In particular, for the
$\Delta(1232)~\!3/2^+$ the dynamical width leads to an $\approx 30\%$ reduction
of the (positive) correction relative to the zero-width case across all
$\varepsilon$, and a smaller but non-negligible increase in the (negative)
$N(1520)~\!3/2^-$ contribution at backward angles.

For the higher-mass resonances, the contributions again enter with oscillating
signs, producing a net effect of the width in the total TPE cross section ratio
$\delta_{\rm tot}$, including nucleon elastic and all excited resonance states,
that is very small across all $\varepsilon$ values for both $Q^2=1$ and
3~GeV$^2$ [\Cref{fig.Width2}(e)-(f)] for all three width prescriptions. The kink
in the zero-width result at $\varepsilon \approx 0.4$ for $Q^2=1$~GeV$^2$ arises
from threshold effects in the third resonance region [see Table~\ref{tab.cusp}
and Fig.~\ref{fig.Width1}(h)]. As for the $\Delta(1232)$ and $N(1520)$, the kink
is eliminated by the tail effects of the resonances for either the
constant-width or dynamical-width approximation, producing a smooth, monotonic
result. At the higher $Q^2=3$~GeV$^2$ value the effects of the finite widths are
negligible. Since the differences between the constant- and dynamical-width
results are generally not large, for computational simplicity we employ the
constant decay width approximation as the default throughout this work.

\subsection{Spin, isospin and parity dependence}
\label{ssec.spin}

\begin{figure}[t]
\graphicspath{{Images/}}
\includegraphics[width=8.15cm]{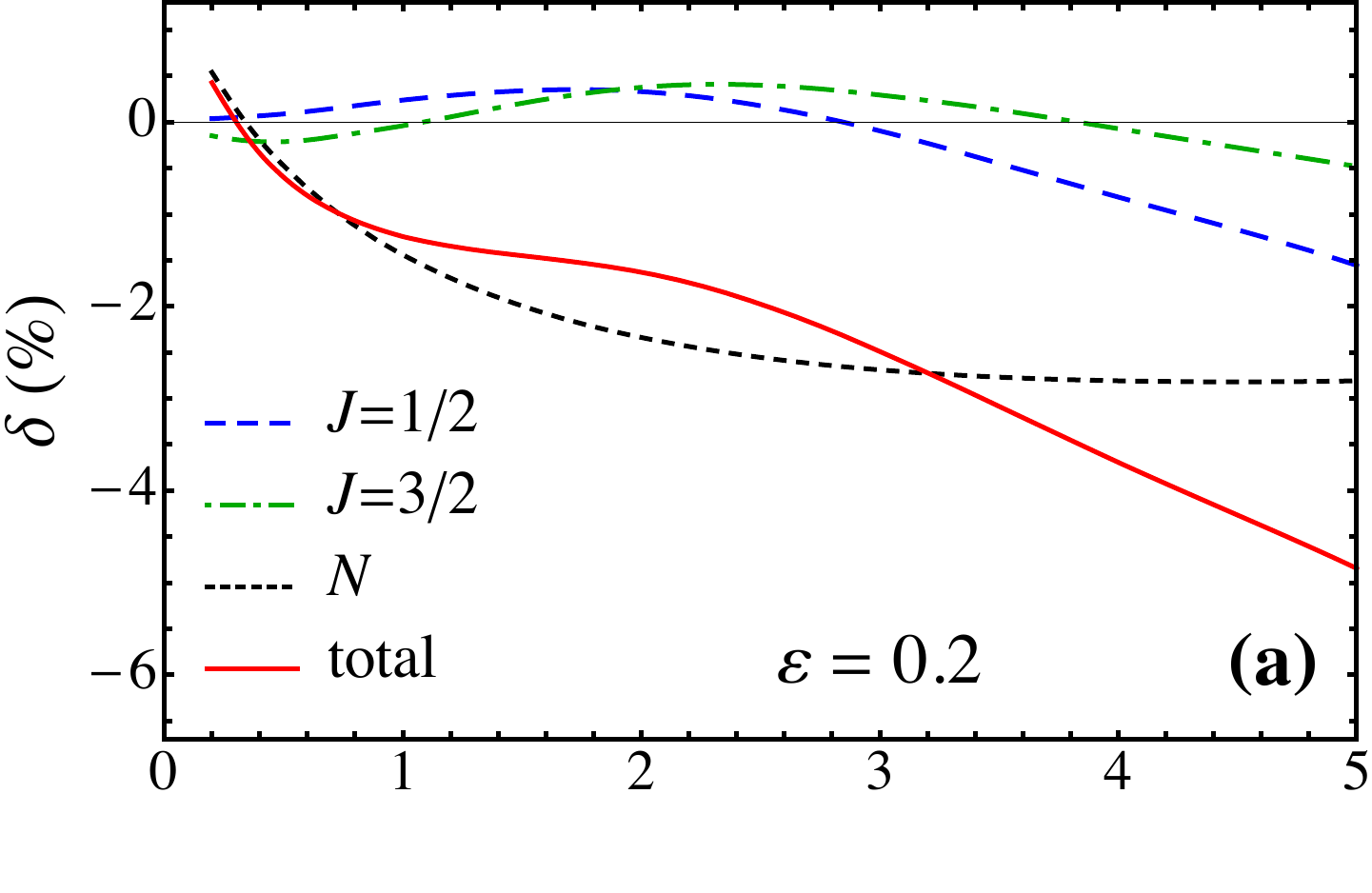} \hspace*{-0.5cm}
\includegraphics[width=8.40cm]{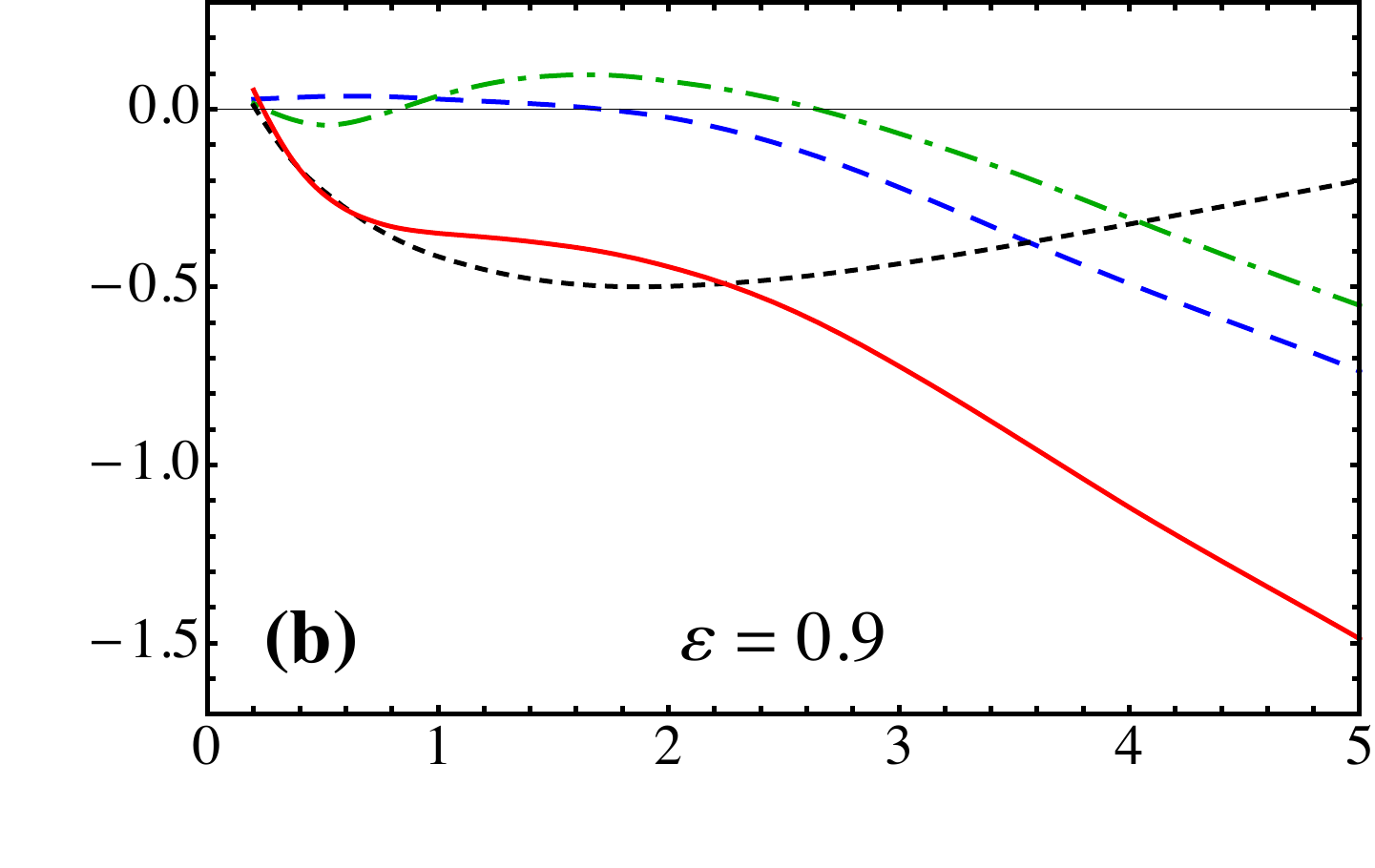} \\ [-0.7em]
\hspace*{-0.1cm}
\includegraphics[width=8.20cm]{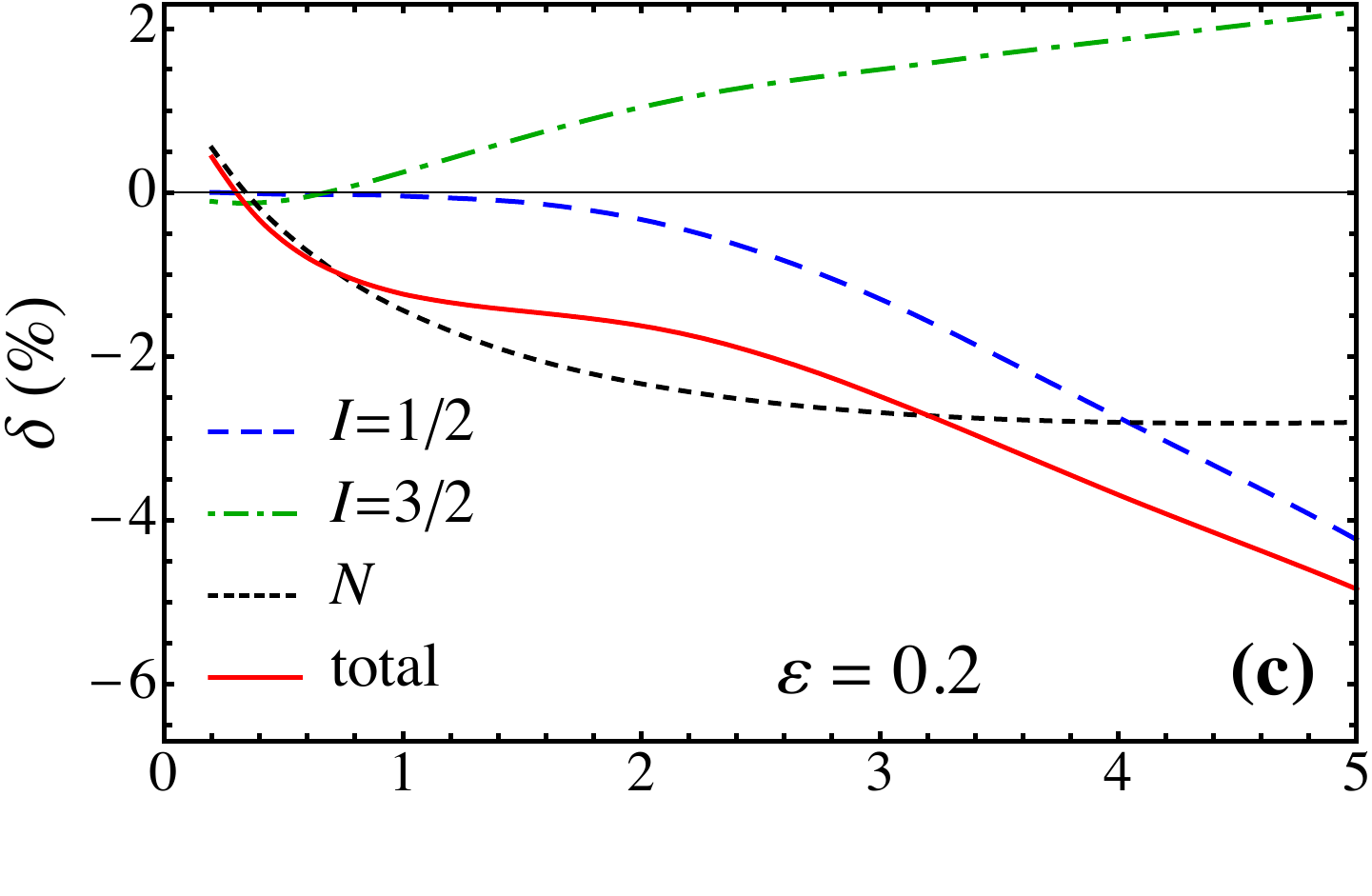} \hspace*{-0.5cm}
\includegraphics[width=8.45cm]{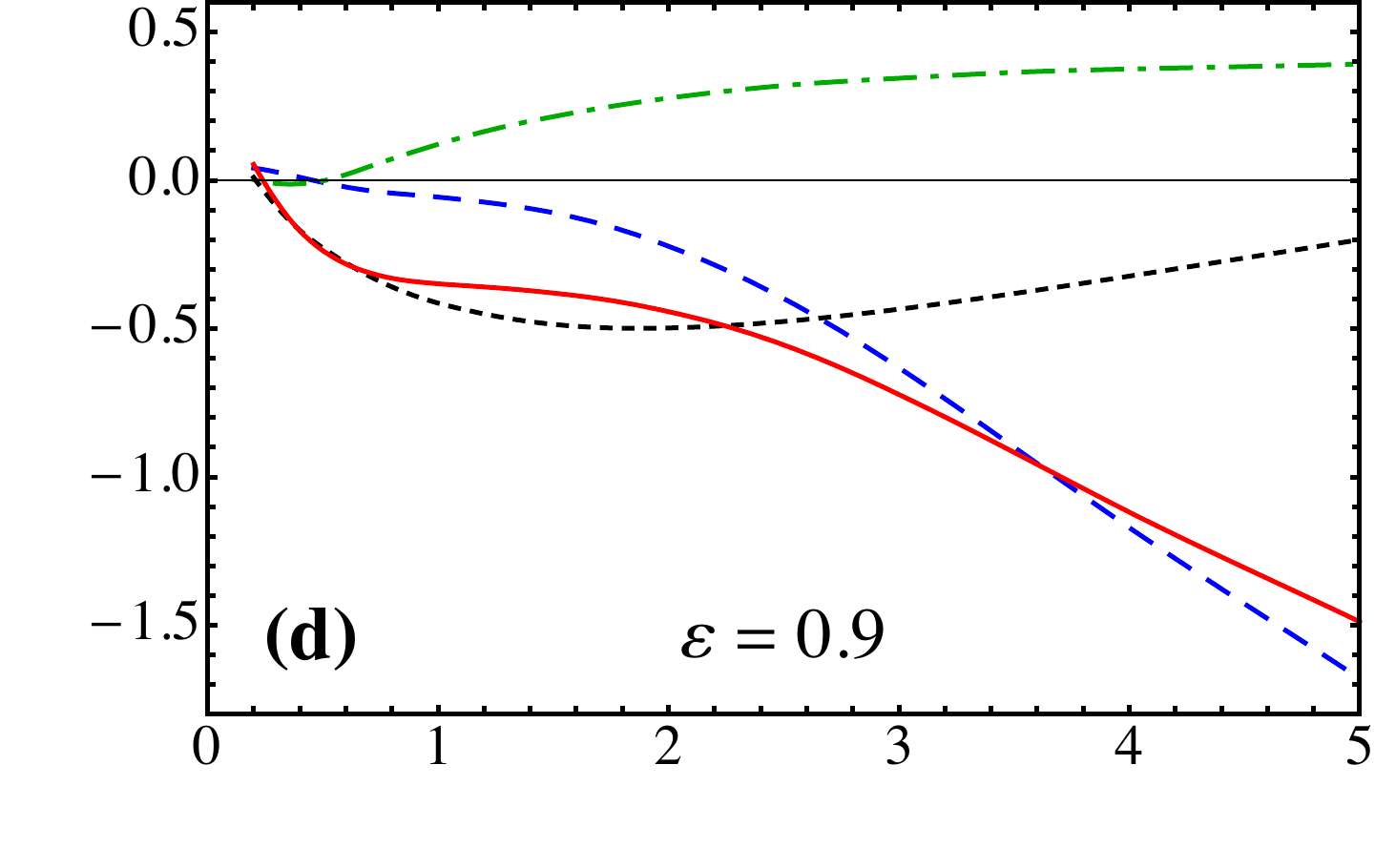} \\ [-0.7em]
\hspace*{-0.1cm} 
\includegraphics[width=8.20cm]{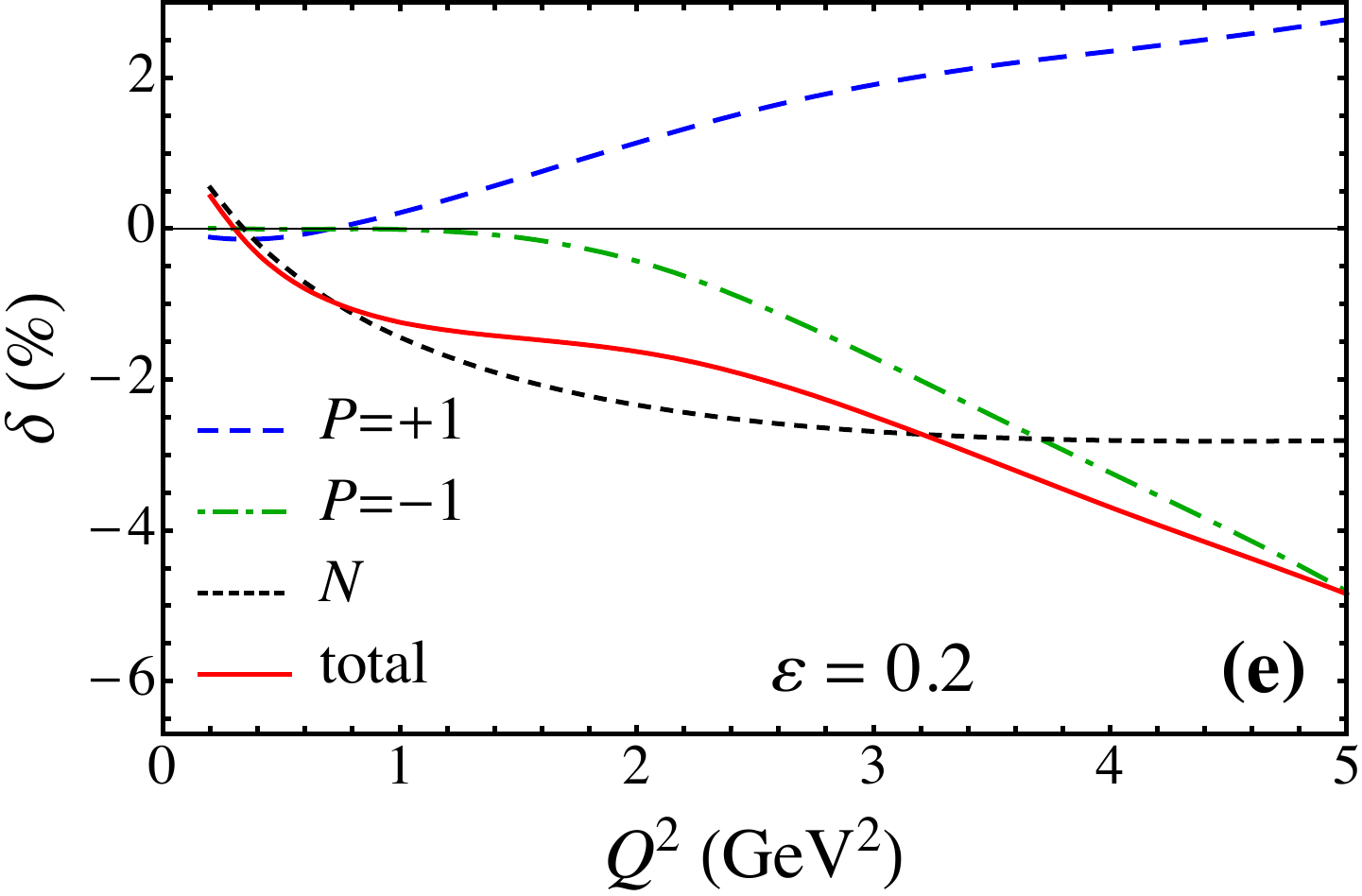} \hspace*{-0.5cm}
\includegraphics[width=8.45cm]{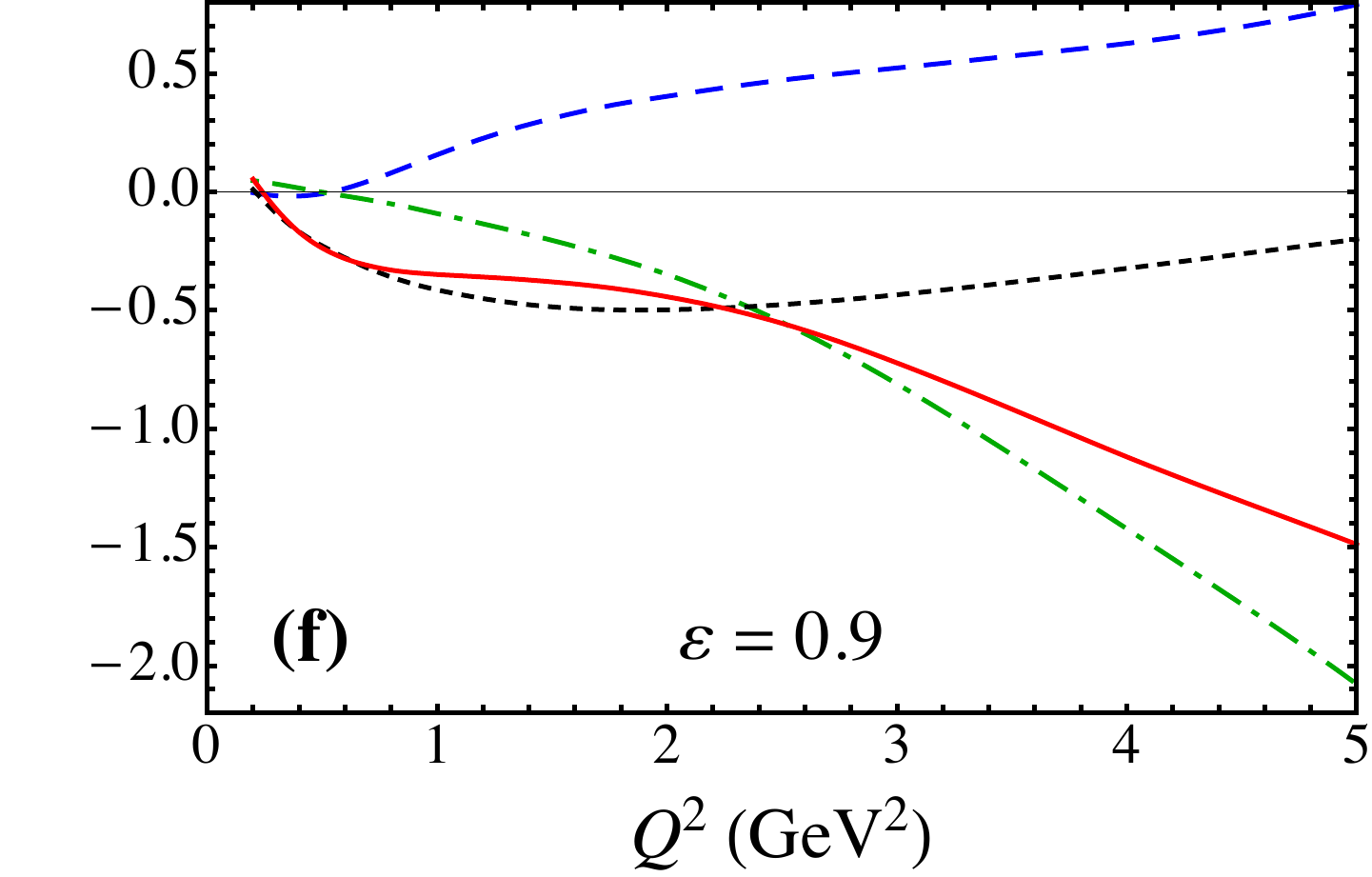}
\caption{Comparison between the contributions to the TPE correction $\delta$ (in
\%) from intermediate state resonances with spin $J=1/2$ and $J=3/2$ [{\bf (a),
(b)}], isospin $I=1/2$ and $I=3/2$ [{\bf (c), (d)}], and even parity $P=+1$ and
odd parity $P=-1$ [{\bf (e), (f)}], for $\varepsilon=0.2$ (left columns) and
$\varepsilon=0.9$ (right columns). The nucleon-only contribution (black dotted
lines), which is not included in the other curves, and the total (red solid
lines) are shown for comparison in each panel.}
\label{fig.isospin}
\end{figure}

To further investigate the systematics of the TPE corrections from various
intermediate states resonances, we compare the relative contributions from
resonances with similar spin $J$, isospin $I$ and parity $P$. In
Fig.~\ref{fig.isospin} we show the combined effects of the different groupings
versus $Q^2$ for two representative values of $\varepsilon$, where the TPE
effects are relatively large (backward angles, $\varepsilon=0.2$) and where they
are relatively small (forward angles, $\varepsilon=0.9$). To contrast the impact
of the exicted states, we show the resonance contributions separately from the
nucleon elastic channel and the total (both of which are the same in the left
and right columns).

For the resonance contributions with different spin,
Fig.~\ref{fig.isospin}(a)-(b) shows qualitatively similar effects from excited
states with spin $J=1/2$ and those with spin $J=3/2$. The sum of the resonances
in both channels is significantly smaller than the nucleon elastic at low values
of $Q^2$, and only starts to become non-negligible for larger $Q^2$, $Q^2
\gtrsim (3-4)$~GeV$^2$, with the relative impact somewhat greater at high
$\varepsilon$ than at low $\varepsilon$. The total TPE correction $\delta$ is
therefore well approximated by the elastic term alone for $Q^2 \lesssim
3$~GeV$^2$ at $\varepsilon=0.2$ and $Q^2 \lesssim 2$~GeV$^2$ at
$\varepsilon=0.9$.

The decomposition into contributions from different isospins in
Fig.~\ref{fig.isospin}(c)-(d) is rather more dramatic. Large cancellations occur
between the (negative) isospin $I=1/2$ intermediate states and the (positive)
$I=3/2$ states. At lower $Q^2$, $Q^2 \lesssim 2$~GeV$^2$, the $I=3/2$
transitions are dominant, while at larger $Q^2$ the $I=1/2$ intermediate states
become more important, rendering the TPE effect more negative compared with the
nucleon elastic term alone and contributing to the rapid increase in magnitude
of the (negative) total TPE correction with $Q^2$. This qualitative behavior is
similar at low and high $\varepsilon$.

Interestingly, a similar cancellation is found between the parity-even ($P=+1$)
and parity-odd ($P=-1$) intermediate states in Fig.~\ref{fig.isospin}(e)-(f). In
this case the $P=+1$ contributions to $\delta$ are positive while the $P=-1$
contributions are negative, with the latter becoming more important with
increasing $Q^2$. The qualitative behavior of the curves for each of the spin,
isospin and parity decompositions can be understood from the results illustrated
in Fig.~\ref{fig.sig}, where numerically the largest positive contribution is
seen to be from the $\Delta(1232)~\!3/2^+$ and the negative of that from
$N(1520)~\!3/2^-$ states. The former dominates the isospin 3/2 and even-parity
channels, while the latter dominates the isospin 1/2 and odd-parity channels,
but since both have spin 3/2 and enter with opposite signs, their combined
contributions largely cancel, leaving the spin-1/2 channel as the relatively
more important one phenomenologically.

\subsection{Generalized TPE form factors}
\label{ssec.GFF}

\begin{figure}[t]
\graphicspath{{Images/}}
\includegraphics[width=8.2cm]{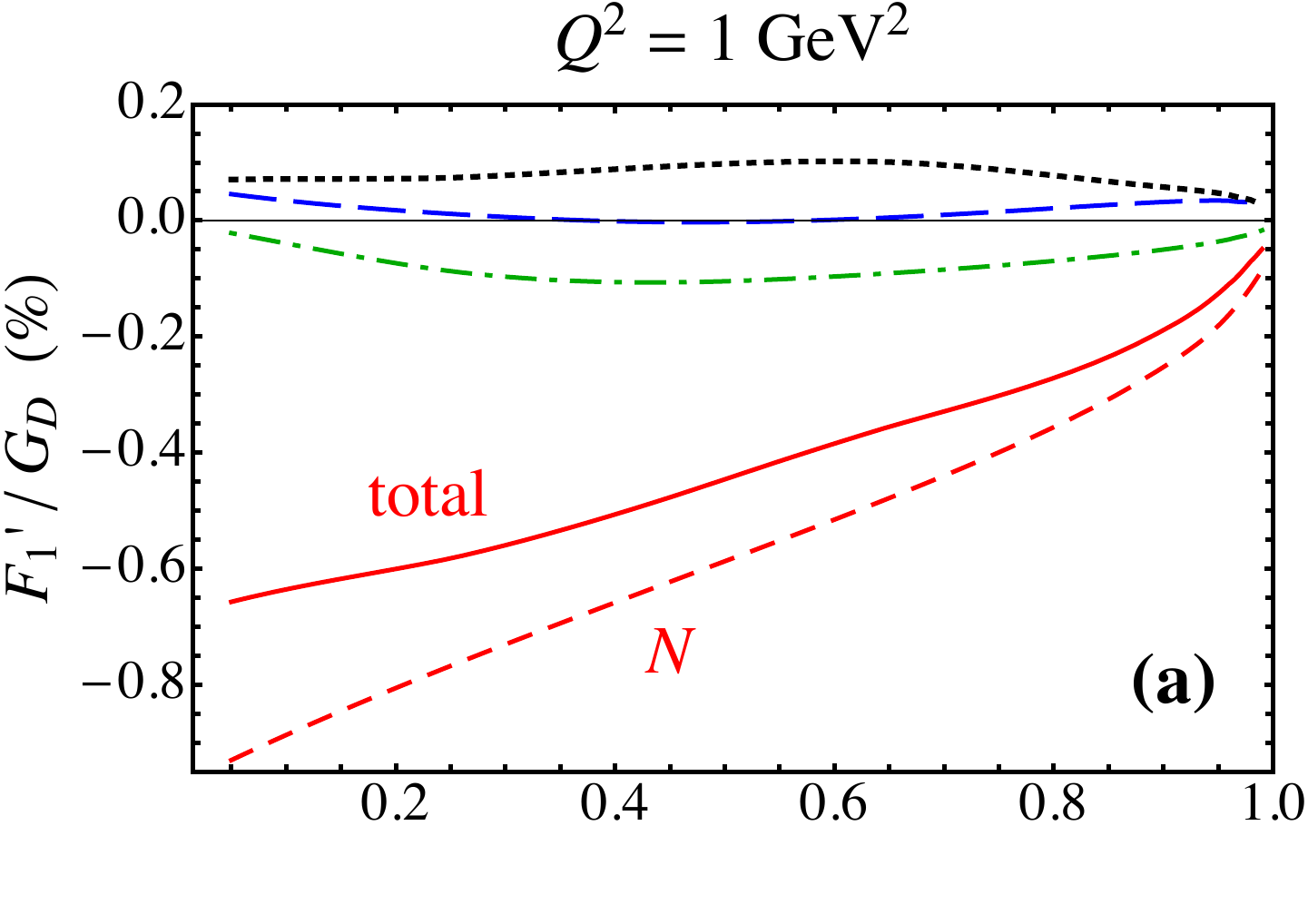} \hspace*{-0.6cm}
\includegraphics[width=8cm]{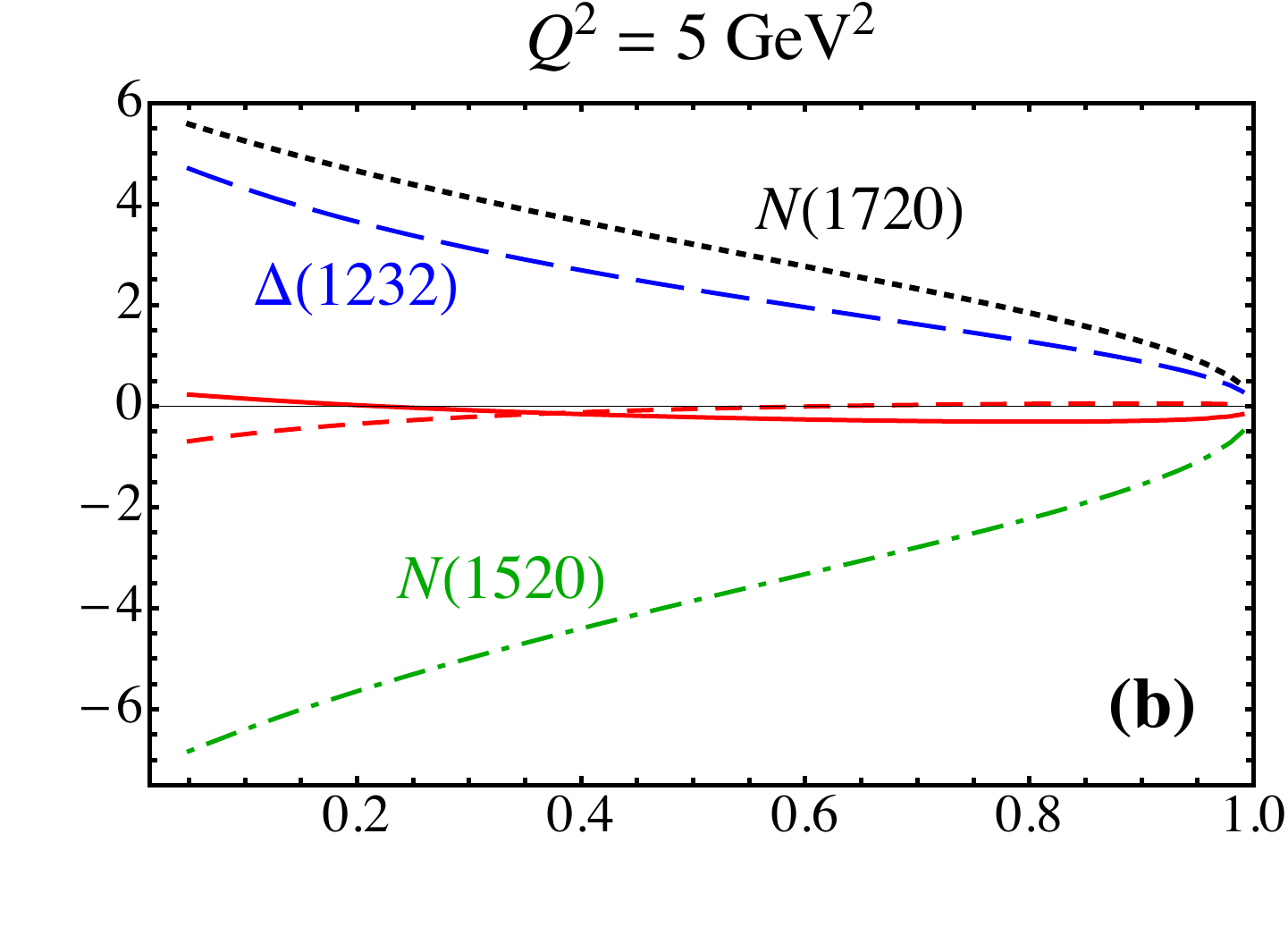} \\[-0.7em]
\hspace*{-0.15cm} \includegraphics[width=8.2cm]{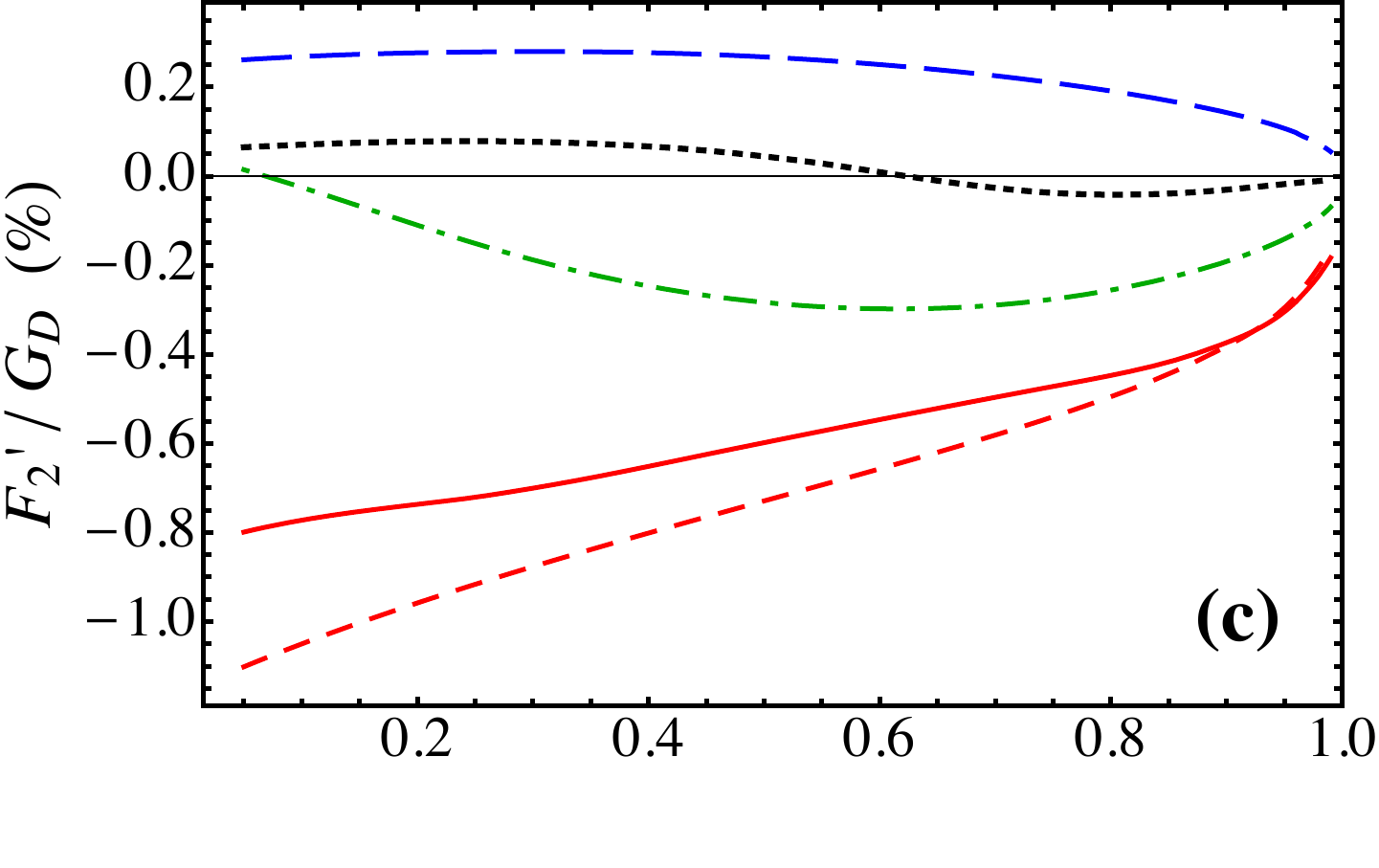} \hspace*{-0.75cm}
\includegraphics[width=8.18cm]{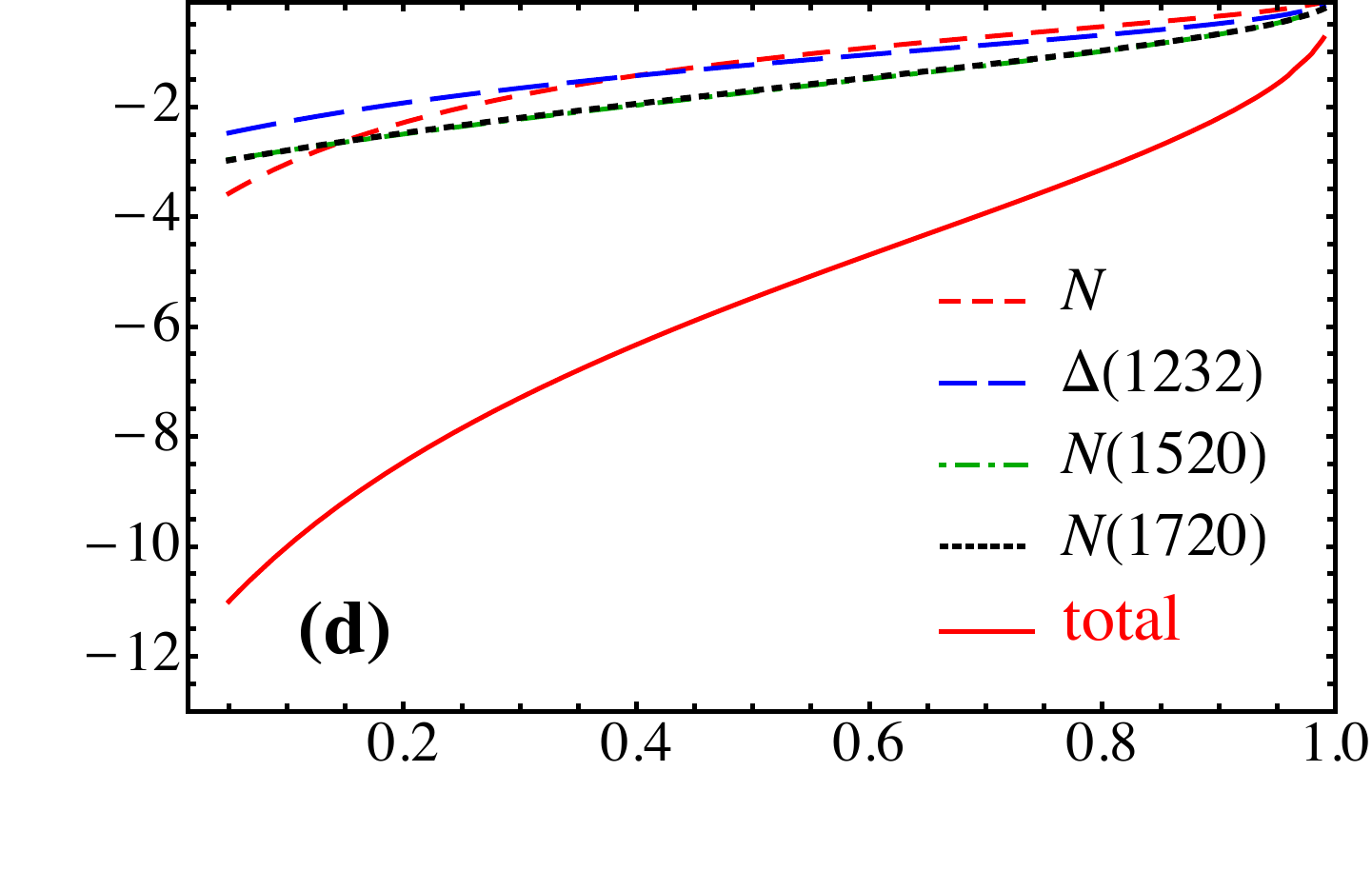} \\[-0.7em]
\hspace*{-0.15cm} \includegraphics[width=8.22cm]{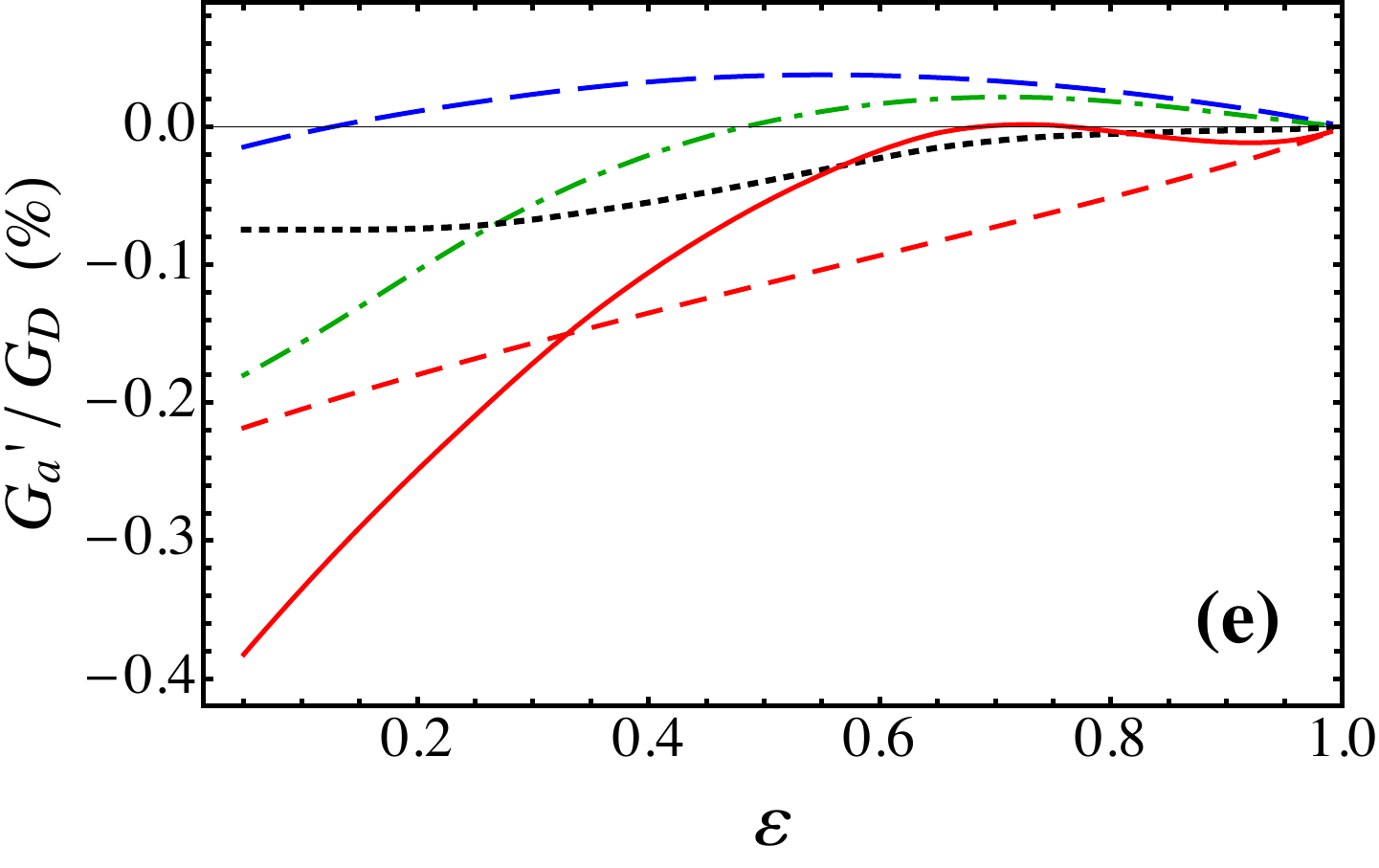} \hspace*{-0.6cm}
\includegraphics[width=8cm]{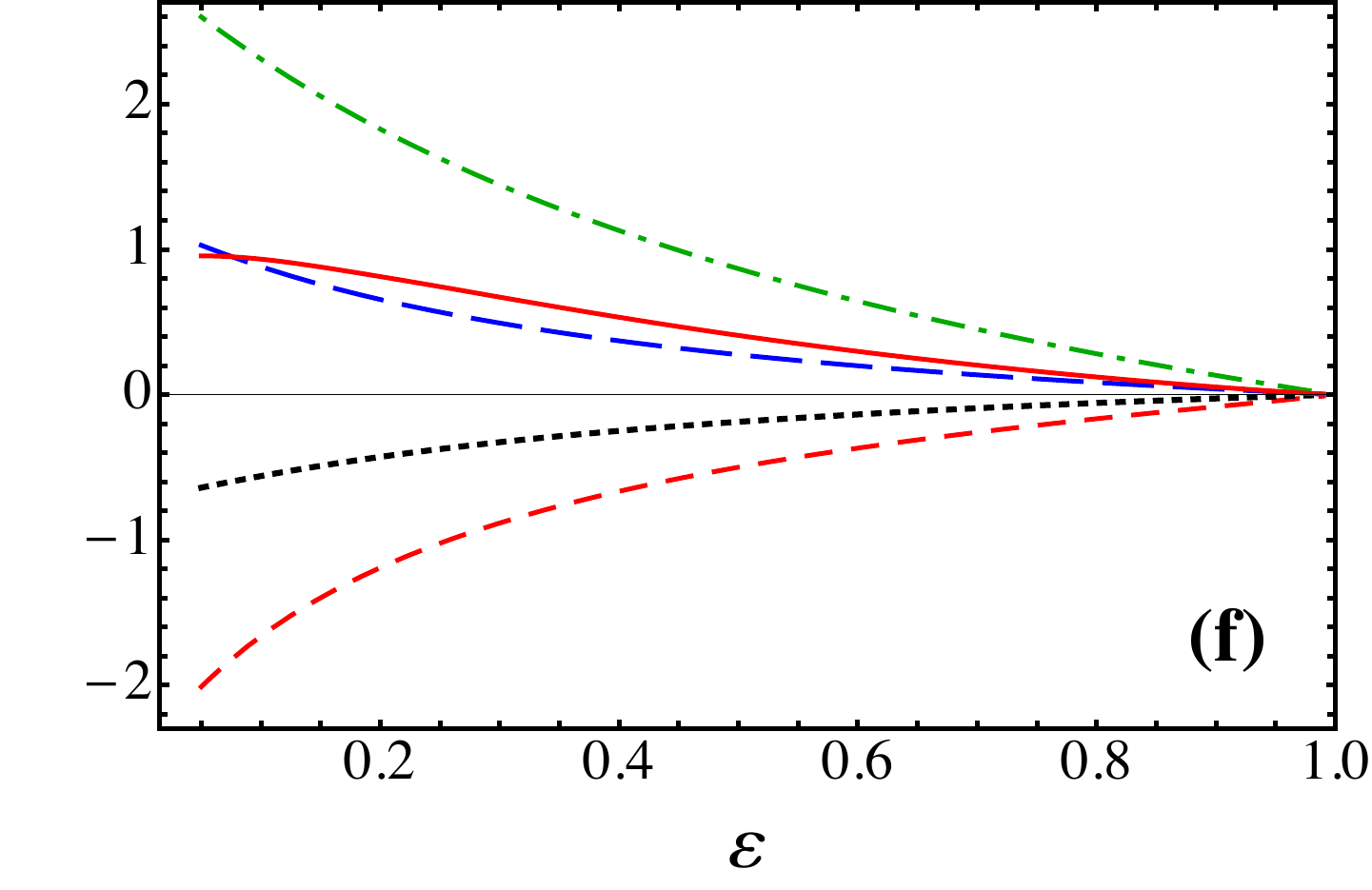}
\caption{Generalized TPE form factors 
$F'_1$ [{\bf (a), (b)}],
$F'_2$ [{\bf (c), (d)}], and
$G'_a$ [{\bf (e), (f)}],
scaled by the dipole form factor $G_D$, versus $\varepsilon$ at fixed
$Q^2=1$~GeV$^2$ (left column) and 5~GeV$^2$ (right column) for the nucleon
elastic (red dashed lines),
$\Delta(1232)~\!3/2^+$ (blue long-dashed lines),
$N(1520)~\!3/2^-$ (green dot-dashed lines),
$N(1720)~\!3/2^+$ (black dotted lines),
and total TPE (red solid lines) contributions.}
\label{fig.GFF-vs-eps}
\end{figure}

Before proceeding to the quantitative comparison of the calculated full cross
sections with experimental observables sensitive to TPE effects, in this section
we present the TPE results in terms of the generalized TPE form factors
introduced in Sec.~\ref{sssec.genFF}. In Fig.~\ref{fig.GFF-vs-eps} we present
the $\varepsilon$ dependence of the TPE form factors $F'_1$, $F'_2$ and $G'_a$
at fixed values of $Q^2=1$~GeV$^2$ and 5~GeV$^2$, scaled by a dipole form factor
$G_D$,
\bea
G_D(Q^2) &=& \bigg( \frac{\Lambda^2}{Q^2 + \Lambda^2} \bigg)^2\ ,
\label{eq:dipFF}
\eea
with mass $\Lambda=0.84$~GeV. Illustrated are the individual contributions from
the nucleon elastic intermediate state and the 3 most prominent resonance
states, namely, the $\Delta(1232)~\!3/2^+$, $N(1520)~\!3/2^-$, and the
$N(1720)~\!3/2^+$, as well as the total.

Clearly evident for the $F_1'$ TPE form factor is that at $Q^2=1$~GeV$^2$ this
contribution is negative at all $\varepsilon$ values and is dominated by the
nucleon elastic state. The higher-mass resonance contributions grow rapidly with
increasing $Q^2$, but there is a strong cancellation between the (positive)
$J^P=3/2^+$ and (negative) $J^P=3/2^-$ states, rendering the total effect to be
very small and close to zero at $Q^2=5$~GeV$^2$.

For the Pauli $F_2'$ TPE form factor, a similar pattern repeats as for the Dirac
form factor, namely, at $Q^2=1$~GeV$^2$ the cancellations between the various
resonance contributions leave the total TPE form factor to be negative and
dominated by the nucleon elastic intermediate state. In contrast to the $F_1'$
case, however, at larger $Q^2$ the main resonance contributions grow in
magnitude but remain negative, so that the net effect is a coherent enhancement
of the TPE form factor up to $\sim 10\%$ of the dipole at $Q^2=5$~GeV$^2$ for
backward angles.

For the axial $G'_a$ TPE form factor, the magnitude of the various resonance
contributions is generally smaller than for the other two TPE form factors, with
the nucleon elastic state giving negative contributions at both low and high
$Q^2$. Once again a high degree of cancellation occurs between the (positive)
$\Delta(1232)$ and $N(1520)$ states and the (negative) nucleon elastic and
$N(1720)$ states, leaving an overall small positive total correction to $G'_a$.

\begin{figure}[t]
\graphicspath{{Images/}}
\includegraphics[width=8.4cm]{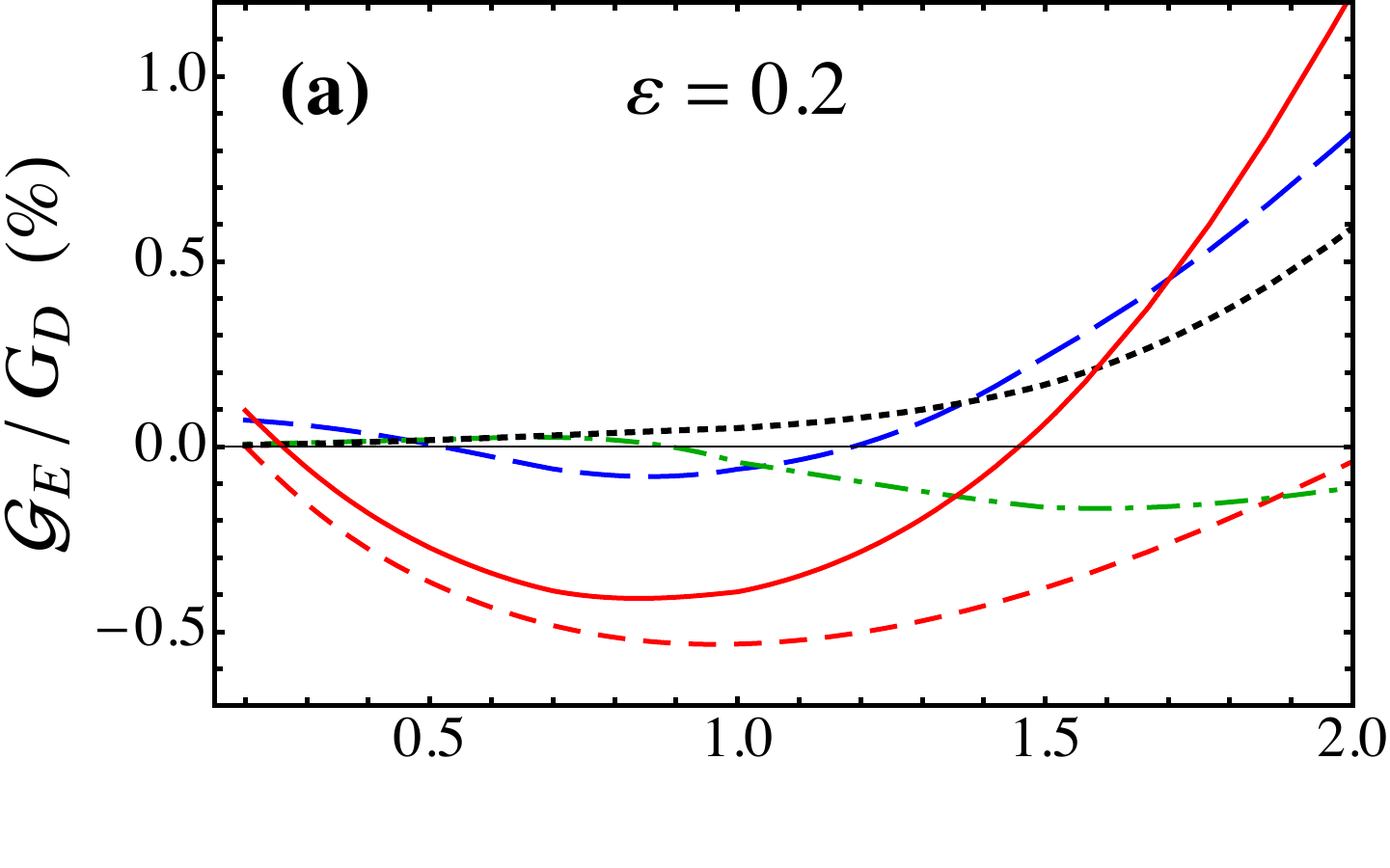} \hspace*{-0.6cm}
\includegraphics[width=8.0cm]{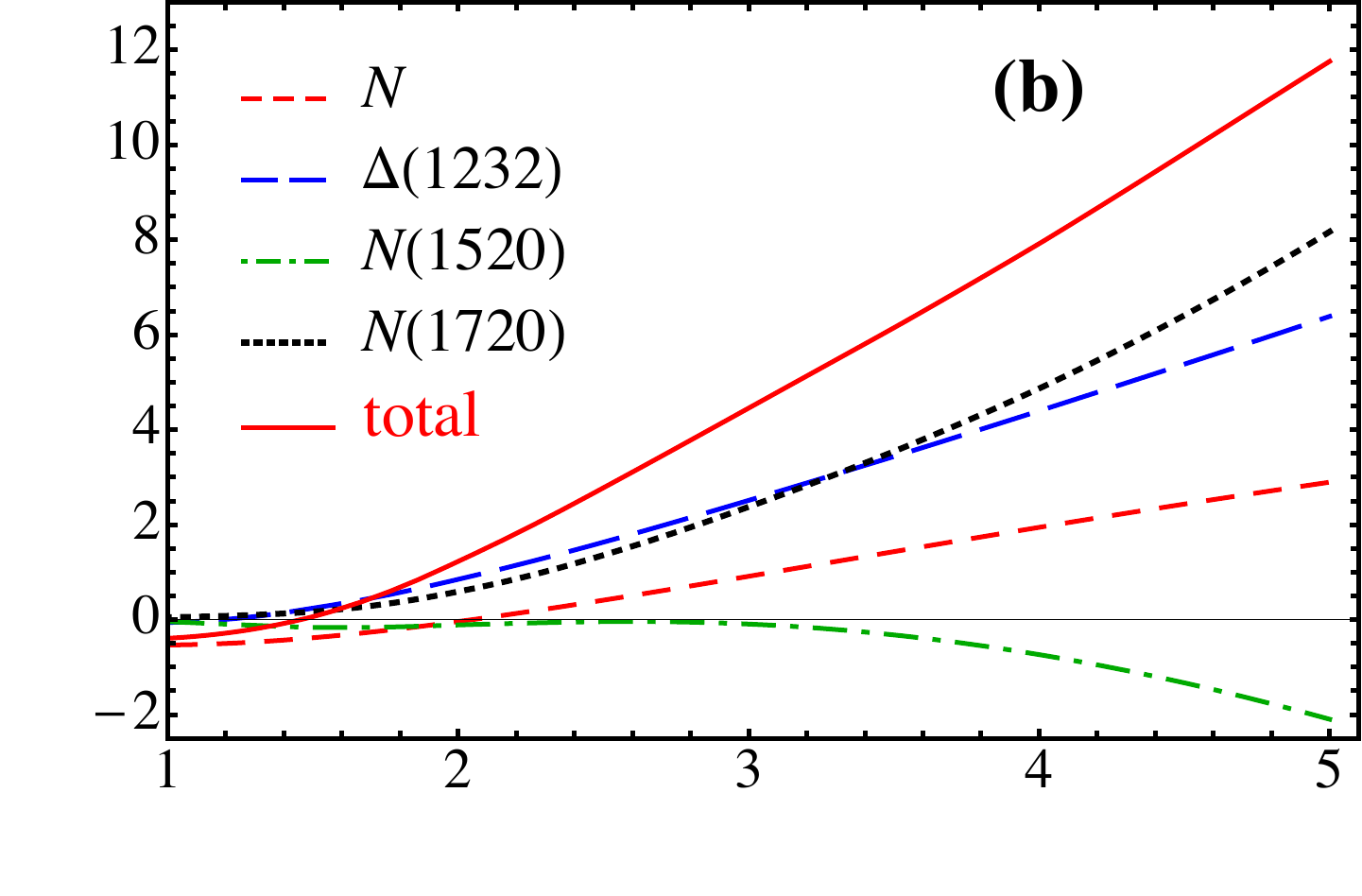} \\[-0.7em]
\hspace*{0.1cm} \includegraphics[width=8.15cm]{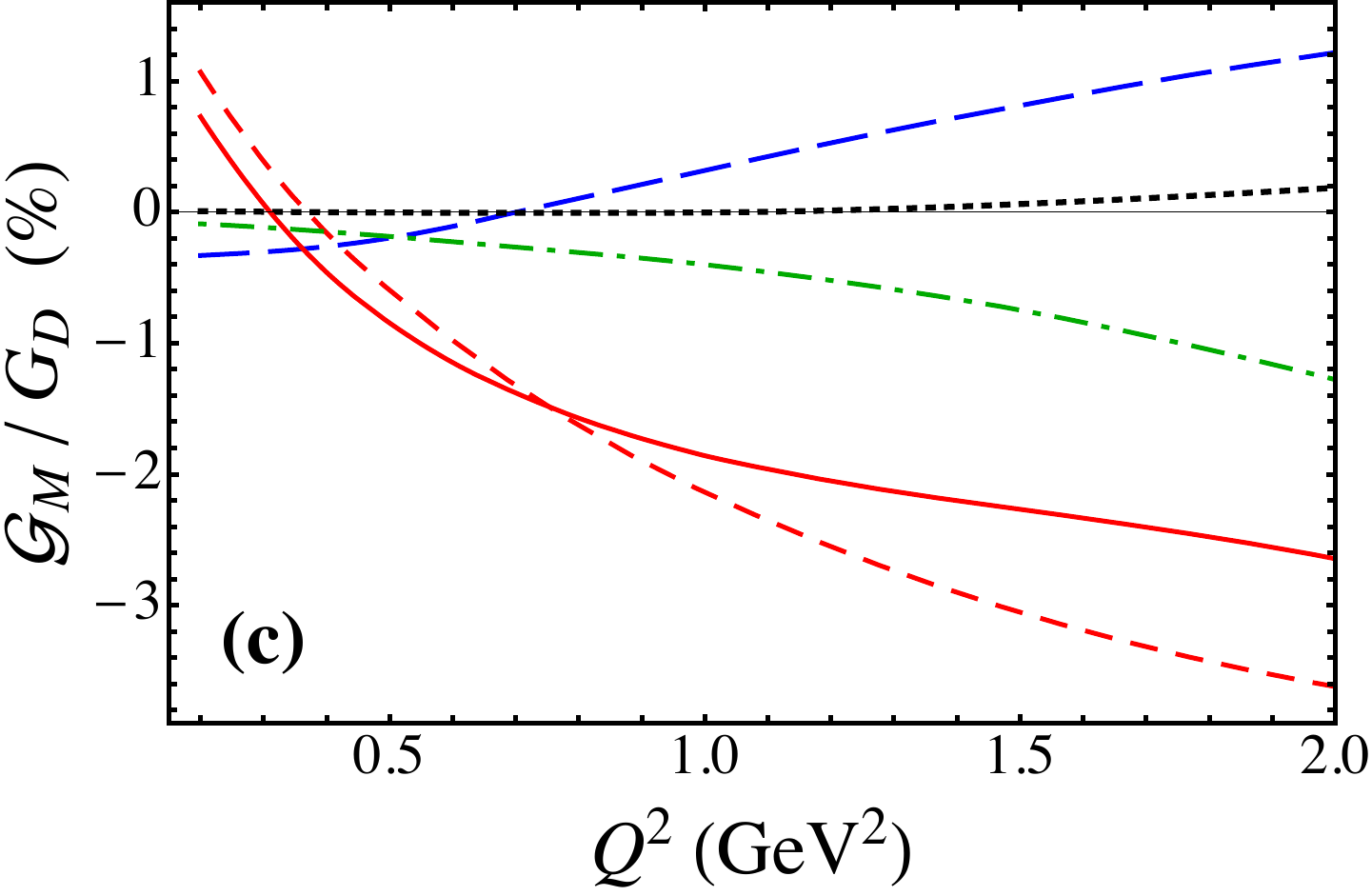} \hspace*{-0.65cm}
\includegraphics[width=8.0cm]{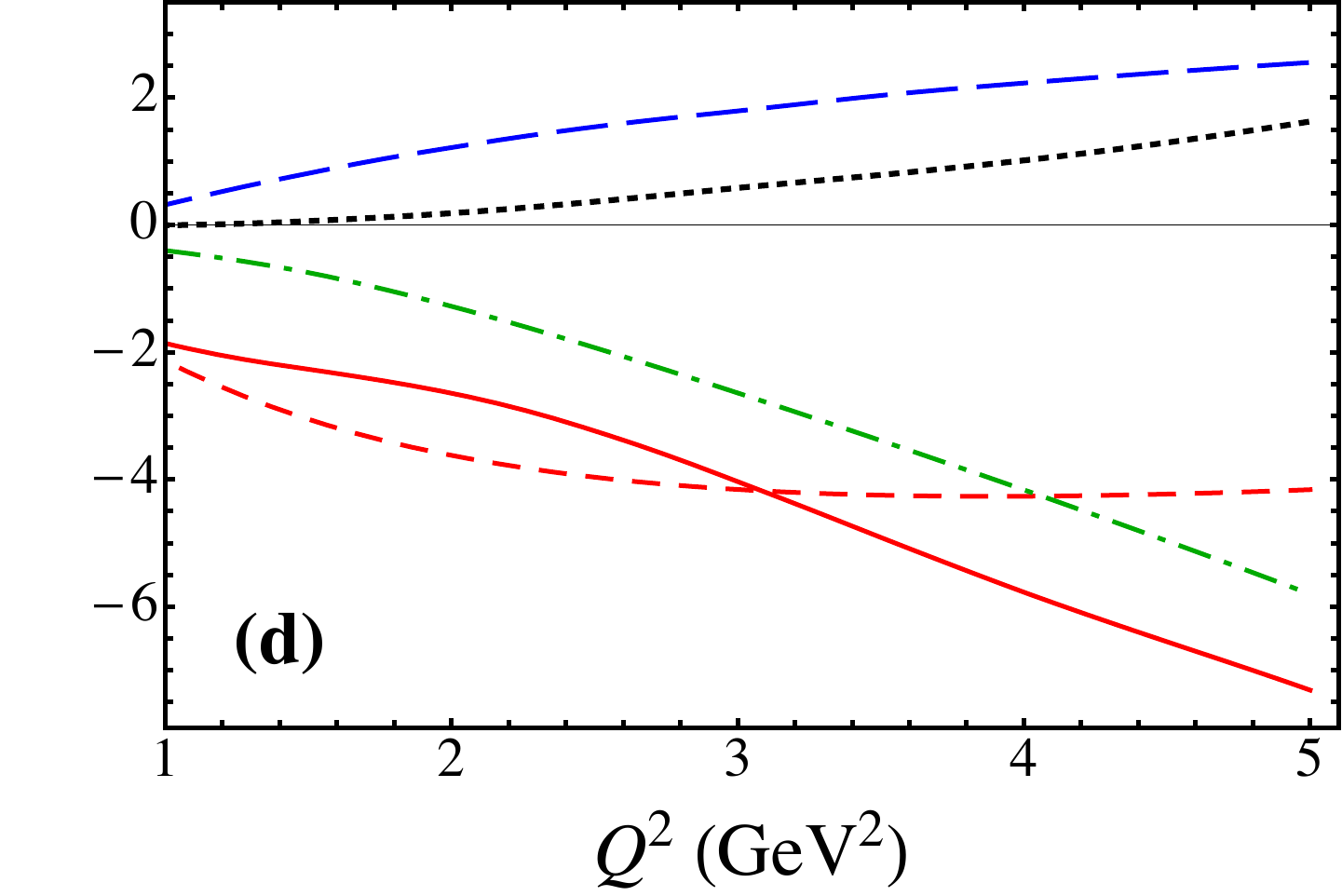}
\caption{Generalized TPE form factors ${\cal G}_E$ [{\bf (a), (b)}] and ${\cal
G}_M$ [{\bf (c), (d)}], scaled by the dipole form factor $G_D$, at fixed
$\varepsilon=0.2$ for low $Q^2$ ($Q^2 \leq 2$~GeV$^2$, left column) and high
$Q^2$ ($1 \leq Q^2 \leq 5$~GeV$^2$, right column), for the nucleon elastic (red
dashed lines),
$\Delta(1232)~\!3/2^+$ (blue long-dashed lines),
$N(1520)~\!3/2^-$ (green dot-dashed lines),
$N(1720)~\!3/2^+$ (black dotted lines),
and total TPE (red solid lines) contributions.}
\label{fig.GFF-vs-Q2}
\end{figure}

In fact, as observed by Borisyuk and Kobushkin \cite{borisyuk2008}, it is quite
natural to combine the small $G'_a$ contribution with the $F'_1+F'_2$ form
factor combination into an effective ``magnetic'' TPE form factor ${\cal G}_M$
as in Eq.~(\ref{eq.calGM}). Observing that the TPE FFs in
Fig.~\ref{fig.GFF-vs-eps} do not in general show strong variation with
$\varepsilon$, in Fig.~\ref{fig.GFF-vs-Q2} we display the $Q^2$ dependence of
both the ``electric'' and ``magnetic'' TPE form factor ${\cal G}_E$ and ${\cal
G}_M$, scaled by the dipole form factors, at a fixed value of $\varepsilon=0.2$,
where the TPE effects are not suppressed.

For $Q^2 \gtrsim 2$~GeV$^2$ one observes that the magnitude of both the
generalized electric ${\cal G}_E$ and magnetic ${\cal G}_M$ TPE form factors
rises linearly with $Q^2$. The positive sign of ${\cal G}_E$ and the negative
sign of ${\cal G}_M$ result in corrections to the effective Born level form
factors that render the $G_E/G_M$ ratio smaller than that naively extracted from
cross section data without TPE corrections. This would make it more compatible
with the $G_E/G_M$ ratio extracted from the polarization transfer data, which
suggest a strong fall-off of the ratio with $Q^2$ above $Q^2 \sim 1$~GeV$^2$,
resolving the discrepancy with the Rosenbluth cross section results.

At low $Q^2$, $Q^2 \lesssim 1$~GeV$^2$, the TPE form factors are dominated by
the nucleon elastic contribution, as already indicated in the $Q^2$ dependence
of the total TPE correction $\delta$ in Fig.~\ref{fig.DsigQ}. For higher $Q^2$
values, $Q^2 \gtrsim 2$~GeV$^2$, the magnitudes of the various excited state
contributions grow, with the $\Delta(1232)$ and $N(1720)$ contributions to both
${\cal G}_E$ and ${\cal G}_M$ remaining positive and the $N(1520)$ states
negative.

More specifically, while the $N(1520)$ resonance state gives rather small
corrections to ${\cal G}_E$ at most values of $Q^2$, its contribution to ${\cal
G}_M$ becomes even more important than the nucleon elastic for the largest
$Q^2$, $Q^2 \gtrsim 4$~GeV$^2$. Because of the $\tau$ factor in
Eq.~(\ref{eq.geTPE_delG}), the magnetic contribution to the total cross section
dominates at high $Q^2$, so that the $N(1520)$ state plays the most significant
role in the TPE cross section at high $Q^2$. At high $Q^2$ the negative sign of
the ${\cal G}_M$ TPE form factor is driven by the nucleon elastic and $N(1520)$
states, while the positive sign of the ${\cal G}_E$ TPE form factor is due
mostly to the $\Delta(1232)$ and $N(1720)$.

\section{TPE-sensitive observables}
\label{sec.observables}

Having described the features of the TPE corrections from excited intermediate
states to elastic $ep$ scattering cross sections in the previous sections, in
the remainder of this paper we will discuss the impact of these corrections on
observables sensitive to the TPE effects. In particular, we analyze the
numerical effects of the  calculated TPE corrections on the elastic $e^+ p$ to
$e^- p$ cross section ratio measured recently by the CLAS~\cite{CLASTPE2017},
VEPP-3 \cite{VEPP2015} and OLYMPUS \cite{OLYMPUS2017} experiments, as well as
with polarization transfer data from the GEp$2\gamma$ experiment~\cite{gep2011}
in Hall~C at Jefferson Lab. In addition, we investigate the effect of the
resonance contributions to the TPE on the proton $G_E/G_M$ form factor ratio
discrepancy between the LT and PT data~\cite{blunden2003, blunden2005,
guichon2003, chen2004}.

\subsection{$e^+ p$ to $e^- p$ elastic scattering ratio}
\label{ssec.e+e-}

Perhaps the most direct consequence of TPE in lepton scattering is the deviation
from unity of the ratio of $e^+ p$ to $e^- p$ elastic scattering cross sections.
The interference of the Born amplitude and the TPE amplitude here depends on the
sign of the lepton charge, so that the ratio
\be
R_{2\gamma}\
=\ \frac{\sigma(e^+ p)}{\sigma(e^- p)}\
\approx\ 1 - 2\, \delta_{\gamma\gamma},
\ee
where $\sigma(e^\pm p) \equiv d\sigma(e^\pm p \to e^\pm p)/d\Omega$, is a direct
measure of the TPE correction $\delta_{\gamma\gamma}$. Early measurements of
$R_{2\gamma}$ in the 1960s at SLAC \cite{SLAC1965, SLAC1968}, Cornell
\cite{cornell1966}, DESY \cite{desy1967} and Orsay \cite{orsay1968} obtained
some hints of nonzero TPE effects, however, since the data were predominantly at
low $Q^2$ and forward angles the deviations of $R_{2\gamma}$ from unity  were
small and within the experimental uncertainties. The more recent experiments at
Jefferson Lab~\cite{CLASTPE2017}, Novosibirsk~\cite{VEPP2015} and
DESY~\cite{OLYMPUS2017} have attempted more precise determinations of
$R_{2\gamma}$ over a larger range of $Q^2$ and $\varepsilon$ values than
previously available.

\begin{figure}[t]
\graphicspath{{Images/}}
\includegraphics[width=8.0cm]{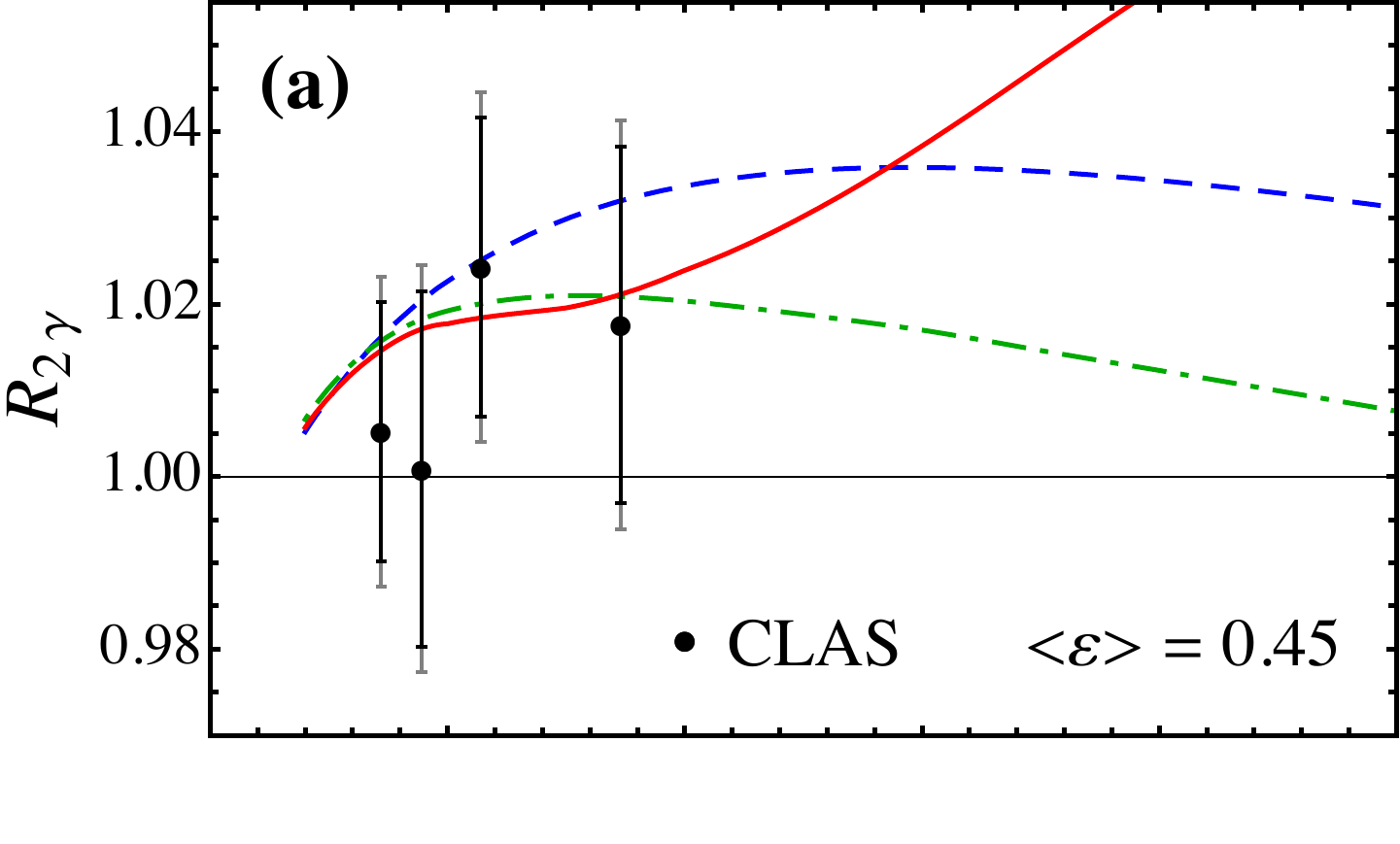} \hspace*{-0.3cm}
\includegraphics[width=8.0cm]{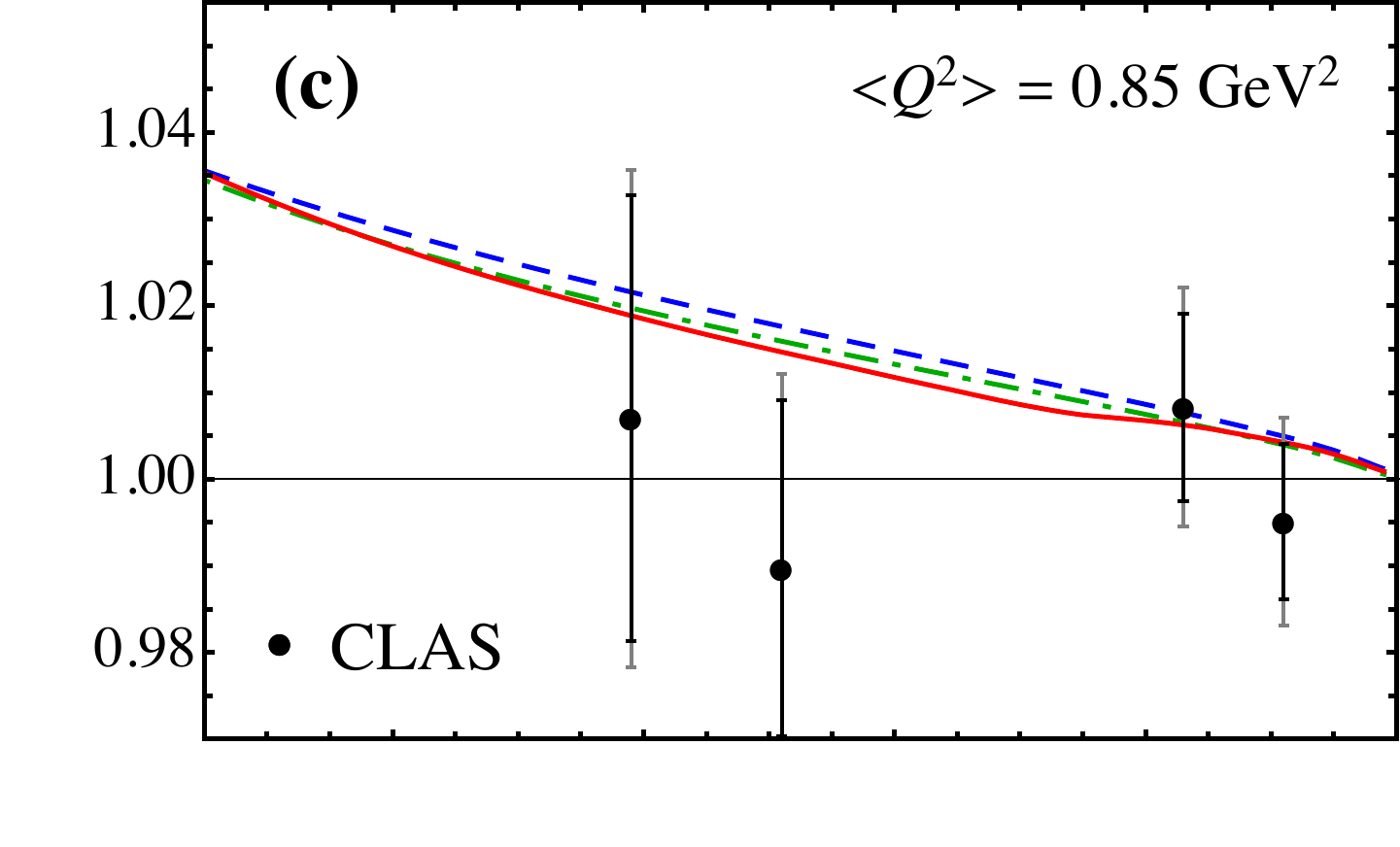} \\ [-0.9em]
\hspace*{0.07cm} \includegraphics[width=8.1cm]{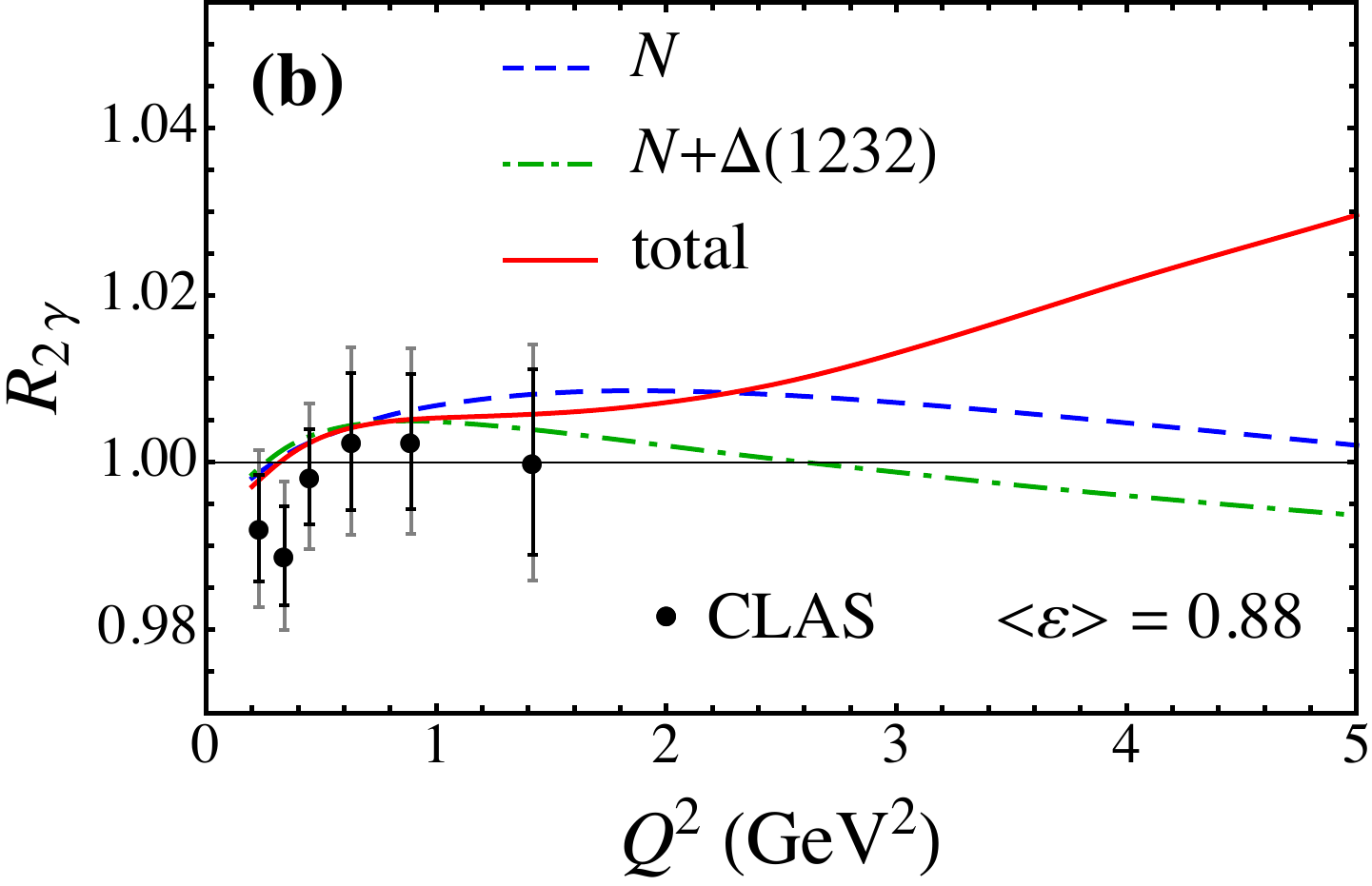} \hspace*{-0.4cm}
\includegraphics[width=8.2cm]{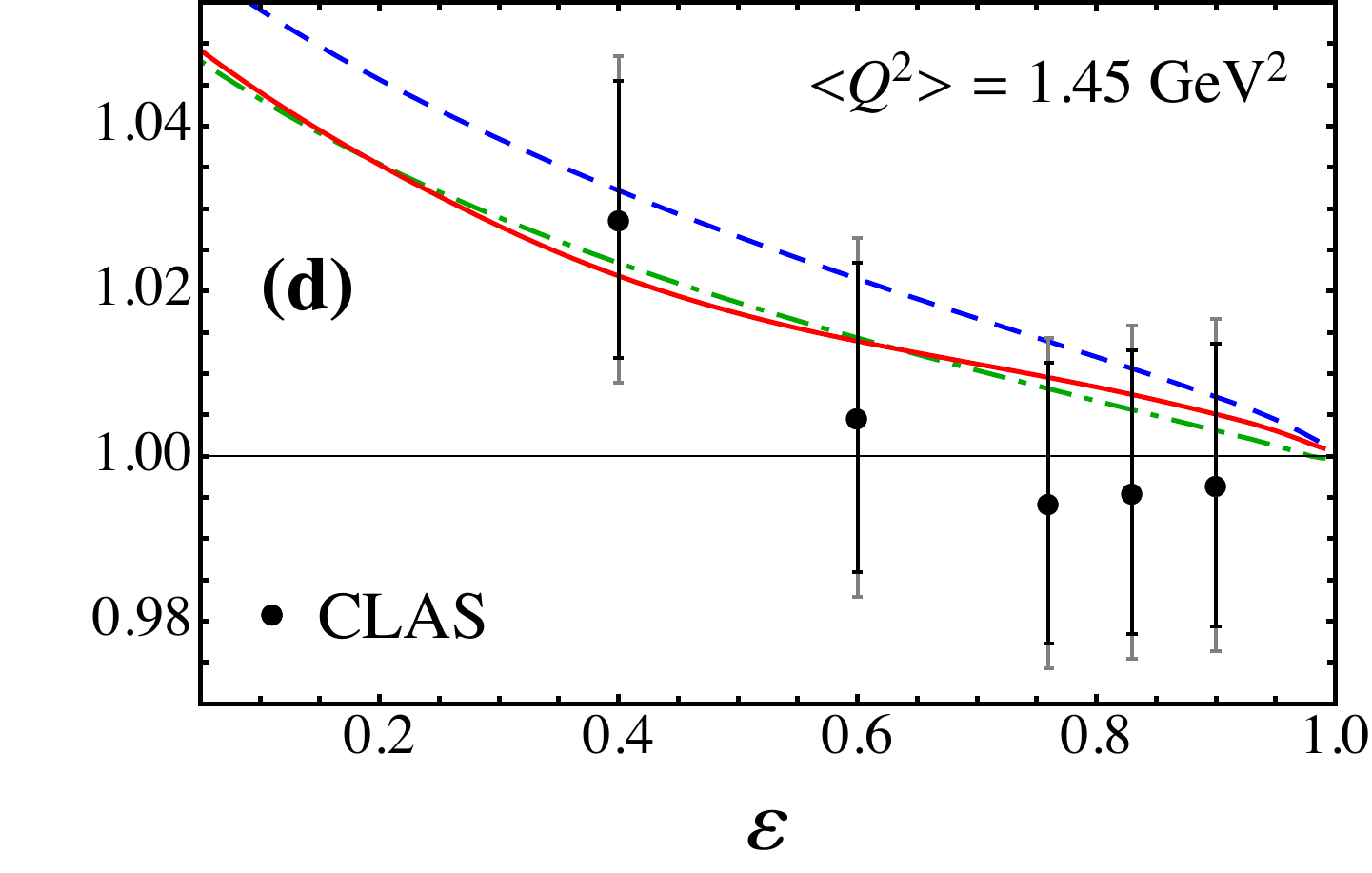}
\caption{Ratio $R_{2\gamma}$ of $e^+ p$ to $e^- p$ elastic cross sections from
CLAS~\cite{CLASTPE2017}
{\bf (a)} versus $Q^2$ for fixed averaged $\langle\varepsilon\rangle = 0.45$ and
{\bf (b)} $\langle\varepsilon\rangle = 0.88$,
{\bf (c)} versus $\varepsilon$ for fixed averaged $\langle Q^2 \rangle = 0.85$~GeV$^2$ and 
{\bf (d)} $\langle Q^2\rangle =1.45$~GeV$^2$,
compared with the nucleon only (blue dashed lines), sum of nucleon and
$\Delta(1232)$ (green dot-dashed lines), and sum of all intermediate state
contributions (red solid lines). The experimental statistical and systematic
uncertainties are indicated by the (black) inner and (gray) outer error bars,
respectively.}
\label{fig.CLAS}
\end{figure}

The $R_{2\gamma}$ ratio from the CLAS experiment \cite{CLASTPE2017} is shown in
Fig.~\ref{fig.CLAS} versus $Q^2$ at fixed averaged $\varepsilon$ values,
$\langle\varepsilon\rangle = 0.45$ and 0.88 [Fig.~\ref{fig.CLAS}(a), (b)], and
versus $\varepsilon$ for fixed averaged $Q^2$, $\langle Q^2 \rangle = 0.85$ and
1.45~GeV$^2$ [Fig.~\ref{fig.CLAS}(c), (d)].
The deviations from unity of the measured ratios are relatively small, with most
of the data points consistent with no TPE effects within the relatively large
experimental uncertainties. (Note that in Fig.~\ref{fig.CLAS} and in subsequent
data comparisons, the statistical and systematic uncertainties are shown
separately as inner and outer error bars, respectively.) The data are also
consistent, however, with the calculated TPE corrections, which are $\lesssim
2\%$ in the measured region, but increase at lower $\varepsilon$ and higher
$Q^2$. A significant contribution to the cross section ratio is observed from
the nucleon elastic intermediate state, with the $\Delta(1232)~\!3/2^+$
resonance canceling some of the deviation from unity. The higher mass resonances
have little impact in the experimentally measured regions of $\varepsilon$ and
$Q^2$, but their contributions become more significant at higher $Q^2$ in
particular, $Q^2 \gtrsim 3$~GeV$^2$.

\begin{figure}[t]
\graphicspath{{Images/}}
\includegraphics[width=8.15cm]{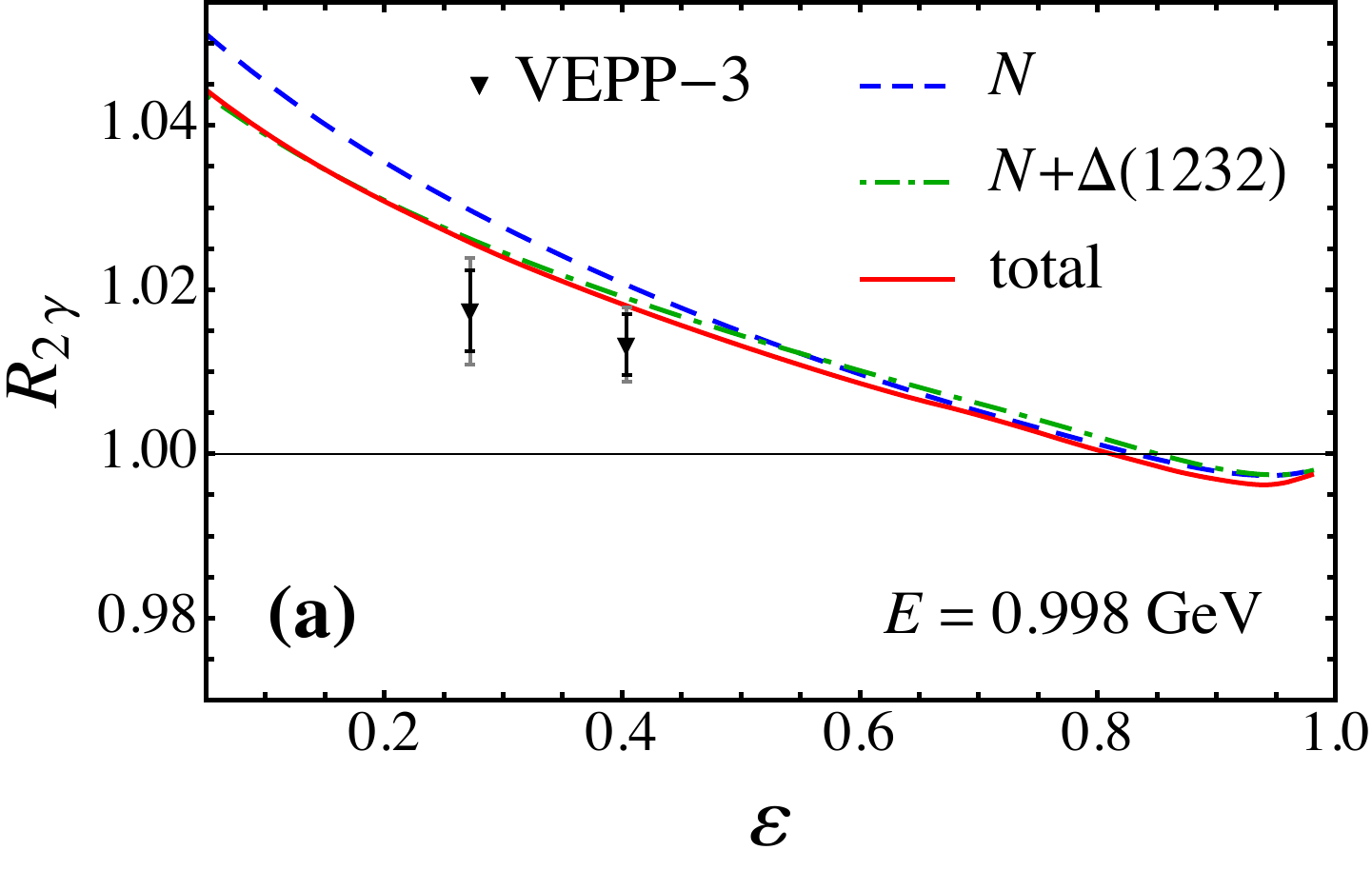} \hspace*{-0.5cm}
\includegraphics[width=8.15cm]{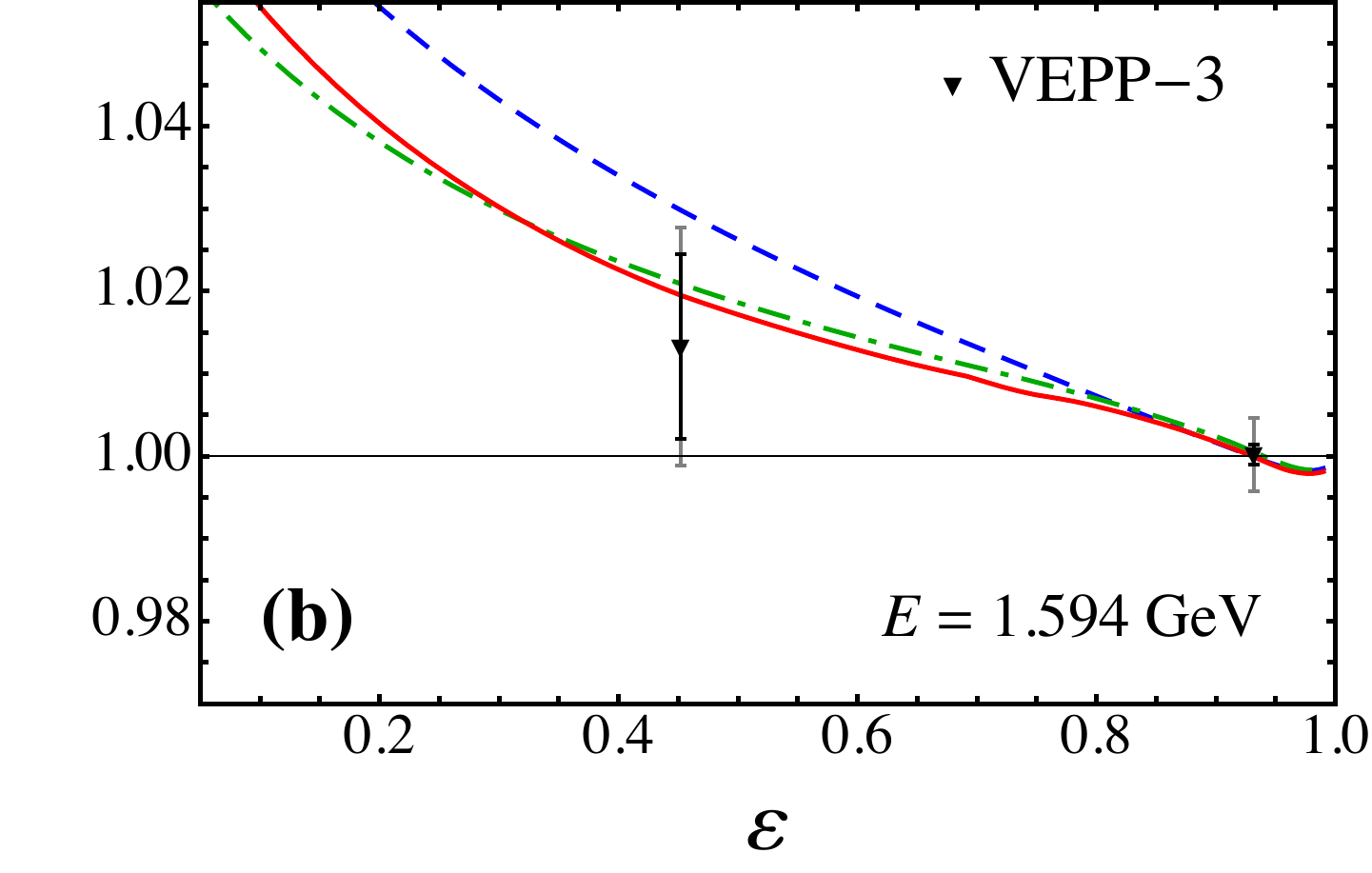}
\caption{Ratio $R_{2\gamma}$ of $e^+ p$ to $e^- p$ elastic cross sections versus
$\varepsilon$ from the VEPP-3 experiment~\cite{VEPP2015} for beam energy 
{\bf (a)} $E=0.998$~GeV and 
{\bf (b)} $E=1.594$~GeV, compared with the nucleon only (blue dashed lines), sum
of nucleon and $\Delta(1232)$ (green dot-dashed lines), and sum of all
intermediate state contributions (red solid lines). The experimental statistical
and systematic uncertainties are indicated by the (black) inner and (gray) outer
error bars, respectively.}
\label{fig.V1}
\end{figure}

A similar comparison of the calculated $R_{2\gamma}$ ratio with data from the
VEPP-3 experiment at Novosibirsk \cite{VEPP2015} is shown in Fig.~\ref{fig.V1}.
The experiment scattered electrons at fixed beam energy $E=0.998$~GeV
[Fig.~\ref{fig.V1}(a)] and $E=1.594$~GeV [Fig.~\ref{fig.V1}(b)], for
$\varepsilon$ down to $\approx 0.3$. This corresponds to a $Q^2$ range between
$\approx 0.3$~GeV$^2$ and 1.5~GeV$^2$. At these $Q^2$ values the nucleon elastic
intermediate state gives the largest contribution, with again the $\Delta(1232)$
canceling some of the effect, and bringing the calculation with the TPE
corrections in better agreement with the data. The contributions of the higher
mass resonances at the kinematics of this experiment are negligible.

\begin{figure}[t]
\graphicspath{{Images/}}
\includegraphics[width=9cm]{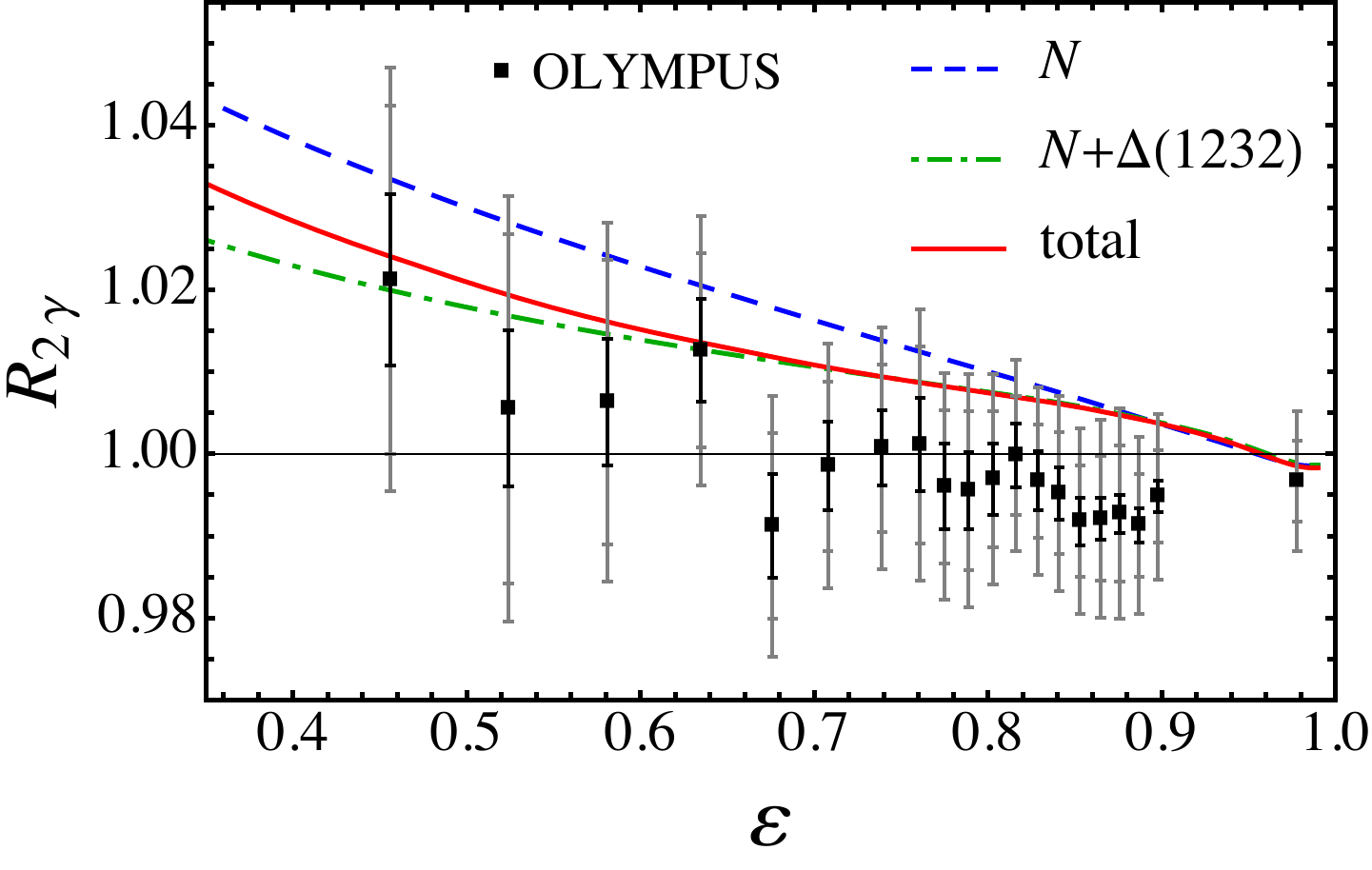}
\caption{Ratio $R_{2\gamma}$ of $e^+ p$ to $e^- p$ elastic cross sections versus
$\varepsilon$ from the OLYMPUS experiment~\cite{OLYMPUS2017} with beam energy $E
= 2.01$~GeV, compared with the nucleon only (blue dashed lines), sum of nucleon
and $\Delta(1232)$ (green dot-dashed lines), and sum of all intermediate state
contributions (red solid lines). The experimental statistical and systematic
uncertainties are indicated by the (black) inner and (gray) outer error bars,
respectively.}
\label{fig.Olympus}
\end{figure}

The most recent OLYMPUS experiment at DESY~\cite{OLYMPUS2017} measured the ratio
$R_{2\gamma}$ over a range of $\varepsilon$ from $\approx 0.46$ to 0.9 at an
electron energy $E \approx 2$~GeV, with $Q^2$ ranging up to $\approx 2$~GeV$^2$.
The results, illustrated in Fig.~\ref{fig.Olympus}, indicate an enhancement of
the ratio at $\varepsilon \lesssim 0.6$ and a dip below unity at $\varepsilon
\gtrsim 0.7$, although still compatible with no deviation from 1 within the
combined statistical and systematic uncertainties. The suppression of the ratio
at large $\varepsilon$ is in slight tension from other measurements, but again
the effect is consistent within the errors~\cite{blunden2017}. Inclusion of the
$\Delta(1232)$ intermediate state reduces the effect of the nucleon elastic
contribution away from the forward scattering region, but the effect of the
higher mass resonances is very small for all $\varepsilon$ shown. The overall
agreement between the TPE calculation and the OLYMPUS data is reasonable within
the experimental uncertainties, although there is no indication in our model for
a decrease of the ratio below unity at large~$\varepsilon$.

\subsection{Polarization observables}
\label{ssec.polobs}

In addition to the unpolarized $e^+ p$ to $e^- p$ cross section ratio, other
observables that are directly sensitive to the presence of effects beyond the
Born approximation involve elastic scattering of longitudinally polarized
electrons from unpolarized protons, with polarization transferred to the final
state proton, $\vec e\, p \to e\, \vec p$. The relevant observables are the
transverse and longitudinal polarizations, $P_T$ and $P_L$, defined relative to
the proton momentum in the scattering plane as
\begin{subequations}
\bea
P_T
&=& -\dfrac{\sqrt{2\tau \varepsilon(1-\varepsilon)}}
           {\sigma_R}
    \left[ G_E G_M + G_M \Re{\cal G}_E
         + G_E \Re\Big({\cal G}_M + \dfrac{\nu\varepsilon}{\tau} G_a'\Big)
    \right],        \\
P_L
&=& \dfrac{\tau \sqrt{1-\varepsilon^2}}
          {\sigma_R}
    \left[ G_M^2
         + 2 G_M \Re
            \Big({\cal G}_M + \dfrac{\nu\varepsilon^2}{\tau(1+\varepsilon)} G'_a \Big) 
    \right],
\eea
\end{subequations}
where $\sigma_R = \sigma_R^{\rm Born}(1+\delta_{\gamma \gamma})$, and the
reduced Born cross section $\sigma_R^{\rm Born}$ is given in
Eq.~(\ref{eq.sigred}). The ratio of the transverse to longitudinal polarizations
is then given by
\be
R_{TL}
= -\mu_p \sqrt{\dfrac{\tau(1+\varepsilon)}{2\varepsilon}}\dfrac{P_T}{P_L}.
\ee
In the Born approximation, $R_{TL}$ reduces to the ratio of electric to magnetic
form factors, $\mu_p G_E/G_M$, and becomes independent of $\varepsilon$. Any
observed $\varepsilon$ dependence of $R_{TL}$ would therefore be an indication
of TPE effects. 

\begin{figure}[t]
\graphicspath{{Images/}}
\includegraphics[width=8.1cm]{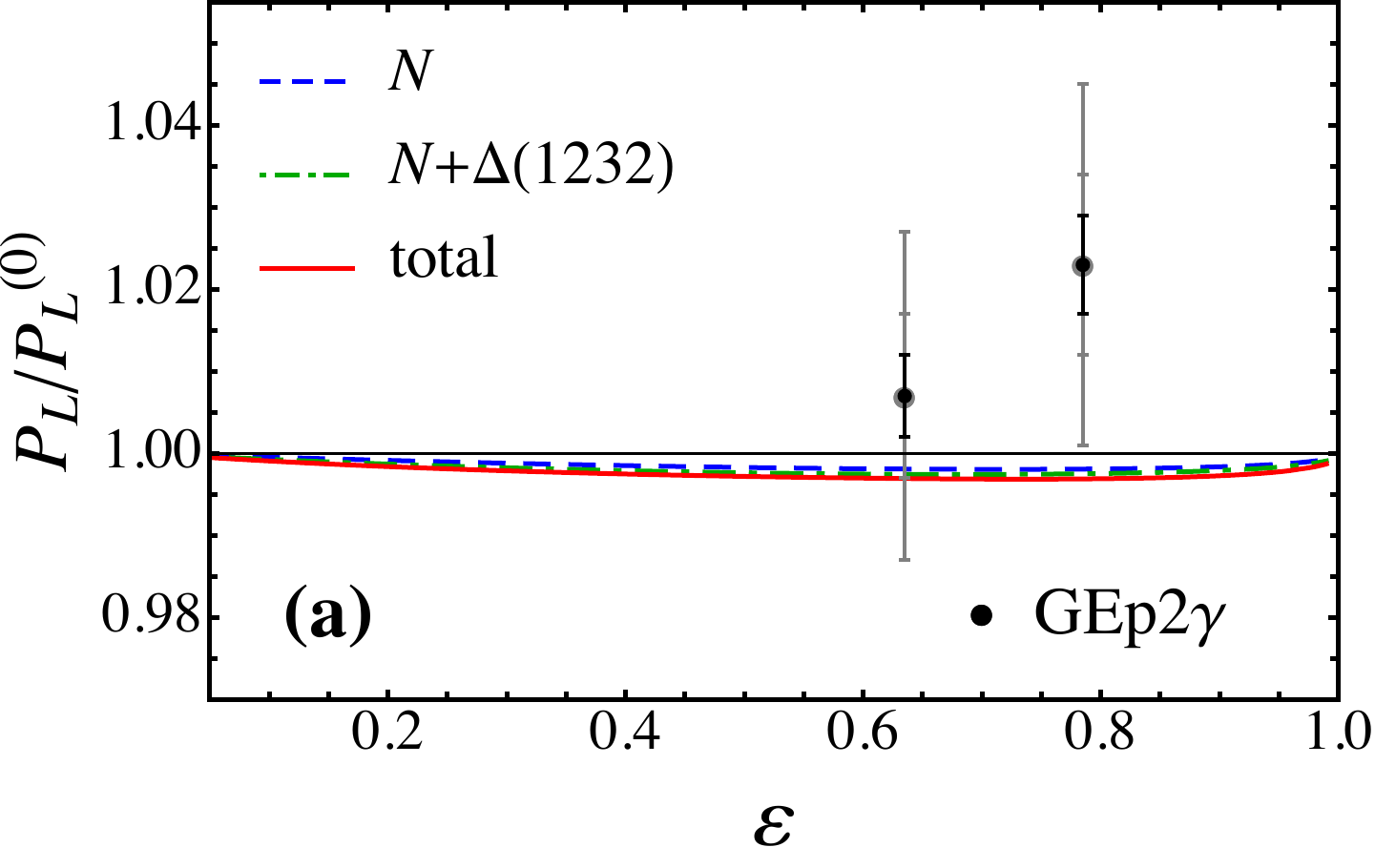} \hspace*{0cm}
\includegraphics[width=8.1cm]{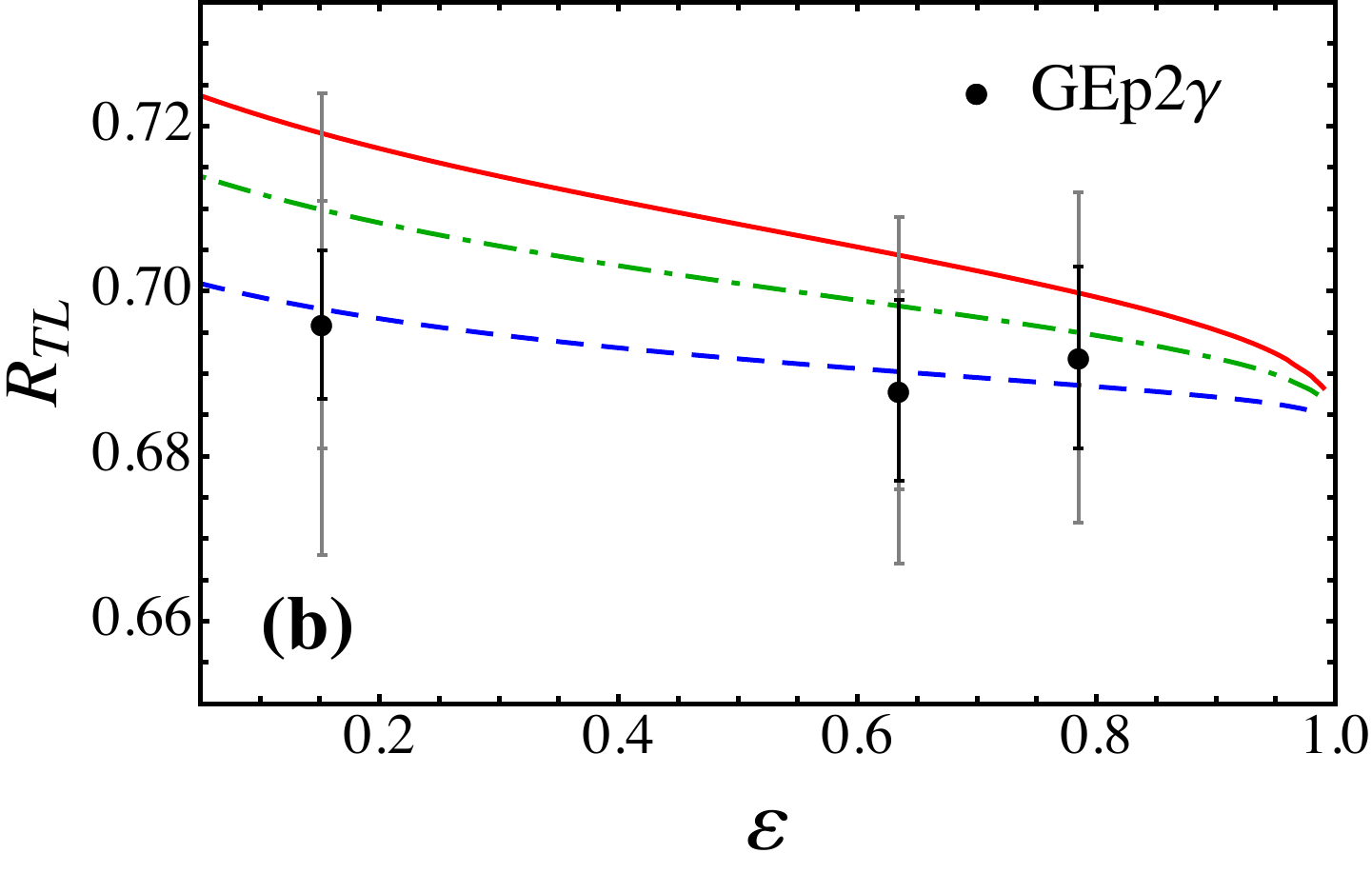}
\caption{Effect of TPE corrections on polarization observables from the
GEp2$\gamma$ experiment at Jefferson Lab~\cite{gep2011} for
{\bf (a)} longitudinal polarization $P_L$ relative to the Born level result
$P_L^{(0)}$, and
{\bf (b)} polarization transfer ratio $R_{TL}$ at $Q^2=2.49$~GeV$^2$, compared
with calculations including nucleon only (blue dashed lines), sum of nucleon and
$\Delta(1232)$ (green dot-dashed lines), and sum of all intermediate state
contributions (red solid lines). The experimental statistical and systematic
uncertainties are indicated by the (black) inner and (gray) outer error bars,
respectively.}
\label{fig.RTL}
\end{figure}

Data on the transverse and longitudinal polarizations were obtained from the
GEp$_{2\gamma}$ experiment at Jefferson Lab~\cite{gep2011}, and are shown in
Fig.~\ref{fig.RTL} for the ratio $P_L/P_L^{(0)}$, where $P_L^{(0)}$ is the Born
level longitudinal polarization, and the ratio $R_{TL}$ versus $\varepsilon$ at
an average value of $Q^2=2.49$~GeV$^2$. The calculated TPE effect in our model
is almost negligible for the longitudinal polarization, giving very little
additional $\varepsilon$ dependence in the ratio $P_L/P_L^{(0)}$ in
Fig.~\ref{fig.RTL}(a), and consistent within 1$\sigma$ with the data. A larger
TPE effect is found for the transverse polarization, where the nucleon alone
gives a small slope in $\varepsilon$, with the effects of the $\Delta(1232)$ and
higher mass intermediate states enhancing the TPE correction to $\approx 3\%$
effect at $\varepsilon \approx 0.2$. For the nucleon intermediate state this was
already concluded in the earlier analysis in Ref.~\cite{blunden2005}. The data
do not show any clear evidence for an $\varepsilon$ dependence within the
experimental uncertainties, although the calculated effect is also compatible
with the data within 1$\sigma$ errors.

\subsection{Electric to magnetic form factor ratio $\mu_p G_E/G_M$ }
\label{ssec.GeGm}

Perhaps the most well-known consequence of TPE that has been identified in the
last two decades is the ratio of the electric to magnetic form factors extracted
from elastic scattering cross sections using the LT separation
method~\cite{blunden2003}. Longitudinal-transverse separation requires
measurements of cross sections as a function of $\varepsilon$ (or scattering
angle) at fixed values of $Q^2$. In the Born approximation, the reduced cross
section $\sigma_R^{\rm Born}$ in Eq.~(\ref{eq.sigred}) is a linear function of
$\varepsilon$, which allows the form factors $G_M^2$ and $G_E^2$ to be extracted
from a linear fit to the reduced cross section data.

As observed in the preceding sections, the TPE correction induces an additional
shift in the $\varepsilon$ dependence, which alters the effective slope of the
reduced cross section versus $\varepsilon$. Furthermore, since the $\varepsilon$
dependence of the TPE effect is not restricted to be linear, any nonlinearity
introduced through radiative corrections could potentially complicate the form
factor extraction via the LT analysis, especially at higher values of $Q^2$.

In Secs.~\ref{ssec.e+e-} and \ref{ssec.polobs} we compared the available data to
calculations incorporating TPE effects. However, to extract $G_E$ and $G_M$ it
is more appropriate to correct the data for TPE contributions at the same level
as other radiative corrections in order to obtain the genuine Born contribution,
$\sigma_R^{\rm Born}$. The measured and Born cross sections can be related by
\bea
\sigma_R^{\rm meas}\,
=\, C_{\rm RC}^{\rm old}\, \left(\sigma_R^{\rm Born}\right)^{\rm old}\,
=\, C_{\rm RC}^{\rm new}\, \left(\sigma_R^{\rm Born}\right)^{\rm new},
\eea
where $C_{\rm RC}^{\rm old}$ is the radiative correction (RC) factor applied in
the original analyses~\cite{walker1994, Andivahis1994}, and $C_{\rm RC}^{\rm
new}$ incorporates any improvements, including the new TPE effects. For the RC
factor $C_{\rm RC}$ we adopt the definition used by Gramolin and
Nikolenko~\cite{Gramolin2016},
\begin{subequations}
\bea
C_{\rm RC} &=& C_L \exp\left(\delta_{\rm RC} + \delta\right), \\
\delta_{\rm RC} &=& \delta({\rm MTj}) + \delta_{\rm VP} + \delta_{\rm brems},
\eea
\end{subequations}
where $C_L$ is the correction factor for ionization losses in the target,
$\delta({\rm MTj})$ represents the standard RCs of Maximon and
Tjon~\cite{maximon2000}, $\delta_{\rm VP}$ are vacuum polarization corrections
not included in $\delta({\rm MTj})$, $\delta_{\rm brems}$ are hard photon
internal and external bremsstrahlung corrections not accounted for in
$\delta({\rm MTj})$, and $\delta$ is the hard TPE correction $\delta =
\delta_{\gamma\gamma}-\delta_{\rm IR}({\rm MTj})$ in
Eq.~(\ref{eq.delta_finite}). Although exponentiation is strictly only justified
for the soft photon emission correction, it is conventionally applied to all
RCs.

Gramolin and Nikolenko~\cite{Gramolin2016} reanalyzed the SLAC
data~\cite{walker1994, Andivahis1994}, which used the standard RCs of Mo and
Tsai~\cite{mo1969}, to include improvements to $\delta_{\rm brems}$ as well as
the use of the standard RCs of Maximon and Tjon~\cite{maximon2000}. Their Born
cross section can be written in terms of that given in
Refs.~\cite{walker1994,Andivahis1994} as
\bea
\left(\sigma_R^{\rm Born}\right)^{\rm new}\,
=\, \frac{C_{\rm RC}^{\rm old}}{C_{\rm RC}^{\rm new}}
    \left(\sigma_R^{\rm Born}\right)^{\rm old}.
\eea
The ratio $C_{\rm RC}^{\rm old}/C_{\rm RC}^{\rm new}$ is tabulated for the SLAC
data in Ref.~\cite{Gramolin2016}, to which we add our calculated TPE
contribution $\delta$. For the Super-Rosenbluth data~\cite{qattan2005} details
of the RCs that were applied are not available, so the improvements made to
$\delta_{\rm RC}$ are restricted to using $\delta_{\rm IR}({\rm MTj})$ instead
of $\delta_{\rm IR}({\rm MTs})$.

\begin{figure}[t]%
\graphicspath{{Images/}}
\includegraphics[width=10.cm]{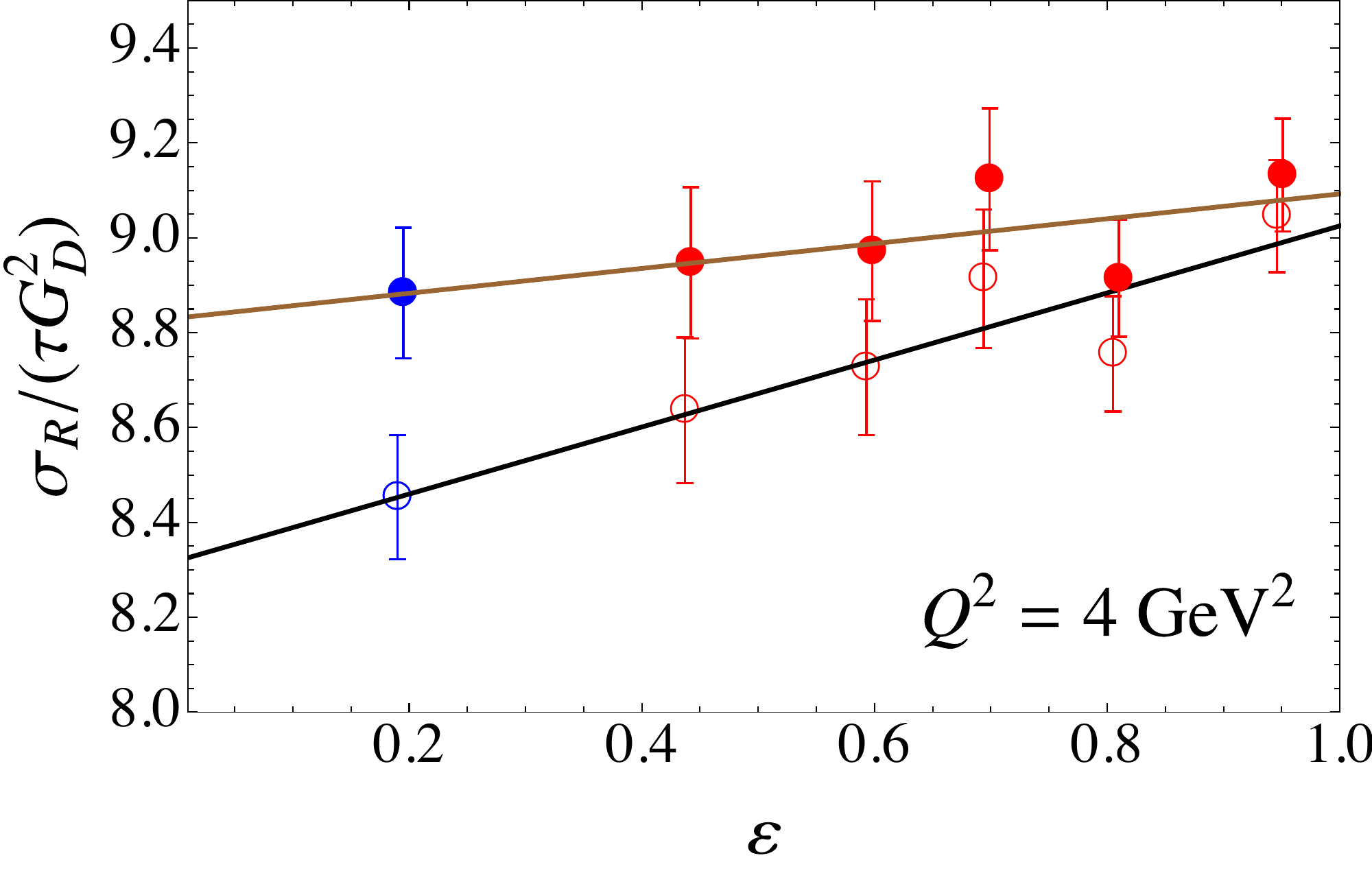}
\caption{Reduced cross section $\sigma_R^{\rm Born}$ at $Q^2=4$~GeV$^2$, scaled
by $\tau$ times the dipole form factor squared $G_D^2$. Open circles are the
original data points from Ref.~\cite{Andivahis1994}. Filled circles (slightly
offset for clarity) include improved standard RCs from Ref.~\cite{Gramolin2016},
together with the TPE corrections from the present work. The weighted least
squares fits (solid lines) determine $G_E^2$ and $G_M^2$. Data points from the
8~GeV spectrometer are shown in red, while the data point from the 1.6~GeV
spectrometer (which is separately normalized~\cite{Andivahis1994}) is shown in
blue.}
\label{fig.SigRedQ4}
\end{figure}

\begin{figure}[t]
\graphicspath{{Images/}}
\includegraphics[width=8.cm]{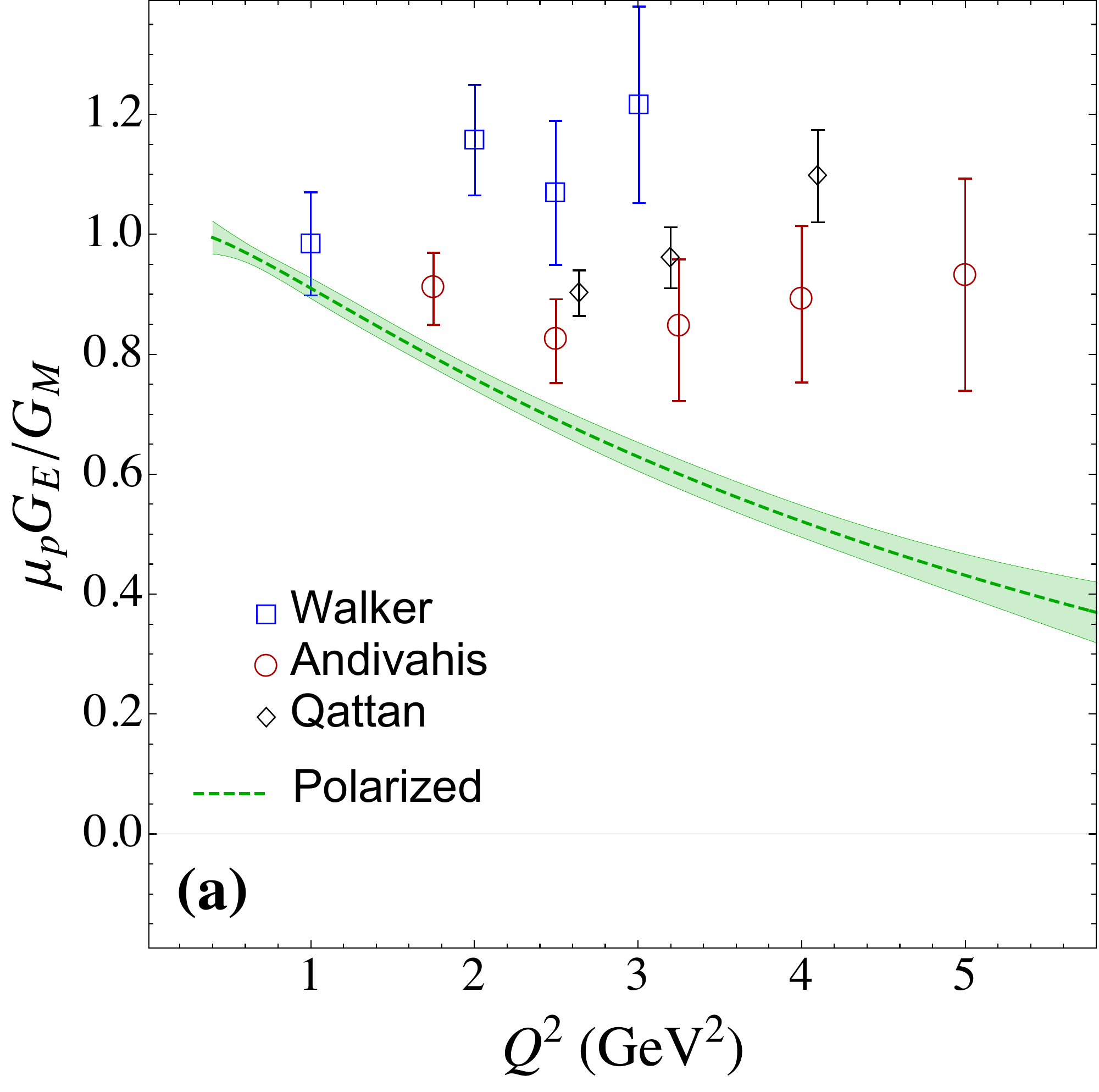} \hspace*{0.cm}
\includegraphics[width=8.cm]{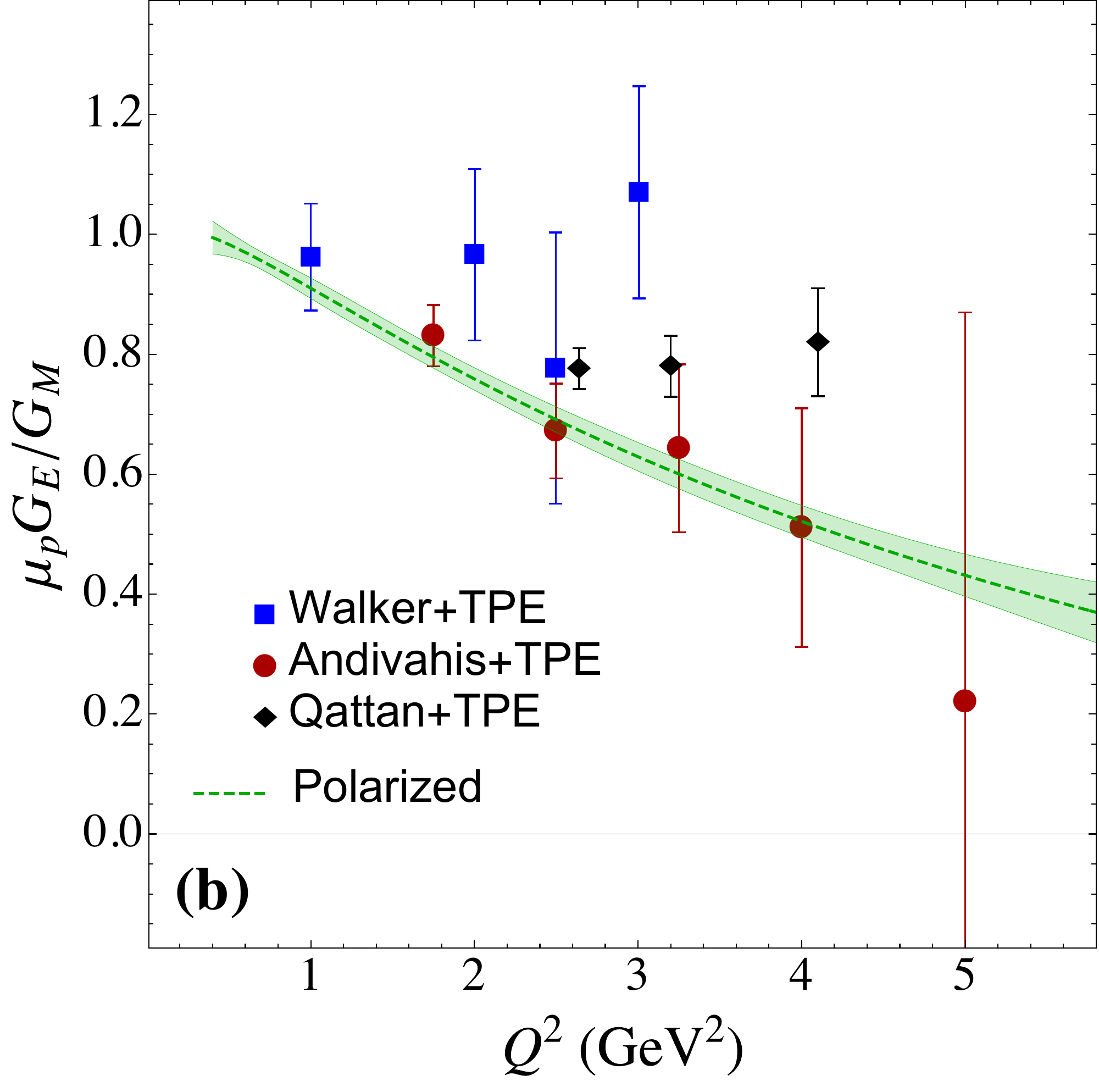}
\caption{{\bf (a)} Ratio of the proton electric to magnetic form factors,
$\mu_p\, G_E/G_M$, versus $Q^2$, extracted using LT separation
data~\cite{walker1994, Andivahis1994, qattan2005}. A nonlinear fit to the
combined PT results~\cite{jones2000g, gayou2002, punjabi2005, puckett2010,
puckett2012} at the 99\% confidence limit is shown by the green band.
{\bf (b)} The ratio $\mu_p\, G_E/G_M$ extracted from a reanalysis of the LT data
using improved standard RCs from Ref.~\cite{Gramolin2016}, together with the TPE
effects from the present work.}
\label{fig.GeGmRatio}
\end{figure}

A comparison of the original reduced cross sections and the results with the
improved RCs of Ref.~\cite{Gramolin2016} plus our TPE is shown in
Fig.~\ref{fig.SigRedQ4} for the $Q^2=4$~GeV$^2$ data from
Ref.~\cite{Andivahis1994}. We note that the original and the TPE-corrected data
are equally well described by a linear dependence on $\varepsilon$, and no
nonlinearity effects are apparent.

In Fig.~\ref{fig.GeGmRatio} we show the $G_E/G_M$ ratio extracted from our
analysis for the SLAC~\cite{walker1994, Andivahis1994} and Jefferson Lab
Super-Rosenbluth~\cite{qattan2005} experiments up to $Q^2=5$~GeV$^2$. To avoid
clutter, the PT data from Refs.~\cite{jones2000g, gayou2002, punjabi2005,
puckett2010, puckett2012} are shown as a band, which is a nonlinear fit at the
99\% confidence limit. The original analysis, shown in
Fig.~\ref{fig.GeGmRatio}(a), is consistent with $\mu_p G_E/G_M \approx 1$, while
a progressively larger effect of TPE with increasing $Q^2$ for all LT data sets
is seen in Fig.~\ref{fig.GeGmRatio}(b), with a commensurate increase in the
uncertainty of $G_E$. 
In particular the LT data of Andivahis~{\it et al.}~\cite{Andivahis1994}
are striking in their consistency with the PT band, with a near linear falloff of
$G_E/G_M$ with $Q^2$. These results provide compelling evidence that there is no
inconsistency between the LT and PT data once improvements in the RCs and TPE
effects are made.

\section{Conclusions}
\label{sec.conclusion}

In this study we have applied the recently developed dispersive formalism of
Ref.~\cite{blunden2017} to compute the TPE corrections to elastic
electron-proton cross sections, including for the first time contributions from
all $J^P = 1/2^\pm$ and $3/2^\pm$ excited intermediate state resonances with
mass below 1.8~GeV. For the resonance electrocouplings at the hadronic vertices
we employed newly extracted helicity amplitudes from the analysis of CLAS meson
electroproduction data at $Q^2 \lesssim 5$~GeV$^2$ \cite{HillerBlin:2019hhz,
mokeev2012, mokeev2009}.

To assess the model dependence of the resonance calculations, we investigated
the effects of finite Breit-Wigner resonance widths, comparing the TPE results
for the pointlike, constant width and variable width approximations. We found
that for the pointlike case kinematical thresholds produce artificial cusps at
specific values of $Q^2$ and $\varepsilon$, however, these are effectively
smoothed out across all kinematics when a nonzero width is introduced. The
effect of using a constant or dynamical width was less dramatic, with the latter
reducing somewhat the magnitude of some of the low-lying resonances, such as the
$\Delta(1232)$, at low $Q^2 \sim 1$~GeV$^2$ and at backward angles.

We also examined the spin, isospin and parity dependence of the resonance
contributions to the TPE amplitudes, finding large cancellations between the
(negative) isospin $I=1/2$ and the (positive) $I=3/2$ intermediate states, as
well as between the parity-even and parity-odd contributions. This behavior is
mostly driven by the dominance of the (positive) $\Delta(1232)~\!3/2^+$ and
(negative) $N(1520)~\!3/2^-$ contributions to the TPE amplitudes, especially at
larger $Q^2$ values.

More specifically for the individual hadronic intermediate states, at low $Q^2$,
$Q^2 \lesssim 1$~GeV$^2$, the nucleon elastic state dominates, with
contributions from excited states there mostly negligible. For $Q^2 \approx$ (1
-- 2)~GeV$^2$, the $\Delta(1232)$ resonance starts to play a more important
role, and here the sum of $N+\Delta(1232)$ provides a good approximation to the
total TPE amplitude. At still larger~$Q^2$, the $N(1520)$ gives the largest
contribution among the higher-mass resonances, exceeding even the nucleon
component for $Q^2 \gtrsim 4$~GeV$^2$. The higher-mass resonances each grow with
increasing $Q^2$, but enter with different signs and largely cancel each other's
contributions. Compared to the nucleon elastic component alone, the resonance
excitations give rise to an overall enhancement of the TPE cross section
correction for $Q^2 \gtrsim 3$~GeV$^2$.

The excited state resonance contributions generally provide some improvement of
the phenomenological description of observables that are sensitive to TPE
corrections, such as the ratios of $e^+ p$ to $e^- p$ elastic cross sections
measured recently in dedicated experiments at Jefferson Lab~\cite{CLASTPE2017},
Novosibirsk~\cite{VEPP2015} and DESY~\cite{OLYMPUS2017}. Unfortunately, most of
these data are in kinematic regions where resonance contributions are not large,
and in some cases the results are consistent with no TPE effect within the
experimental uncertainty. On the other hand, the resolution of the $G_E/G_M$
ratio discrepancy with the inclusion of the TPE corrections, especially for the
cross section data of Andivahis {\it et al.}~\cite{Andivahis1994}, compels a
global reanalysis of the LT data with inclusion of all other radiative
corrections and TPE at the same level.

Improvements on the theoretical front should involve exploration of the effects
from spin-5/2 intermediate resonant states, as well as incorporation of
nonresonant contributions \cite{tomalak2017} at larger $Q^2$ values. Future
precision measurements at higher $Q^2$ values and backward angles (small
$\varepsilon$), where the TPE effects are expected to be most significant, would
be helpful for better constraining the TPE calculations. This would provide a
more complete understanding of the relevance of TPE in the resolution of the
proton's $G_E/G_M$ form factor ratio puzzle, and better elucidate the role of
multi-photon effects in electron scattering in general.

\acknowledgements

We thank V.~Mokeev for communications about the CLAS electrocoupling data. This
work was supported by the Natural Sciences and Engineering Research Council
(Canada), and the US Department of Energy contract DE-AC05-06OR23177, under
which Jefferson Science Associates, LLC operates Jefferson Lab. JA acknowledges
funding from the University of Manitoba Graduate Fellowship and the Sir Gordon
Wu Scholarship.

\bibliography{Ref}

\end{document}